\documentclass[12pt,twoside]{book}
\usepackage{epsfig,latexsym,fancyheadings,amssymb,pifont,shadow}
\renewcommand{\chaptermark}[1]%
               {\markboth{#1}{#1}}
\renewcommand{\sectionmark}[1]%
               {\markright{\thesection\ #1}}
\def\ap#1#2#3{Ann.\ Phys.\ (NY) #1 (19#3) #2}
\def\cmp#1#2#3{Comm.\ Math.\ Phys.\ #1 (19#3) #2}

\def\np#1#2#3{Nucl.\ Phys.\ B#1 (19#3) #2}
\def\zp#1#2#3{Z.\ Phys.\ C #1 (19#3) #2}
\def\pl#1#2#3{Phys.\ Lett.\ #1B (19#3) #2}
\def\pr#1#2#3{Phys.\ Rev.\ D #1 (19#3) #2}
\def\prb#1#2#3{Phys.\ Rev.\ B #1 (19#3) #2}
\def\prep#1#2#3{Phys.\ Rep.\ #1 (19#3) #2}

\def\lnc#1#2#3{Lettere al Nuovo Cimento #1 (19#3) #2}
\def\jmp#1#2#3{J.\ Math.\ Phys.\ #1 (19#3) #2}
\def\nc#1#2#3{Il Nuovo Cimento #1A (19#3) #2}
\def\apa#1#2#3{Acta Phy. Austriaca #1 (19#3) #2}

\def\lnc#1#2#3{Lettere al Nuovo Cimento #1 (19#3) #2}

\def\re#1{(\ref{#1})}
\def\beq{\begin{equation}}
\def\eeq{\end{equation}}
\def\beeq{\begin{eqnarray}}
\def\beeqn{\begin{eqnarray*}}
\def\eeeq{\end{eqnarray}}
\def\eeeqn{\end{eqnarray*}}
\def\bit{\begin{itemize}}
\def\eit{\end{itemize}}
\def\ben{\begin{enumerate}}
\def\een{\end{enumerate}}
\def\nome#1{{\label{#1}}}
\def\ltap{\raisebox{-.4ex}{\rlap{$\sim$}} \raisebox{.4ex}{$<$}}

\def\Maxlp{\raisebox{-1.5 ex}{\rlap{\tiny $\;\;p_i^2\le c\l^2$}}
\raisebox{0ex} {$\; \mbox{Max}\;\;\;\,$}}

\def\a{\alpha}
\def\b{\beta}
\def\g{\gamma}                  \def\G{\Gamma}
\def\de{\delta}                 \def\D{\Delta}
\def\eps{\varepsilon}

\def\l{\lambda}                 \def\L{\Lambda}
\def\m{\mu}
\def\n{\nu}
\def\ta{\tau}
\def\r{\rho}
\def\s{\sigma}                  \def\S{\Sigma}
\def\th{\theta}

\def\z{\zeta} 
\renewcommand{\AA}{{\cal A}}
\newcommand{\DD}{{\cal D}}
\renewcommand{\SS}{{\cal S}}
\newcommand{\OO}{{\cal O}}
\newcommand{\WW}{{\cal W}}
\newcommand{\LL}{{\cal L}}
\def\d4#1{\frac {d^4 {#1} }{(2\pi)^4}}
\newcommand{\zzint}{\int d^4x\; d^2\th\,d^2\bt\;}
\newcommand{\cint}{\int d^4x\; d^2\th\;}
\newcommand{\acint}{\int d^4x\;d^2\bt\;}
\newcommand{\zint}{\int_{z}}
\newcommand{\spint}{\int_{p}}
\newcommand{\spqint}{\int_{pq}}
\newcommand{\sppint}{\int_{p'}}
\newcommand{\skint}{\int_{k}}
\newcommand{\sqint}{\int_{q}}

\newcommand{\lp}{\left(}
\newcommand{\rp}{\right)}
\renewcommand{\lq}{\left[}
\renewcommand{\rq}{\right]}
\newcommand{\lgr}{\left\{}
\newcommand{\rgr}{\right\}}
\newcommand{\identity}{1\hspace{-0.4em}1}
\newcommand{\no}{\nonumber}
\newcommand{\ph}{\phantom} 
\def\abs#1{\left|#1\right|}
\def\tr{\,\mbox{Tr}\,}
\def\frac#1#2{ {{#1} \over {#2} }}
\def\half{\mbox{\small $\frac{1}{2}$}}
\def\p{\partial}
\def\partder#1{{\partial   \over\partial #1}}
\newcommand{\dpad}[2]{{\displaystyle{\frac{\partial #1}{\partial #2}}}}
\newcommand{\dfud}[2]{{\displaystyle{\frac{\delta #1}{\delta #2}}}}
\def\ie{\hbox{\it i.e.}{ }}      

\def\eg{\hbox{\it e.g.}{ }}
      
\newcommand{\sde}{\de^8}
\newcommand{\ad}{{\dot\a}}  \newcommand{\bd}{{\dot\b}}  
\newcommand{\BD}{{\bar{\D}}}
\newcommand{\hD}{{\hat{\D}}}
\newcommand{\bJ}{{\bar{J}}}
\def\bc{\bar c}
\def\cp{c_{+}}
\def\cm{c_{-}}
\def\bcp{\bar c_{+}}
\def\bcm{\bar c_{-}}
\def\bchi{\bar \chi}
\def\bxi{\bar\xi}
\def\bp{\bar p}

\def\bpsi{\bar \psi}
\def\bphi{\bar \phi}
\def\bl{\bar \lambda}

\def\bs{\bar \sigma}

\def\tgv{\tilde\g_{\scriptscriptstyle{V}}}
\def\gv{\g_{\scriptscriptstyle{V}}}
\def\bs{\bar\s}
\def\bt{\bar\th}
\def\bD{\bar{D}}
\newcommand{\smuaad}{\s^\m_{\a\ad}}
\def\ei{\varepsilon_i}
\def\ej{\varepsilon_j}
\def\ek{\varepsilon_k}
\def\el{\varepsilon_l}

\def\Wi{W^{\mbox{\scriptsize{int}}}}
\def\se{S_{\mbox{\footnotesize{eff}}}}
\def\set{S_{\mbox{\scriptsize{eff}}}}

\def\ser{S_{\mbox{\footnotesize{eff,rel}}}}
\def\st{S_{\mbox{\footnotesize{tot}}}}
\def\si{S_{\mbox{\scriptsize{int}}}}
\def\scl{S_{\mbox{\scriptsize{cl}}}}
\def\sgf{S_{\mbox{\scriptsize{gf}}}}
\def\sfp{S_{\mbox{\scriptsize{FP}}}}
\def\sbrs{S_{\mbox{\scriptsize{BRS}}}}

\def\Pir{\Pi_{\mbox{\footnotesize{rel}}}}

\def\Piinv{\Pi_{\mbox{\footnotesize{inv}}}}
\def\Pit{\tilde\Pi}
\def\bG{\bar\Gamma}
\def\Gr{\G_{\mbox{\footnotesize{rel}}}}

\def\Gir{\G_{\mbox{\footnotesize{irr}}}}
\def\Gi{\G^{\mbox{\scriptsize{int}}}}
\def\Gio{\G^{\mbox{\scriptsize{int}}(0)}}
\def\De{\D_{\mbox{\footnotesize{eff}}}}
\def\Der{\D_{\mbox{\footnotesize{eff,rel}}}}

\def\DG{\D_{\G}}

\def\DGi{\D_{\G,\mbox{\footnotesize{irr}}}}
\def\DGb{\bar{\D}_{\G}}
\def\DGh{\hat{\D}_{\G}}
\def\DGhr{{\hat{\D}}_{\G,\mbox{\footnotesize{rel}}}}

\def\mdm{\m\,\p_\m}
\def\LdL{\L\partial_\L}

\def\UV{$\L_0\to\infty\;$}
\def\IR{$\L\to 0\;$}
\def\K{K_{\L\L_0}}
\def\Ki{K_{0\L}}
\def\Kiu{K_{0\L_{0}}}
\def\Kin{K_{\L_{0}\infty}}

\def\gm{\gamma_{\mu}}
\def\gn{\gamma_{\nu}}
\def\gr{\gamma_{\rho}}
\def\gs{\gamma_{\sigma}}

\def\gc{\gamma_5}
\def\ds#1{\ooalign{$\hfil/\hfil$\crcr$#1$}}
\def\e#1#2#3#4{\varepsilon^{#1 #2 #3 #4}}
\def\t,#1{t^{#1}}
\def\f,#1#2#3{f^{#1 #2 #3}} 
\def\bom#1{\mbox{\boldmath$#1$}}

\begin{document}



\begin{titlepage}


\vspace{-9cm}

\begin{center}
{\LARGE \bf Federica Vian} \\[0.5cm]

\end{center}

\vskip 2truecm

\begin{center} 
\LARGE \bf
Quantum Field Theories with Symmetries

in the 

Wilsonian Exact Renormalization Group
\end{center}

\vskip 2truecm
\begin{center}
{\large \rm  Ph.D. Thesis } \\[1.5cm]
\end{center}
\begin{figure}[htbp]
\epsfysize=4cm
\begin{center}
\epsfbox{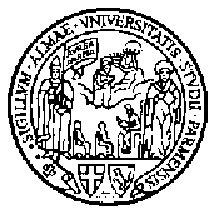}
\end{center}
\end{figure}

\vspace{1cm}

\begin{center}
{\Large \bf Universit\`a degli Studi di Parma }

{\large   PARMA - 1998 } \\[0.5cm]

\end{center}

\clearpage

\end{titlepage}


\mainmatter

\tableofcontents

\addcontentsline{toc}{chapter}{Introduction}

\chapter*{Introduction}

The main goal of renormalization from a traditional point of view is
to  determine when and how  the cancellation of ultraviolet (UV)
divergences in quantum field theory occurs. Such a cancellation is
essential if a theory has to yield quantitative physical
predictions. What is not obvious is how the quantum fluctuations
associated with extremely short distances (\ie very high momenta)
can be so harmless to affect a theory only through the values of a few
of its parameters, typically the bare masses and coupling constants or
the counterterms in renormalized perturbation theory.

Another picture for renormalization can be conceived, and the idea is
due to K. Wilson \cite{w}. He suggested that all of the parameters of a
renormalizable field theory can be thought of as scale dependent
objects and such a scale dependence is described by 
differential equations, the so-called {\it renormalization group} (RG)
equations or flow equations.
The RG method is based on the functional integral approach to field theory
and the origin of the ultraviolet (UV) divergences is perused by
isolating in the functional integral the short-distance degrees of
freedom of the field. Actually in the generating functional $Z[J]$ the
basic integration variables are the Fourier components $\phi(k)$ of
the field, namely $Z[J]$ is expressed by
$$
Z[J]= \int \DD\phi e^{i\int[\LL+J\phi]}= 
\lp \prod_k \int d\,\phi(k)\rp \,e^{i\int[\LL+J\phi]}\,.
$$
In order to cure the ultraviolet divergences, a sharp UV cutoff $M$ is
imposed. This means we integrate only over the fields  $\phi(k)$ with
$\abs{k}\le M $ and set $\phi(k)=0 $ for $\abs{k}\ge M $ so that in the real
space the fields are defined on a lattice of spacing $2\,\pi/M$.
According to Wilson, the fundamental fields are replaced by their averages
over a certain space-time volume (blockspin transformations) and thus
defined on a coarser lattice. By this averaging process small scale
fluctuations which correspond to high frequencies are eliminated.

Rather than in the discrete Wilson RG, we are interested in the
continuous Wilsonian RG \cite{p}-\cite{b}. The idea behind it is very 
similar to that of
the discrete case: in the generating functional (partition function in
the Euclidean) we do not integrate over all momenta in one go, but we
first integrate out modes between a cutoff scale $\L_0$  (UV) and a very
much lower energy scale $\L$. What is left over ---integration between 
$\L$ and zero--- may still be expressed as a generating functional but
the bare action is replaced by a very complicated effective action,
$\se$, containing an infinite series of non-local terms. This is
naturally interpreted as the high frequency modes of the fields 
generating  effective couplings for the low-energy modes. However, we
expect the behaviour at small scales to be controlled only by a finite
number of parameters, \ie the ``relevant parameters'' (with
non-negative mass dimension). Deviations from locality will be of
order $\L/\L_0$. Thus, when the relevant parameters have been fixed at
low energies, the dependence on $\L_0$ will be given by powers of 
 $\L/\L_0$ at any order in perturbation theory. 
Moreover, by requiring the physical Green functions to be independent
of the cutoff $\L$, it follows the functional $\se$ obeys an evolution
equation. Hence, the evolution equation with a suitable set of
boundary conditions ---which encode both renormalizability and the
renormalization conditions--- can be thought of  as an alternative
definition of a theory.

From what we have seen so far we should be driven to view the 
RG formulation as a natural setting for the analysis of effective
theories \cite{pich}. Effective theories are very popular nowadays: Chiral
Perturbation Theory \cite{wein},  Heavy Quark Effective Theory
\cite{eich}, low energy $N=2$ Super Yang Mills \cite{seib} are just a
few examples. 
Even though the dream of modern physics is to achieve a simple 
understanding of
all the observed phenomena in terms of some fundamental dynamics
(unification), 
assuming a theory of everything appeared at some point, the description
of nature at all physical scales would have little to do with a
quantitative analysis at the most elementary level. Therefore, in
order to study a particular physical system in a huge surrounding
world,
the key issue is to identify and  pick up the most appropriate
variables. 

Usually,  a physics problem involves widely separated energy
scales. The basic idea is to identify those parameters which are very
large (small) compared with the energy scale of the system and to set
them to infinity (zero).  A sensible description of the system would
obviously consider the corrections induced by the neglected energy
scales as small perturbations.
Effective field theories are the appropriate theoretical tool to
examine low-energy physics, where low is referred to some energy scale
$M$ ($\L_0$ in the RG). Only the relevant degrees of freedom, 
\ie the states with $k<< M$
 are kept, whereas the heavier excitations with  $k>> M$ are
integrated out from the action. The by-product of such integration is
a bunch of non-renormalizable interactions among the light states,
which can be expanded in powers of $E/M$, $E$ being the energy.
Thus the information on the heavier degrees of freedom is  stored in
the couplings of the low-energy Lagrangian.
Although an effective field theory contains an infinite  number of
interactions, renormalizability can be trusted since, at a given order
in the expansion in $E/M$ ($\L/\L_0$ in the RG), the low-energy 
theory is specified by a
finite number of couplings.

We naturally expects the effective theory keeps track of the
symmetries of the fundamental theory. Global symmetries, such as
Lorentz  invariance, isotopic spin invariance and so on, are
automatically maintained in the RG method. 
It is certainly not so for gauge symmetries. In fact the division of
momenta into large or small (according to some scale $\L$) ---which is
fundamental in the RG approach--- is not preserved by gauge
transformations, since in the momentum space the field is mapped into
a convolution with the element of the gauge group.
We are forced to conclude that the symmetry of the fundamental theory
is lost 
at the effective level.
Nevertheless, a remainder of the original invariance survives  in the
form of an effective symmetry which constraints the flow of all the
couplings of the theory at the scale $\L$. Unfortunately, the task of
solving the relations  among the couplings coming from those constraints 
is impossible to carry out, due to non-linearity.
Therefore one is left with two options: either work in
non-perturbative field theory by means of an analytic
approximation or solve equations in the perturbative regime.
In the former case we have to face the unpleasant aspect that there is
no known truncation consistent with gauge invariance and the best one
can do is to give a numerical estimate
of the symmetry breaking term by using effective Ward identities
\cite{ellw}.
In this thesis we will choose the latter option and the implementation
of symmetries in perturbation theory will be extensively treated.

Even though the topics of the thesis will be discussed at the
perturbative level, we must recall for completeness that the RG
formulation is in principle  non perturbative. Clearly analytic
approximation methods must be employed in non-perturbative quantum
field theory where there are no small parameters to expand in. In this
direction much progress has been made. Let us just mention the
applications to chiral symmetry breaking, phase transitions, 
finite temperature, large $N$ limit and to many other
sectors. For a review see \cite{morbib}. 
Two major problems affect non-perturbative RG.  Of the first of these,
that any known truncation violates   gauge invariance, we have
already said. 
The second problem is the lack of a
recipe to evaluate errors in a certain approximation scheme. 

We now present the outline of the thesis.
In the first chapter we will introduce the Wilsonian Exact
Renormalization Group for a general theory (\ie containing both bosons
and fermions, scalars and vectors). The procedure of integrating out
the modes with frequency above $\L^2$ and below $\L_0^2$ will be
performed  multiplying the quadratic part of the classical action by a
cutoff function which is one between  $\L^2$ and $\L_0^2$ and rapidly vanishes
outside this interval. We will then derive the RG flow  by requiring
the generating functional of the theory to be independent of the
infrared cutoff $\L$. 

As an example of how the RG method works, in
chapter 2 we will apply it to the massless scalar theory. We
will see in details how an iterative solution of the  flow equation,
together with a set af suitable boundary conditions, provides the
usual loop expansion. Furthermore we will explicitly compute the
one-loop two-point and four-point vertices. For this theory we will also prove
perturbative renormalizability, \ie the existence of the \UV limit,
and infrared finiteness, that is the vertex functions at
non-exceptional momenta are finite order by order in perturbation
theory.

The third chapter will be devoted to establishing the Quantum Action Principle
(QAP) in the RG and we will show that the Slavnov-Taylor (ST) identities, which
completely characterize the classical theory, can be directly formulated
for the cutoff effective action at any $\L$. Afterwards we will use
these effective identities to fix the couplings in the bare action.

In the fourth chapter the QAP will be exploited to analyse the
breaking of dilatation invariance  occurring in the scalar theory in 
the RG approach. An analogue of the
Callan-Symanzik equation will be derived for the cutoff
effective action and from the effective Ward identities of dilatation
the one-loop beta function for such a theory will be reproduced.

In the fifth chapter we will address $SU(N)$ Yang-Mills theory. After 
deriving the evolution equation, we will treat the
key issue of boundary conditions which, in this case, have also to
ensure restoration of symmetry for the physical theory when the
cutoffs are removed (in the limits \IR and \UV). We will then use the 
effective ST identities to  derive some of the bare couplings (fine-tuning) at
the first loop.
The next step will be  the extension of the RG method to chiral gauge
theory. 
This will be the subject of chapter 6. 
Since in the RG formulation the space-time dimension is four, there is
no ambiguity in the definition of the matrix $\g_5$ and in the
regularized action left and right fermions will not be coupled. 
Therefore the solution of the fine-tuning procedure we be simpler
than in the standard case (global chiral symmetry is preserved) and
will be explicitly performed.
We will then show how the chiral anomaly can be obtained in the RG.

Having gone through non-supersymmetric theories, chapter 7 will be
dedicated to  extending the RG formulation  to supersymmetric
theories.
Regularization will be implemented in such a way that supersymmetry is
preserved.  Actually, it  suffices to write  the classical action
in terms of superfields and multiply the propagators by the same cutoff
function. In components this corresponds to use the same
cutoff for all fields. We will start with the Wess-Zumino
model to set up the formalism and then, in chapter 8, we will approach
supersymmetric gauge theories. We will solve the fine-tuning equation
at the first loop and show how the gauge anomaly can be derived.
Finally, the appendix contains the supersymmetric conventions.

\chapter{Wilson Renormalization Group}

\section{Wilson effective action}

To start with, we will acquaint with the Wilson renormalization
group (RG) ---or exact renormalization group--- formulation of a theory.
The basic idea of Wilson RG \cite{w} 
is to regard an interacting field theory 
as an effective theory, that is to consider the high frequency modes of the 
fields of the theory as generating 
effective couplings for the low energy modes. 
In this picture one introduces an UV cutoff $\L_0$ to make the Green
functions of the theory finite in the ultraviolet region. 
Then a scale $\L$ is introduced and the
frequencies between $\L$ and $\L_0$ are viewed as generating
interactions for the
frequencies lower than $\L$. 

As we will see throughout the following chapters, the RG approach can
be applied to a wide variety of theories, so that a general formulation
is in order. To do so, 
we will denote with $\Phi_a=\{\phi,\psi,\bpsi\}$ the fields of the 
theory (the $\phi$'s are commuting fields while the $\bpsi$, $\psi$ are
anticommuting, fermions or ghosts) and $J_a=\{j,\bchi,-\chi\}$ the 
corresponding sources in the generating functional.
Let us suppose the classical action $\scl$~\footnote{Here we mean  the classical
action contains the gauge-fixing term and the corresponding
Faddeev-Popov term.} is 
 invariant under the
infinitesimal local transformation of the field $\Phi$
\beq\nome{gaugetr}
\de\Phi_a(x)=\eta \,P_a[\Phi(x)]\,,
\eeq
where the $P_a[\Phi]$ are (anticommuting)
polynomials in the fields
and $\eta$ is an anticommuting parameter. Introducing the sources
$\g_a$, associated to the composite operators defining the symmetry 
transformations of $\Phi_a$, we can write the BRS action
$$
\sbrs [\Phi,\g]=\scl[\Phi]+\int_x \, \g_a\, P_a[\Phi]  \,.
$$

In order to quantize the theory one needs a regularization procedure
of the ultraviolet divergences. Such divergences are regularized by
assuming that in the path integral only the fields with frequencies
smaller then a given UV cutoff $\L_0$ are integrated. This procedure
is equivalent to assume the free propagators vanish for $p^2 >
\L_0^2$.
The generating functional is 
\beq\nome{Z}
Z[J,\g]=e^{iW[J,\g]}=\int {\mathcal D}\Phi \, \exp i {\left\{
-\half(\Phi, \,D^{-1}\Phi)_{0\L_0}+
(J,\,\Phi)_{0\L_0} +\si[\Phi,\g;\L_0]
\right\}}
\,,
\eeq
where the free propagators are collected in the matrix
$D^{-1}_{ab}$
and, more generally, we have introduced the cutoff scalar
product between fields and sources~\footnote{Summations over internal indices are understood.}
\beeq\nome{scalarprod}
&&(\Phi, \,D^{-1}\Phi)_{\L\L_0}
\equiv
\int_p\, K^{-1}_{\L\L_0}(p)\,\Phi_a(-p)\,D^{-1}_{ab} \Phi_b(p)
\,,
\;\;\;\;\;\;\;\;\;\;\;\;\;\;\;\;\;\;
\int_p \equiv \int \frac{d^4p}{(2\pi)^4}\,,\nonumber \\
&&(J, \,\Phi)_{\L\L_0}\equiv
\int_p\, K^{-1}_{\L\L_0}(p)\,J_a(-p)\,\Phi_a(p)\,.
\eeeq
The cutoff function $K_{\L\L_0}(p)$ is one for $\L^2\le p^2\le\L_0^2$
and rapidly vanishes outside this interval and its introduction can be
responsible for a possible loss of the symmetry. Hence the UV action
$\si$ must contain all possible local, renormalizable interactions which are
needed to compensate the symmetry breaking induced by the
regularization.

According to Wilson we integrate over the high energy modes, \ie the
fields with frequencies $\L^2 < p^2 < \L_0^2$ and obtain 
\beq\nome{Z'}
Z[J,\g]=N[J,\g;\L,\L_0] \;\int {\mathcal D}\Phi \,
\exp i{\biggl\{
-\half(\Phi, \,D^{-1}\Phi)_{0\L}
+(J,\,\Phi)_{0\L}
+\se[\Phi,\g;\L,\L_0]
\biggr\}}\,,
\eeq
where the coefficient $N$ is given by
$$
\log N[J,\g;\L,\L_0] = -\frac i 2  (J, \,D J)_{0\L_0}
+\frac i 2 (J, \,D J)_{0\L} \,.
$$
The functional $\se$ is the Wilsonian effective action and contains 
the effective interaction coming from the frequencies $p^2>\L^2$.
We now prove  that this functional is equivalent to a generalization
of \re{Z}, in which the free propagators contain $\L$ as an infrared
cutoff \cite{bdm,mo}. 
The generating functional \re{Z} can be rewritten as
\beeq\nome{Z12}
&&N[J,\g;\L,\L_0] 
\int {\cal D}\Phi{\cal D}\Phi_1 \,
\exp
i\biggl\{-\half (\Phi,\, D^{-1}\Phi)_{0\L}
-\half (\Phi_1,\, D^{-1} \Phi_1)_{\L\L_0}
+(J,\Phi)_{0\L} \no \\
&&\phantom{N[J,\g;\L,\L_0] \int {\cal D}\Phi{\cal D}\Phi_1 \, \exp i\bigl\{,}
+ \si[\Phi+\Phi_1,\g;\L_0]\biggr\}\,,
\eeeq
where
$$
K_{0\L_0}(p)\;= K_{0\L}(p)\;+\;K_{\L\L_0}(p)\,.
$$
This can be easily seen by performing  the change of
variables $\Phi_1=\Phi'-\Phi$ in \re{Z12}, which gives
\beeq\nome{Z''}
&&N[J,\g;\L,\L_0] \;
\int {\cal D}\Phi'
\exp{i\{-\half (\Phi',\, D^{-1} \Phi')_{\L\L_0}+\si[\Phi',\g;\L_0]\}} \\
&& \phantom{N[J,\g;}
\times
\int {\cal D}\Phi \,
\exp{i\{-\half (\Phi,\, D^{-1} \Phi)_{0\L}
-\half (\Phi,\, D^{-1} \Phi)_{\L\L_0}
+(J_1,\Phi)_{\L\L_0}
+(J,\Phi)_{0\L}\}}\no \,,
\eeeq
where the source $J_1(p)$ is
$$
J_1(p)=D^{-1}(p)\, \Phi'(p)\,.
$$
Then we carry out the integration over the field $\Phi$, which is
gaussian, and get \re{Z}.  On the other hand, by integrating over the
field $\Phi'$ in equation \re{Z''} we get back the definition of the
Wilsonian effective action $\se$ given in \re{Z'}.  

The comparison
between \re{Z'} and \re{Z''} provides us with a further definition of
$\se$
\beeq\nome{esse}
&&\exp {i\{\half (\Phi, \,D^{-1}\Phi)_{\L\L_0} +
\se[\Phi,\g;\L,\L_0]\}}\nonumber \\
&&=
\int {\cal D}\Phi' \,
\exp i\{-\half (\Phi', \,D^{-1} \Phi')_{\L\L_0}+
(J', \Phi' )_{\L\L_0}+\si[\Phi',\g;\L_0]\} \,,
\eeeq
where the source is
\beq\nome{j'}
J'(p)=D^{-1}(p)\,\Phi (p)\,.
\eeq
Finally, 
\re{Z} and \re{esse} put together allow us to see  the functional $\se$
in a new perspective, that is to say as a 
generator of  the connected Green functions ---apart for the tree-level two-point functions---
in which the internal
propagators have frequencies in the range
$\L^2<p^2<\L_0^2$. In other words, the functional~\footnote{Here and
in the following we explicitly write only the dependence on the cutoff
$\L$, since we expect the theory to be renormalizable and we are
interested in the limit \UV.}
\beeq\nome{ZL}
W[J',\g;\L] &\equiv&
\half (\Phi, \,D^{-1} \Phi)_{\L\L_0} + \se[\Phi,\g;\L]\\
J_a'(-p)&=&\K^{-1}(p)\,\Phi_b(-p)\, D^{-1}_{ba}(p) \nonumber
\eeeq
is the generating functional of the connected
amputated cutoff Green function, since the factor $K_{\L\L_0}^{-1} \,
D^{-1}$ in $W$ cancels out the external free propagators of the Green
functions.

\section{The RG flow}
By integrating out the modes over a fixed scale $\L$ we have come to
the definition of the effective action $\se$. If we consider such an
action as pertaining to a low-energy theory, we are naturally driven
to analyse the evolution of $\se$ in the infrared cutoff $\L$.

The requirement that the generating functional \re{Z'} is independent 
of the IR cutoff $\L$ gives rise to a differential equation for the
Wilsonian effective action, the so-called exact RG equation
\cite{w,p,b}
\beq\nome{erg1}
\LdL \frac{\se[\Phi,\g;\L
]}\hbar=(2\pi)^8\, \frac{\hbar}2 \int_p\, 
\LdL  \Ki (p) \, e^{-i\frac{\set}{\hbar}} D_{ab}(p)\,\frac{\de^2}{
\de\Phi_a(-p) \, \de\Phi_b(p)}\,e^{i\frac{\set}{\hbar}} \,,
\eeq
which can be translated into an equation for $W[J,\g;\L]$ 
\beq\nome{eveqW}
\L\p_{\L} W[J,\g;\L]=\half\spint \L\p_{\L} \K^{-1}(p)\,D^{-1}_{ab}(p)
\lp
\frac{\de W}{\de J_a(-p)}\frac{\de W}{\de J_b(p)}
-i \,\frac{\de^2 W}{\de J_a(-p) \de J_b(p)}
\rp
\,.
\eeq
This equation can be more easily understood taking into account that
$\L$ enters as an IR cutoff in the internal propagators of the cutoff
Green functions.  Furthermore, it is non-perturbative and, together
with a set of suitable boundary conditions, can be thought as an
alternative definition of the theory.  

As far as one is concerned with
its perturbative solution, the usual loop expansion is recovered by
solving iteratively \re{erg1} or \re{eveqW}. 
The solution of \re{erg1} is possible since the
evolution equation for the vertex $S_n(p_1,p_2, \cdots,p_n)$ of $\se$
at a given loop $\ell$ involves  lower loop vertices or, at
worse, vertices $S_m$ of the same order but with $m<n$. 
Therefore, in order to perform any perturbative study 
a filtration \cite{bbbcd}
(\ie the introduction of a field-counting operator) 
in the space of vertices is required and the analysis at any loop order
must be done by starting from the vertices with lower number of external 
fields.
Unfortunately this twinned recursive
procedure ---in the perturbative order and in the number of fields---
proves rather cumbersome. 
We can get rid of those troublesome vertices
of the same loop order by realizing they are 1P-reducible and so
disappear in the analysis of the generator of the cutoff 1PI functions.  

As one expects, from a technical point of view it is simpler  to study
the Legendre transform of $W[J,\g;\L
]$ 
\beq\nome{Leg}
\G[\Phi,\g;\L,\L_0]=W[J,\g;\L,\L_0]-\int_p J\Phi\,,
\eeq
which 
we call ``cutoff effective action'' and is a generalization
of the usual quantum effective action, since it contains the infrared
cutoff $\L$ in the free propagators \cite{we,bdm,mo}.
 The functional $\G$ generates the cutoff vertex functions in which the 
internal propagators have frequencies in the range $\L^2<p^2<\L_0^2$ 
and reduces to
the physical quantum effective action in the limits $\L\to 0$ and \UV.
In the following
we will show in the scalar case that both these limits
can be taken in perturbation theory \cite{p,b,bdm}.
For this reason   the dependence on
the ultraviolet cutoff $\L_0$ has been  understood.

The evolution equation for the functional $\G[\Phi,\g;\L]$ can be derived
from \re{eveqW} by using \re{Leg} and inverting the functional
$\frac{\de^2 W}{\de J\de J}$.  This inversion can be performed
isolating the full two-point contributions $\G_2$ in the
functional $\G[\Phi,\g;\L]$
$$
(2\pi)^8 \frac{\de^2\G}{\de\Phi_b(p)\,\de\Phi_c(k)}
= (2\pi)^4 \G_{2\;cb}(k;\L) \, 
\de (k+p) + \Gi_{cb}[\Phi,\g;k,p;\L] 
$$
and  $W_2$ in $W[J,\g;\L]$
\beq\nome{inversionw}
(2\pi)^8 \frac{\de^2 W}{\de J_c(-k)\,\de J_{a}(q)}
= (2\pi)^4 W_{2\;ac}(k;\L) \,
\de (q-k) + \Wi_{ac}[J;q,-k;\L]\,.
\eeq
Then making use of the identity 
\beeq
\dfud{\Phi_a(-q)}{\Phi_b(p)} &=& 
\de (q+p) \, \de_{ab}
\nonumber \\
&=& (2\pi)^8 \int_k 
\frac{\de^2 W}{\de J_c(-k) \,\de J_{a}(q)}\,
\frac{\de^2\G}{\de\Phi_b(p) \, \de\Phi_c(k)}\no
\eeeq
we can express $\Wi_{ab}$ in \re{inversionw} as a functional of 
$\Phi$ and $\g$ obtaining
\beq \nome{wint}
\Wi_{ab}[J(\Phi,\g);q,\,p;\L]= -\G_{2\;db}^{-1}(p;\L)\,
\bG_{cd}[\Phi,\g;q,\,p;\L]\,
\G_{2\;ac}^{-1}(q;\L)
\,,
\eeq
where the auxiliary functional $\bG$ satisfies the recursive equation
\beq \nome{gammab}
\bG_{ab}[\Phi,\g;q,p;\L]= (-)^{\de_b}\Gi_{ab}[\Phi,\g;q,p;\L]-\int_k 
\,\Gi_{cb}[\Phi,\g;k,p;\L]\,\G_{2\;dc}^{-1}(k;\L)\,\bG_{ad}[\Phi,\g;q,-k;\L]
\eeq
which gives $\bG$ in terms of the proper vertices of $\G$.  The
grassmannian parity $\de_a$ is one  for a fermionic field and zero otherwise
and the factor $(-)^{\de_b}$  has been introduced to take into 
account the  possible anti-commuting nature of the fields.
A graphical representation of the functional $\bG$ is given in fig.~1.
\begin{figure}[htbp]
\epsfysize=2.5cm
\begin{center}
\epsfbox{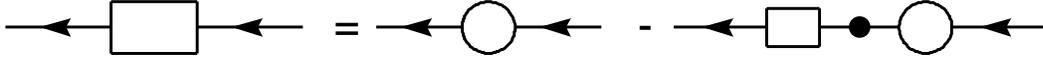}
\end{center}
\caption{\small{Graphical representation of the auxiliary functional 
$\bG$. The box and the blob represent the functionals $\bG$ and 
$\Gi$, respectively. The dot indicates a cutoff full propagator.}}
\end{figure}
\newline
Finally, inserting \re{inversionw}  in \re{eveqW} and using \re{wint},
we obtain the evolution equation for the functional $\G [\Phi,\g; \L]$ 
\beeq \nome{eveq}
&&\L\p_{\L}\lq \G [\Phi,\g; \L]-
\half \,\int_p\, \K^{-1}(p)\, \Phi_a(-p)
\,D^{-1}_{ab}(p)\,\Phi_b(p)
\rq
=-\frac i 2 \int_q \L\p_{\L} \K^{-1}(q)
\nonumber \\ 
&&\;\;\;\;\;\;\;\;\;\;\;\;\;\times\,
\G_{2\;db}^{-1}(q;\L)\;
D^{-1}_{ba}(q)\;
\G_{2\;ac}^{-1}(q\;\L)
\bG_{cd}[\Phi,\g;-q,\,q;\L]\,.
\eeeq
In this case a perturbative solution  of \re{eveq} is simpler since
the l.h.s. at a given loop order depends only on lower loop vertices.

In order to integrate the RG equations ---either \re{erg1} or
\re{eveq}--
we  have to supply the boundary conditions. 
For this reason it is useful to distinguish between 
relevant  couplings and irrelevant vertices according to their mass dimension.
One performs a Taylor expansion of the cutoff vertices around vanishing 
momenta or  around 
non-vanishing subtraction points in case there are massless fields (to
avoid infrared divergences).
The coefficients of decreasing dimension appearing in the expansion 
 are the couplings of the theory.
The ``relevant'' part is obtained by  keeping the terms with coefficients
having non-negative dimension (relevant couplings)
\beq\nome{pir}
\Pir[\Phi,\g;\s_i(\L)]=\sum_i \s_i(\L)\,P_i[\Phi,\g]\,,
\eeq
where 
\beq\nome{defpi}
\Pi[\Phi,\g;\L]=\G[\Phi,\g;\L]+\half (\Phi,\, D^{-1}\Phi)_{\L\L_0}-
(\Phi,\, D^{-1}\Phi)_{0\L_0}\,.
\eeq
With such a definition, in the \UV limit the two-point function 
does not contain the cutoff function. Should instead the relevant part of the
cutoff effective action  be given, we should isolate
it only from the interaction part since  the two-point function of
$\G[\Phi,\g;\L]$ contains the cutoff function.
The couplings $\s_i(\L)$ can be expressed either in terms of the 
cutoff vertices at vanishing momenta ---if all the fields are massive,
or at a given subtraction point when massless fields are present.
When this is the case, for the two-field components the subtraction 
point is assumed
at $p^2=\mu^2$,
while for the $N$-field components it is assumed at the symmetric point
$NSP$ defined by
\beq\nome{sp}
\bp_i\bp_j=\frac{\mu^2}{N-1}(N\de_{ij}-1)\,,\;\;\;\;\;\;\;\;N=3,4,...\,.
\eeq
The operation of extracting the relevant part of a
functional of a multicomponent massless scalar field $\psi_i$ in four
space-time dimensions can be performed via the operator
$T_4^{(\mu)}$ \cite{bdm3}-\cite{bdm5} 
\beeqn
T_4^{(\mu)}\,F[\psi] &\equiv& F[0]+\int d^4x \, \psi_i(x) \biggl\{
\bigl[\frac{\de F}{\de\psi_i(0)}\bigr]_{\psi=0} +
\frac 1 2 \psi_j(x) \bigl[
\frac{\de^2 F}{\de{\tilde \psi}_j(0)\de\psi_i(0)}\bigr]_{\psi=0} \\
&+& \frac i 2 \partial_\mu \psi_j(x) \bigl[ \partder{p_\mu}
\frac{\de^2 F}{\de{\tilde \psi}_j(p)\de\psi_i(0)}\bigr]_{p=0,\psi=0} +
\frac 1 6 \psi_j(x) \psi_k(x) \bigl[
\frac{\de^3 F}
{\de{\tilde \psi}_j(0)\de{\tilde \psi}_k(0)\de\psi_i(0)}\bigr]_{\psi=0}
\\ &-&\frac 1 2 \partial^2 \psi_j(x) \bigl[ \partder{p^2}
\frac{\de^2 F}{\de{\tilde \psi}_j(p)\de\psi_i(0)}\bigr]_{p^2=\mu^2,\psi=0}
\\ &+& \frac i 6 \psi_j(x) \partial_\mu \psi_k(x) \bigl[ \partder{p_{2\mu}}
\frac{\de^3 F}{\de{\tilde \psi}_j(p_1)\de{\tilde \psi}_k(p_2)
\de\psi_i(0)}\bigr]_{p_i=3SP,\psi=0}
\\ &+& \frac{1}{24} \psi_j(x) \psi_k(x) \psi_h(x) \bigl[
\frac{\de^4 F}
{\de{\tilde \psi}_j(p_1)\de{\tilde \psi}_k(p_2)\de{\tilde \psi}_h(p_3)
\de\psi_i(0)}\bigr]_{p_i=4SP,\psi=0} \,,
\eeeqn
with $\tilde \psi_i$ the Fourier transform  of $\psi_i$. 
For the analogous operation at vanishing momenta see ref.~\cite{b}.

The remaining part of the cutoff effective action 
 is called ``irrelevant''.
Since we expect the theory to be renormalizable, for $\L\sim\L_0$
the dimension of the irrelevant couplings should be given only by 
negative powers of $\L_0$. Thus the simplest boundary condition for the 
irrelevant part of the cutoff effective action is
\beq\nome{irr}
\Gir[\Phi,\g;\L=\L_0]=0\,.
\eeq
For $\L=\L_0$, then, the cutoff effective action becomes local and
corresponds to the bare action $\si$ in \re{Z}.

The relevant couplings are naturally set in the infrared, when most of
the degrees of freedom have been integrated out (in particular at the
physical point $\L=0$, where the cutoff effective action becomes the
physical one, so that the relevant couplings are related to 
measurable quantities). In the usual field theory language this
corresponds giving the physical renormalization conditions. In the
language of the Wilson RG, this means that the flow in the infrared
is controlled by the relevant couplings. 
In this way some of the relevant couplings
are related to physical couplings such as the wave function
normalizations and the couplings $g_i$ at a subtraction point
$\mu$. The remaining are fixed imposing the symmetry at the physical
point. This procedure is highly not trivial since one has to analyse
non-local functionals.  Alternatively we can discuss the symmetry at
the ultraviolet scale and determine $\s_i(\L=\L_0)$.  In this case the
discussion is simpler, since all functionals are relevant, but we have
to perform a perturbative calculation (\ie to solve the RG equations)
to obtain the physical couplings. Notice that while the physical
couplings are independent of the cutoff function, the bare action is
generally not.

An example of how the prescription of the boundary conditions works
will be given in the next chapter, when we implement the RG
formulation in our toy model, the scalar massless theory.


\chapter {The massless scalar case}

\section{The RG flow}

We consider a four-dimensional massless scalar field theory
with a four point-coupling $g$ at the scale $\mu$, \ie with
the two- and four-point vertex functions satisfying the conditions
\beq          \nome{norm}
\G_2(0)=0 \,,
\;\;\;\;\;\;\;\;
\left.  \frac {d \G_2(p^2)}{dp^2}\right|_{p^2=\mu^2}=1  \,,
\;\;\;\;\;\;\;\;
\G_4(\bp_1,\bp_2,\bp_3,\bp_4)=g  \, ,
\eeq
where $\bp_i$ are the momenta at $4SP$.

The generating functional of Green functions is
\beq\nome{Zscalar}
Z[j]=e^{iW[j]}=\int {\mathcal D}\phi \, \exp i {\left\{
-\half(\phi, \,D^{-1}\phi)_{0\L_0}+
(j,\,\phi)_{0\L_0} +\si[\phi;\L_0]
\right\}}
\,,
\eeq
where $D(p)=1/p^2$ is the free propagator of the massless theory and
$\si [\phi;\L_0]$ is the self-~interaction
\beq                 \nome{intaction}
\si[\phi;\L_0]=\half \int \d4 p \,
 \phi(-p) \lp
\s^{(B)}_1 p^2 +\s^{(B)}_2 \rp
\phi(p) + \frac {\s^{(B)}_3}{4!} \int_x \phi^4(x) \, .
\eeq
According to Wilson we derive $\se$ by integrating over the fields
with frequencies $\L^2<p^2<\L_0^2$ 
\beq\nome{Z'scalar}
e^{iW[j]}=N[j;\L]\int {\cal D}\phi \, 
\exp{i\biggl\{
-\half (\phi, \,D^{-1}\phi)_{0\L}+(j\,\phi)_{0\L}
+\se[\phi;\L]
\biggr\}}
\,.
\eeq
As we have seen previously, 
in order to study the renormalizability and the infrared finiteness of this
theory it is more convenient to consider the proper vertices
$\Gamma_{2n}(p_1,\cdots,p_{2n};\L)$ (see fig.~1) 
\begin{figure}[htbp]
\epsfysize=5cm
\begin{center}
\epsfbox{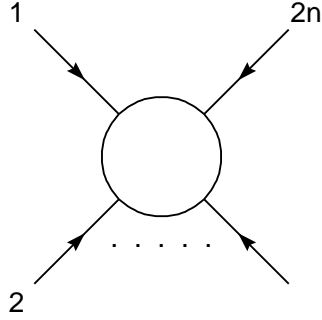}
\end{center}
\caption{\small{
Vertices $\Gamma_{2n}(p_1,\cdots,p_{2n})$.
}}
\end{figure}
\noindent
and their generating 
functional
$$
\G[\phi;\L]=
\sum_{n=1} \frac 1 {(2n)!} \int \prod_{i=1}^{2n} \d4 {p_i} \phi(p_i)
\, \G_{2n}(p_1, \cdots,p_{2n};\L)
\, (2\pi)^4\delta^4(\sum_{i=1}^{2n} p_i) \; ,
$$
which  is given by 
the Legendre transform of $W[j;\L]$
\beq\nome{Legscalar}
\G[\phi;\L]=W[j;\L]-\int_p j\phi\,.
\eeq
We know this functional generates the cutoff vertex functions in which
the internal propagator has frequencies in the range $\L^2<p^2<\L_0^2$
and satisfies the evolution equation \re{eveq}
\beeq\nome{eveqs}
&&\LdL \biggr\{ \G[\phi;\L]
-\half \int_p  D^{-1}_{\L\L_0} (p)\,\phi(p)\phi(-p) \biggr\}
\\
&& \phantom{\LdL \biggr\{ \G[\phi;\L] }= - \half \int_q 
\LdL D^{-1}_{\L\L_0}(q)\biggr[ \frac {1}{\G_2(q;\L)} \biggr]^2\,
\bG[q,-q;\phi;\L]\, , \no
\eeeq
with $ D_{\L\L_0} (p)=\K (p)\, D(p)\,.$
In turn the  functional $\bG$ satisfies \re{gammab}, whose expansion
provides us with the auxiliary vertices
$\bG_{2n+2}(q,p_1, \cdots,p_{2n},q';\L)$ 
in terms of the proper vertices (see fig.~2). 
\begin{figure}[htbp]
\epsfysize=5cm
\begin{center}
\epsfbox{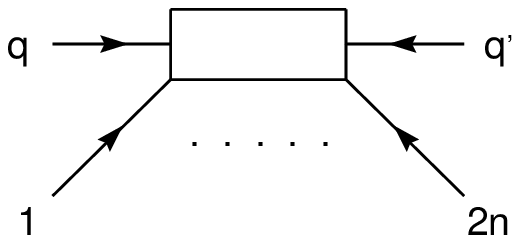}
\end{center}
\caption{\small{
Auxiliary vertices $\bG_{2n+2}(q,p_1, \cdots,p_{2n},q';\L)$.
}}
\end{figure}
\noindent
For $n=1$ we find
$$
\bG_4(q,p_1,p_2,q';\L)=
 \G_4(q,p_1,p_2,q';\L)\,,
$$
and in general (see fig.~3)
\beeq\nome{auxv}
&&
\bG_{2n+2} (q,p_1,\cdots,p_{2n},q';\L)=
\G_{2n+2} (q,p_1,\cdots,p_{2n},q';\L)
\no \\ &&
\phantom{\bG_{2n+2} (q} -\sum
\G_{2k+2} (q,p_{i_1},\cdots,p_{i_{2k}},-Q;\L)
\;\frac{1}{\G_2(Q;\L)}\no \\ && 
\phantom{\bG_{2n+2} (q}
\times
\bG_{2n-2k+2}(Q,p_{i_{2k+1}},
\cdots,p_{i_{2n}},q';\L)\,,
\eeeq
where $Q=q+p_{i_1}+\cdots p_{i_{2k}}  $ and
the sum is over $k=1 \dots n-1$ and over the ${2n\choose 2k}$
combinations of $(i_1 \cdots i_{2n})$.
\begin{figure}
\epsfysize=10cm
\begin{center}
\epsfbox{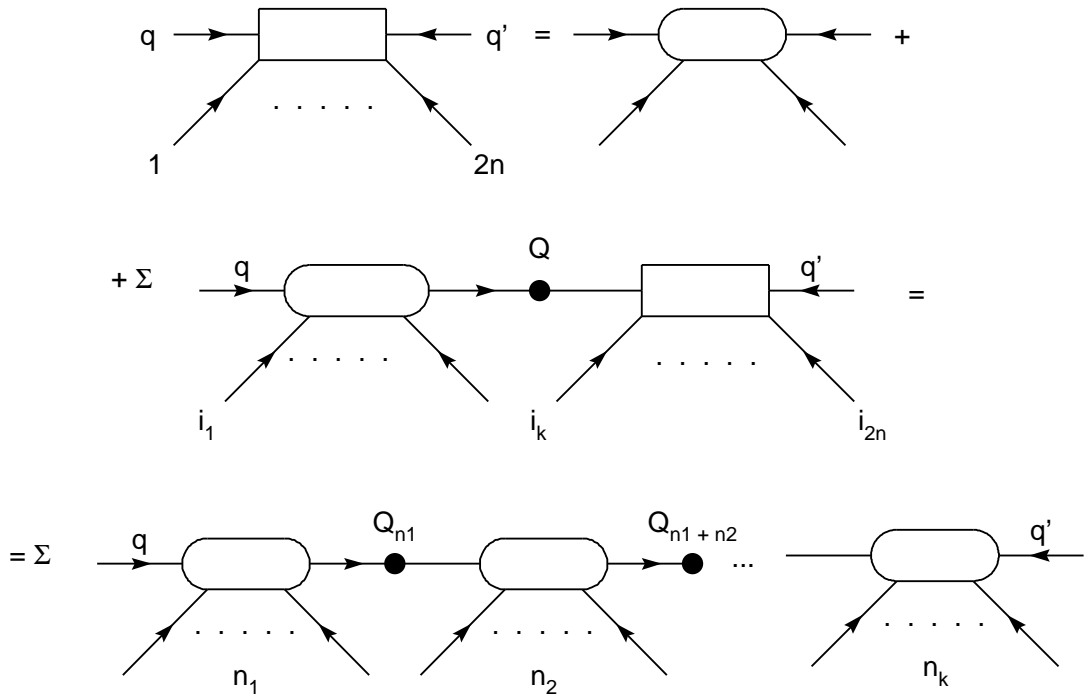}
\end{center}
\caption{\small{
Graphical representation of the equation \re{auxv} defining
the auxiliary vertices $\bG_{2n+2}(q,p_1,\cdots,p_{2n},q')$.
}}
\end{figure}

Let us return to the flow of the cutoff effective action
eq. \re{eveqs}.  
After isolating the interaction part of the
two-point function 
\beq\nome{gamma2}
\G_2(p;\L)=D^{-1}_{\L\L_0}(p)+\Sigma(p;\L)\,,
\eeq
the evolution equation for the proper vertices are 
\beq\nome{eveq1} 
\LdL \Sigma(q;\L) =\half \int \d4 q \frac{S(q;\L)}{q^2} 
\G_{4}(q,p,-p,-q;\L)\,, 
\eeq 
and 
\beq \nome{eveq2} 
\LdL
\G_{2n}(p_1,\cdots,p_{2n};\L) =\half \int \d4 q \frac{S(q;\L)}{q^2}
\bG_{2n+2}(q,p_1,\cdots,p_{2n},-q;\L)\,, 
\eeq 
where
$S(q;\L)$ is given by 
\beq \nome{essep} 
\frac{S(q;\L)}{q^2} \equiv \LdL
D_{\L\L_0}(q) \left[ \frac {1}{1+D_{\L\L_0}(q)\Sigma(q;\L)}
\right]^2\,.  
\eeq 
Notice
\re{eveq1} and \re{eveq2} involve vertices at the infrared cutoff $\L$
with a pair of exceptional momenta $q$ and $-q$.  For \IR these
vertices become singular since we are dealing with a massless theory.
Furthermore, when integrating on $q=-\sum_{I}p_I$ ($I\equiv\;$subset of
the $p_i$'s), we encounter vertices with exceptional momenta. 
We will soon analyse the limit $\l \to \infty$ and show that eqs.~\re{eveq1}, 
\re{eveq2}
allow  to derive vertex functions $\G_{2n}$ with non-exceptional
momenta at the physical point $\L=0$, order by order in perturbation theory.

\subsection{Physical couplings and boundary conditions}
In order to solve the RG flow for the scalar theory we need to supply
boundary conditions.
Dimensional analysis suggests  us the form of the relevant part of the
cutoff effective action
$$
\Pir[\phi;\L]=\half \int_p 
\phi(-p)\bigl[ (1+\s_1(\L)) p^2 + \s_2(\L)\bigr]
\phi(p)+\frac {\s_3(\L)}{4!} \int_x 
\phi^4(x) \,,
$$
where   the ``relevant'' couplings are so defined
$$
\s_1 (\L) = \left.\frac {d\Sigma(p^2;\L)}{dp^2}\right|_{p^2=\mu^2} \,,
\;\;\;\;\;\;\;\;
\s_2 (\L) =\Sigma(0;\L)\,,
\;\;\;\;\;\;\;\;
\s_3 (\L) =\G_4(\bp_1,\bp_2,\bp_3,\bp_4;\L)\,,
$$
and correspond, for $\L=0$, to the physical couplings introduced in
(\ref{norm}).
We then isolate the relevant couplings in the two- and
four-point vertices
\beeq\nome{isol}
&& \quad \quad  \Sigma(p^2;\L)= p^2 \s_1(\L) + \s_2 (\L) +
\S_2(p^2;\L) \,, \no \\
&&\G_4(p_1,p_2,p_3,p_4;\L)= \s_3(\L)+\S_4(p_1,p_2,p_3,p_4;\L)\,,
\eeeq
where, by definition, the vertices $\S_i$ satisfy the conditions
$$\S_2(0;\L)=0\,,\;\;\;\;\;\;\; \left. d\S_2(p^2;\L)/ d p^2
\right|_{p^2=\mu^2} =0\,, \;\;\;\;\;\;\; 
\S_4(p_1,p_2,p_3,p_4;\L)\vert_{4SP} =0 \,.
$$
From dimensional analysis we learn
\beeqn
&&
\s_1 \sim \s_3 \sim (\L)^0\;, \;\;\;\;\;\; \s_2 \sim (\L)^{2}\,,
\\
&&\S_2\sim \S_4 \sim (\L)^{-2}\,.
\eeeqn
Notice that in $\S_2(p^2;\L)$ four powers of momentum are
absorbed by the $p$-dependence required by the two conditions at $p^2=0$
and $p^2=\mu^2$. Similarly in $\S_4$ two powers of momentum are absorbed
by the $p_i$-dependence required by the condition at the symmetric
point.
In both cases the remnant has an inverse power dependence on $\L$, so
that these vertices,  just as $\G_{2n}$ with $n>2$, are irrelevant.

We have already stressed so far the need of boundary conditions to 
get the vertex functions from the evolution equations
\re{eveq1}, \re{eveq2}.
The conditions for the relevant couplings $\s_i(\L)$ must be such to
ensure the physical normalization conditions, \ie the masslessness of
our particle and the value of the self-coupling ($g$). Thus at the
physical value $\L=0$ we impose
\beq\nome{bound1}
\s_1(\L=0) =0\;,\;\;\;\;
\s_2(\L=0) =0\;,\;\;\;\;
\s_3(\L=0) =g\;.
\eeq
The further sensible physical requirement for  the remaining vertex
functions is that they are negligible when the ultraviolet cutoff
$\L_0$ is set to infinity.
The simplest choice is to set all these irrelevant vertices to zero
at $\L=\L_0$
\beq\nome{bound2}
\S_2(p^2;\L_0) =0 \,,
\;\;\;\;\;
\S_4(p_1,p_2,p_3,p_4;\L_0)=0\,,
\;\;\;\;\;
\G_{2n}(p_1,\cdots,p_{2n};\L_0)=0 \,,
\;\;\;\;\; n \ge 3\,.
\eeq
With these prescriptions the functional $\G[\phi;\L]$ has the form 
\re{intaction} with the $\s_i^{(B)}$'s given by the
relevant couplings $\s_i$ evaluated at $\L=\L_0$.
The bare coupling constant is then
$ g^{\mbox\scriptsize{{UV}}}= {\s_3^{(B)} }/{(1+\s_1^{(B)})^2 }$.

The evolution equations \re{eveq1} and \re{eveq2}  together with the boundary conditions
\re{bound1}, \re{bound2} can be converted into a set of integral
equations.
For the three relevant couplings $\s_i$ the boundary conditions
(\ref{bound1}) give
\beeq\nome{inte1}
&&\s_{1}(\L)=
\half \int_q \int_0^{\L}
\frac{d\l}{\l} \frac{S(q;\l)}{q^2}
\frac{\partial}{\partial p^2}
\G_{4}(q,p,-p,-q;\l)|_{p^2=\mu^2}\,,
\no \\
&&\s_{2}(\L)=
\half \int_q \int_0^{\L}
\frac{d\l}{\l} \frac{S(q;\l)}{q^2}
\G_{4}(q,0,0,-q;\l)\,,
\no \\
&&\s_{3}(\L)=g +
\half \int_q \int_0^{\L}
\frac{d\l}{\l} \frac{S(q;\l)}{q^2}
\bG_{6}(q,\bp_1,\ldots,\bp_{4},-q;\l)\,.
\eeeq
As far as the  other vertices are concerned, the prescriptions
\re{bound2} give
\beeq\nome{inte2}
&&\S_{2}(p;\L)=
-\half \int_q \int_{\L}^{\L_0}
\frac{d\l}{\l} \frac{S(q;\l)}{q^2}
\, \D \G_{4}(q,p,-p,-q;\l)\,,
\no \\
&&\S_{4}(p_1 \ldots p_4;\L)=
-\half \int \d4 q \int_{\L}^{\L_0}
\frac{d\l}{\l} \frac{S(q;\l)}{q^2}
\, \Delta \bG_{6}(q,p_1,\ldots,p_{4},-q;\l)\,,\no \\
&&\G_{2n}(p_1\ldots p_{2n};\L)=
-\half \int_q \int_{\L}^{\L_0}
\frac{d\l}{\l} \frac{S(q;\l)}{q^2}
\, \bG_{2n+2}(q,p_1,\ldots,p_{2n},-q;\l)\,,
\eeeq
with $n>2$.
The subtracted vertices $\Delta \G_{4}$ and $\Delta \bG_{6}$ are defined by
\beeq \nome{inte3}
&& \quad\Delta \G_{4}(q,p,-p,-q;\l)
\equiv \,
\G_{4}(q,p,-p,-q;\l) -\G_{4}(q,0,0,-q;\l)
\no \\
&&\phantom{\quad \Delta \G_{4}(q,p,-p,-q;\l)
\equiv }
-p^2 \frac{\partial}{\partial p'^2}
\G_{4}(q,p',-p',-q;\l)|_{p'^2=\mu^2}\,,\\
&&\!\!\!\!\!\Delta \bG_{6}(q,p_1,\ldots,p_{4},-q;\l)
\equiv\,
\bG_{6}(q,p_1,\ldots,p_{4},-q;\l)
-\bG_{6}(q,\bp_1,\ldots, \bp_{4},-q;\l)\,. \no
\eeeq
The subtractions in $\Delta \G_{4}$ and $\Delta \bG_{6}$ follow from
 isolating in eq.~\re{isol} the relevant couplings in the two- and
four-field 
vertices and from the different boundary conditions \re{bound1}, \re{bound2}.
We expect they provide the necessary subtractions to
make  the vertex functions finite for $\L_0\to\infty$ at any order
in perturbation theory.

We should notice the role of the boundary conditions for the relevant
couplings at $\L=0$ and for the irrelevant vertices at $\L=\L_0 \to
\infty$ (see \re{inte1} and \re{inte2}). In the case of the relevant
couplings they act in such a way the $q$-integration is bounded from above by
$\L$.  This is a crucial requisite to obtain a finite result since, as
expected from dimensional counting, the integrands grow with $q^2$.
Hence the bare couplings, recovered by setting $\L=\L_0$, grow with
$\L_0$ and provide the counterterms of the Lagrangian
\re{intaction} in terms of the physical coupling $g$.
On the other hand for the other vertices the $q$-integration is
bounded from above by the
ultraviolet cutoff $\L_0$. In order to show the theory is
renormalizable, we 
must prove that as $q^2 \to \infty$ the vertices in the integrands in
\re{inte2}   vanish sufficiently fast to allow the limit \UV.

If we choose a sharp cutoff like a step function eq.~\re{essep} reads
\beq \nome{esse1}
\frac{S(q;\l)}{q^2} = - \frac {1}{\l} \delta ( \l-\sqrt{q^2} ) s(\l)\,,
\;\;\;\;\;\;
s(\l)=\left[ \frac{1}{1+\frac{1}{\l^2}\Sigma(\l;\l)} \right]^2 \,,
\eeq
which is independent of $\L_0$.

\subsection {Loop expansion}

The iterative solution of  eqs.~\re{inte1} and \re{inte2}
provides us with the usual  loop expansion.
In this section we perform some calculations as an example.
If we set \UV the free propagator becomes 
$$
D_{\L}(q)=K_{\L\infty}(q)/q^2\,,
$$
with $K_{\L\infty}  (q)=1$ for $q^2\geq \L^2$ and vanishing for $q^2<\L^2$.
Clearly  the limit $\L\to 0$ can be taken only for non-exceptional 
momenta.

We start from the zero-loop order (\ie the tree level) in which  the only 
non-vanishing vertex is
$$
\s_3^{(0)} (\L) =g\,,
$$
\begin{figure}[htbp]
\epsfysize=3cm
\begin{center}
\epsfbox{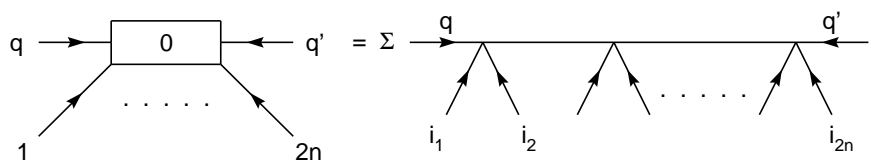}
\end{center}
\caption{\small{
Graphical representation of the auxiliary vertices
at zero loop.
}}
\end{figure}
\noindent
and the auxiliary vertices with $n \ge 2$ are given by (see fig.~4)
\beq \nome{bG2n+20l}
\bG_{2n+2}^{(0)}(q,p_1,\cdots,p_{2n},q';\L)=
-(-g)^n \; \sum_{perm} \prod_{k=1}^{n-1}D_{\L}
 \left( q+\sum_{\ell=1}^{2k}p_{i_\ell} \right) \,,
\eeq
where the sum is over the $(2n)!/2^n$ terms originating from the
permutations of $(p_{i_1},\cdots, p_{i_{2n}})$ and the symmetry of
the four-point coupling.

\subsection{One-loop vertices}

At this order the only non-vanishing contribution for the two-point
function is
$$
\s_2^{(1)}(\L) = \half g \int_q \D_{0\L}(q)
= - \frac{1}{32\pi^2}  g \L^2\,,
$$
where
$$
\D_{0\L}(q)=D_{\L}(q) - D_{0}(q)
= -\frac{1}{q^2}\Theta(\L^2-q^2) \,.
$$
As to the four-point vertex, from \re{bG2n+20l} we have
$$
\s^{(1)}_3(\L)\,  =\, -\frac 3 2 g^2\int_q
\left\{ D_{\L}(q) D_{\L}(q+\bp)-
       D_{0}(q) D_{0}(q+\bp)
\right\}\,,
$$
with  $\bp=\bp_i+\bp_j\,,$ $i \neq j$, and we exploited the symmetry of the
subtraction point \re{sp}.
For large $\L$ the integration range   is bounded by $q^2 \ltap \L^2$
and we get
$$
\s_3^{(1)}(\L) \simeq \frac{3}{16\pi^2} g^2\ln(\L/\mu)\,, \;\;\;\;\;
\mu \ll \L \,,
$$
whereas, for small $\L$ 
$$
\s_3^{(1)}(\L) \sim \L^2\,\;\;\;\;\L \ll \mu\,.
$$
The remnant $\S_4$ of the four-point vertex is found in a
similar fashion
\beeqn
&&\S_4^{(1)}(p_1,\cdots,p_4;\L) = -\half g^2\int_q D_{\L}(q)
\\ && \phantom{\S_4^{(1)}(p_1,\cdots,p_4;\L) =}
\times
\left\{ D_{\L}(q+p_1+p_2)+\cdots -3D_{\L}(q+\bp) \right\}\,,
\eeeqn
where the dots stand for the additional two terms with $p_2$ replaced by
$p_3$ and $p_4$.
Due to the subtractions the integral is convergent as $q^2 \to
\infty$, while 
for a large infrared cutoff it vanishes as $\mu^2/\L^2$
and $P^2/\L^2$, with $P$ a combination of external momenta.
The physical value is reached at $\L=0$ and reads
\beq \nome{onel1}
\G_4^{(1)}(p_1,\cdots,p_4) = \frac{1}{32\pi^2} g^2
\left\{ \ln \left[\frac{(p_1+p_2)^2}{\mu^2}\right] +\cdots \right\}\,.
\eeq

We are now left with the vertices $\G^{(1)}_{2n}$ with $n\ge 2$. For
those we have
\beq \nome{onel2}
\G_{2n}^{(1)}(p_1,\cdots,p_{2n};\L)=
- \frac {(-g)^n} {2n} \int_q \, D_\L(q) \,
\; \sum_{perm} \prod_{k=1}^{n-1}D_{\L}
 \left( q+\sum_{\ell=1}^{2k}p_{i_\ell} \right)\,.
\eeq
It is easy to see the integral is convergent for large $q^2$.
Although at the physical value $\L=0$ these vertex functions become
singular for vanishing momenta, it is known \cite{cw} 
the effective potential resulting of
summing up the vertex functions at vanishing momenta is infrared finite.
We rederive here the one-loop effective potential $V(\phi)$  to
illustrate  the role of the regularization  and the
physical conditions \re{norm} in this framework.
Apart from a volume factor, $V(\phi)$ corresponds  to $\G[\phi]$ 
with the ``classical'' field $\phi(p)=(2\pi)^4\,\delta^4(p) \,\phi$.
When  $\L$ is non-vanishing we get
$$
V(\phi)=\half \s_2^{(1)}\,\phi^2
+\frac {1}{4!}(g+\s_3^{(1)})\, \phi^4 -\half \int_q \Theta(q^2-\L^2)
\left\{ \sum_{n=2}^{\infty} \frac 1 n \left(\frac {-g\phi^2}{2q^2} \right)^n
- \frac {(g\phi^2)^2}{8 q^2(q+\bar p)^2} \right\}\,.
$$
Even if the various terms diverge at $q=0$ for
$\L=0$,  performing the sum and then taking $\L=0$, we have
$$
V(\phi)= \frac {g}{4!}\phi^4 + \half \int_q
\left\{ \ln  \left(1+\frac{g\phi^2}{2q^2}\right)
-\frac{g\phi^2}{2q^2}
+\frac{(g\phi^2)^2}{8 q^2(q+\bar p)^2}
\right\}\,.
$$
This expression is free from infrared singularities in $q=0$
and the integral is convergent at large $q$ (see ref.~\cite{cw}).

Afterwards, using the previous results,  we could go further and
compute the vertex functions at the two-loop order and so on. This was
done in \cite{bdm}. Instead, we will concentrate on proving the
perturbative renormalizability and infrared finiteness of the theory. 

\section{Perturbative renormalizability}
The aim of this section is to prove that the theory is perturbative
renormalizable, namely that in \re{inte2} we can set \UV.  
To be rigorous, we should study the $\L_0$ dependence of the integral
equations and prove the existence of the limit via the Cauchy
criterion, as was done in \cite{p,b,kks,bt}. What we want to do here
is to be less rigorous and stress the role of dimensional analysis by
performing the limit \UV in the integral equations and showing by
induction that the equations so derived produce finite vertex
functions in perturbation theory. This is the reason why the
dependence on the UV cutoff $\L_0$ has been understood so far ---and
will be understood.
As shown
before, the loop expansion develops from iterating eqs.~\re{inte1}
and \re{inte2}. From the vertices $\G_{2n}^{(\ell)}$ we can construct
the integrands at the same order which give the next loop vertices
upon $q$-integration. The convergence of the integrals producing
$\G^{(\ell+1)}_{2n}$ will be ensured by dimensional counting, while
showing $\S_2$ and $\S_4$ are finite will require the subtractions
in $\Delta \G_4$ and $\Delta \bG_6$ introduced in \re{inte3}. The
best way to represent the subtracted vertices $\Delta \G_4$
and $\Delta \bG_6$ consists in a  Taylor expansion as for the Bogoliubov $R$
operators \cite{bo}. Since we are interested in the large $\l$ behaviour 
we will
expand around vanishing momenta. 
Only even derivatives need being considered as odd derivative terms
vanish due to symmetry,
either directly or after integration.
The subtracted vertex $\D\bG_6$ is obtained from the expansion
\beeq\nome{t1}
&& \bG_6(q,p_1,\cdots, p_4,-q;\l) = \bG_6(q, 0,\cdots, 0,-q;\l) 
\no \\
&& \quad\quad +\int_0^1dx(1-x)\left(\sum_{i=1}^3 p_i 
\cdot \partial'_{i,4}\right)^2
\bG_6(q,p'_1,\cdots, p'_4,-q;\l)|_{p'_i=xp_i}
\eeeq
where $\partial'_{i,4}= {\partial}/{\partial p'_i}-
{\partial}/{\partial p'_4}$.
The first term, which represents the most singular contribution, is
cancelled in the subtracted quantity $\Delta \bG_6$.
As to $\D\G_4$ we have to consider the expansion up to four derivatives
\beeq\nome{t3}
&& \D\G_4(q,p,-p,-q;\l)=
\frac 1{3!}\int_0^1dx(1-x)^3(p \cdot \partial')^4
\G_4(q,p',-p',-q;\l)|_{p'=xp} \nonumber \\
&& \qquad -\half (p \cdot \partial')^2
\left\{ \G_4(q,p',-p',-q;\l)|_{p'^2=\mu^2}-
       \G_4(q,p',-p',-q;\l)|_{p'=0} \right\}
\eeeq
where $\partial'=\partial/\partial p'$.  Notice that also the second
term can be expressed in terms of fourth derivatives of
$\G_4$. Similarly the integrand for $\s_1$ can be expressed as the  second derivative of $\G_4$ with respect to the momentum
components.

In order to prove the theory is perturbatively renormalizable we
have to analyse the behaviour for large $\l$ of the vertices in the
integrands and show the integration over $\l$ is convergent for
$\l\to\infty$ (the convergence of the integrals for $\l\to 0$ will be
discussed in the next section).
In this analysis we are not interested in the detailed dependence
of the vertices on the external momenta, except
for the fact that the integration momentum is fixed at $q^2=\l^2$
(see \re{esse1}).
To prove perturbative renormalizability it will  suffice,
as in \cite{p}, to bound the large $\l$ behaviour of the vertices in
which all external momenta do not exceed the cutoff.
Then let us   introduce the following function which depends only on $\l$
\beq\nome{norma}
|f_{2n}|_{\l}\equiv
\Maxlp
|f_{2n}(p_1,\cdots, p_{2n};\l)|
\eeq
where $c$ is some numerical constant and $f(p_1,\cdots, p_n;\l)$ is
$\G_{2n}$, $\bG_{2n+2}$ or one of their derivatives.
Iterating \re{inte1} and \re{inte2} in which we have previously set \UV,
we obtain the following bounds:
\begin{itemize}
\item
for the relevant couplings
\beeq
&&\nome{e1a}
\s^{(\ell+1)}_1(\L)
\ltap \int_0^{\L^2}d\l^2s^{(\ell-\ell')}(\l)
|\partial^2 \G^{(\ell')}_4|_{\l} \;, \\
&&\nome{e1b}
\s^{(\ell+1)}_2(\L)
\ltap \int_0^{\L^2}d\l^2s^{(\ell-\ell')}(\l)
|\G^{(\ell')}_4|_{\l} \;, \\
&&\nome{e1c}
\s^{(\ell+1)}_3(\L)
\ltap \int_0^{\L^2}d\l^2s^{(\ell-\ell')}(\l)
|\bG^{(\ell')}_6|_{\l} \; ;
\eeeq
\item
for the irrelevant vertices
\beeq
&&\nome{e2a}
|\G^{(\ell+1)}_{2n}|_{\L}
\ltap \int_{\L^2}^{\infty }d\l^2s^{(\ell-\ell')}(\l)
|\bG^{(\ell')}_{2n+2}|_{\l} \;, \\
&&\nome{e2b}
|\S_2^{(\ell+1)}|_{\L}
\ltap \L^4 \int_{\L^2}^{\infty }d\l^2s^{(\ell-\ell')}(\l)
|\partial^4 \G^{(\ell')}_{4}|_{\l} \;, \\
&&\nome{e2c}
|\S_4^{(\ell+1)}|_{\L}
\ltap \L^2 \int_{\L^2}^{\infty }d\l^2s^{(\ell-\ell')}(\l)
|\partial^2 \bG^{(\ell')}_{6}|_{\l} \; ;
\eeeq
\item
for the derivatives of vertices
\beeq
\nome{e3a}
&&|\partial^m\G_{2n}^{(\ell+1)}|_{\L}
\ltap \int_{\L^2}^{\infty} d\l^2 s^{(\ell-\ell')}(\l)
|\partial^m\bG_{2n+2}^{(\ell')}|_{\l} \; , \\
&&\nome{e3b}
|\partial^m\S_2^{(\ell+1)}|_{\L}
\ltap \L^4\,\int
_{\L^2}^{\infty} d\l^2 s^{(\ell-\ell')}(\l)
|\partial^{m+4}\S_4^{(\ell')}|_{\l} \; , \\
&&\nome{e3c}
|\partial^m\S_4^{(\ell+1)}|_{\L}
\ltap \L^2\,\int_{\L^2}^{\infty} d\l^2
s^{(\ell-\ell')}(\l)|\partial^{m+2}\bG_6^{(\ell')}|_{\l} \; .
\eeeq
\end{itemize}
where $\partial^m$ stands for $m$ partial derivatives with respect to
external momenta  and the factors $\L^2$ and $\L^4$ in front of
integrals come from maximizing the $p^2$ or $p^4$ factors in
\re{t1} and \re{t3}, respectively. 
Actually $\partial^m$ in \re{e3b}-\re{e3c}
could also act on these $p$ factors. As we will show in the following all
these contributions are of the same order.

Let us now prove, by induction and using the bounds  
in \re{e1a}-\re{e3c}, 
that the theory is perturbatively renormalizable,
namely that the integrals in \re{e2a}-\re{e3c} are convergent for
$\l \to \infty$.

\noindent
(i) {\it Assumptions at loop $\ell$.}

\noindent
The assumption is dimensional counting ---except the
logarithmic corrections, and  concerns the nine quantities above.

\noindent
a) Relevant couplings ($T=\log(\L/\mu)$)
\beq \nome
 {a1}
\s^{(\ell)}_1(\L)= {\cal O}(T^{\ell-1}),\;\;\;\;
\s^{(\ell)}_2(\L)= {\cal O}(\L^2 T^{\ell-1}),\;\;\;\;
\s^{(\ell)}_3(\L)= {\cal O}(T^{\ell})\,.
\eeq
\noindent
b) Irrelevant vertices
\beq \nome {a2}
|\G_{2n}^{(\ell)}|_{\L}={\cal O}(\L^{4-2n}T^{\ell-1}),\;\;\;\;
|\S_2^{(\ell)}|_{\L}={\cal O}(\L^{2}T^{\ell-2}),\;\;\;\;
|\S_4^{(\ell)}|_{\L}={\cal O}(T^{\ell-1})\,.
\eeq
\noindent
c) Derivative vertices
\beq \nome {a3}
|\partial^m\G_{2n}^{(\ell)}|_{\L}={\cal O}(\L^{4-2n-m}T^{\ell-1}),\;\;\;\;
|\partial^m\S_2^{(\ell)}|_{\L}={\cal O}(\L^{2-m}T^{\ell-2}),\;\;\;\;
|\partial^m\S_4^{(\ell)}|_{\L}={\cal O}(\L^{-m} T^{\ell-1})\,.
\eeq
These assumptions are satisfied for $\ell=0$ and $1$.

\noindent
(ii) {\it Iteration to  loop $\ell+1$.}

\noindent
We should notice that the powers of $\L$ in \re{a1}-\re{a3} are
independent of the loop number  since they are dictated by dimensional
counting.
As in the case of the relevant (irrelevant) couplings the integrands
increase (decrease) with $\l$, the integrals are dominated by the
upper (lower) limit $\l=\L$. For the irrelevant couplings we can thus
take the limit \UV, removing the ultraviolet cutoff.
It is easy to see that the integrals in \re{e1a}-\re{e3c}
reproduce at loop $\ell+1$ the same dimensional counting behaviours.
This is just what
we  need to prove perturbative renormalizability, since logarithmic
corrections cannot change  power counting at any finite order.
In fact it is relatively simple to control also the powers of $T$
and in the following we show that the behaviours \re{a1}-\re{a3} are
reproduced by the iteration.

Before discussing the large $\L$ behaviours at loop $\ell+1$,
from \re{a1}-\re{a3} we will
derive  some intermediate results for the
integrands at loop $\ell$.

\noindent
a) From the two-point function and \re{esse1} we have
$$
s^{(\ell)}(\l)\sim t^{\ell-1} \,,
$$
where $t\equiv\log(\l/\mu)$.

\noindent
b) The leading term of the auxiliary vertices $\bG_{2n+2}$
is given by the contribution of fig.~5 which involves only four-point 
vertices
$$
|\bG_{2n+2}^{(\ell)}|_{\l}\sim\l^{2-2n}\prod_1^n\s_3^{(\ell_i)}(\l)
\sim\l^{2-2n}t^{\ell}\,,
$$
where $\sum\ell_i=\ell$ and we have a factor $\l^{-2}$ for each internal
propagator.
All the  contributions coming from higher vertices and from
loop corrections in the intermediate propagators give the same
power in $\l^2$ but a lower power in $t$.

\noindent
c) The leading term of the derivatives of the auxiliary vertices
originates    from the contribution of fig.~5
when the derivatives act on the internal propagators
$$
|\partial^m\bG^{(\ell)}_{2n+2}|_{\l}\sim\l^{2-2n-m}t^{\ell}\,.
$$
\begin{figure}[htbp]
\epsfysize=3cm
\begin{center}
\epsfbox{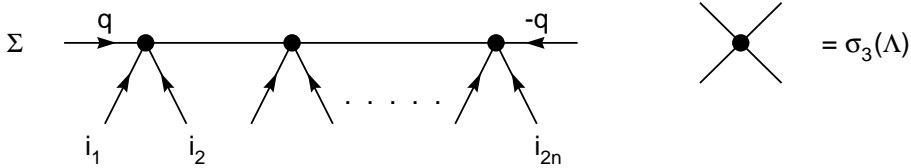}
\end{center}
\caption{\small{
Graphical representation of the leading contribution
of auxiliary vertices for $\L \to \infty$.
}}
\end{figure}
\noindent
Again, the contributions from derivatives of higher vertices or from
loop corrections of internal propagators give lower powers of $t$.

By implementing these results in \re{e1a}-\re{e3c} we reproduce 
at $\ell+1$ loop
order the behaviours in \re{a1}-\re{a3}.
In all cases we have $\ell'=\ell$, \ie loop corrections of the
propagator in $s(\l)$ do not contribute to leading terms.

\section{Infrared behaviour}
In this section we show that for the massless scalar theory we are 
considering
the vertex functions at non-exceptional momenta are finite order by
order in perturbation theory \cite{bdm}.
For a general analysis of  the behaviour in the \IR limit  of vertex 
functions  
at exceptional momenta see refs.~\cite{kk,prr}.
Our aim is to prove that $q$ integration in \re{inte1} and 
\re{inte2}
is convergent in the  limit  $\L \to 0$.
As we have done in the case of renormalizability, this is shown 
by induction in the number of loops.

The integrands in \re{inte1} and \re{inte2} are given by vertices
with one pair $(q, -q)$ of exceptional momenta. Thus, by iteration,
one introduces vertices with any number of soft (\ie of order $\L$) 
exceptional momenta.
In general we say that a  momentum $p_i$ 
in $\G_{2n}(p_1,\cdots,p_{2n};\L)$ is exceptional if 
$p_i={\cal O}(\L)$.
Henceforth we will add  the vertex functions an index
to identify the number  of exceptional momenta. We write
$$
\G_{2n,s}(p_1,\cdots,p_{2n};\L) \equiv
\G_{2n}(p_1,\cdots,p_{2n};\L) 
\,,\;\;\;\;\mbox {for}\;\;\;\;
p_{i_1}\,,\cdots p_{i_s}
= {\cal O}(\L)\,.
$$
where $s=0,\ldots,2n-1$.
Clearly for $s=2n-1$ all pairs of momenta are exceptional and 
 we denote with $\G_{2n,0}$ the vertices  without exceptional
momenta. A similar notation will be employed for the auxiliary vertices
$\bG_{2n+2,s}$ with $s$ soft momenta.

\vskip 0.3 cm \noindent
(i) {\it Assumptions at loop $\ell$}
\vskip 0.2 cm \noindent
As $\L \to 0$ we assume the following behaviours
\beeq
&&\nome{as1}
\G_{2n,0}^{(\ell)}(\L) \to \mbox{finite} \,, \\
&&\nome{as2}
\G_{2n,2}^{(\ell)}(p_1, \cdots, p_{2n} ;\L) = {\cal O}( T^{\ell}) 
 \;\;\;\; n\neq 1\,, \\
&&\nome{as3}
\G_{2n,2s}^{(\ell)}(p_1,\cdots, p_{2n}; \L) = {\cal O}
( \L^{2-2s}T^{\ell-1})
\, \;\;\;\; n \ge 3, \quad s=2, \cdots n-1\,, \\
&&\nome{as3bis}
\G_{2n,2s+1}^{(\ell)}(p_1,\cdots, p_{2n}; \L) 
 \sim  \G_{2n,2s}^{(\ell)}(p_1,\cdots, p_{2n}; \L)
\, \;\; n \ge 2, \, \;\; s=0, \cdots n-1, \\
&&\nome{as4}
\G_{2}^{(\ell)}(p;\L) = {\cal O}( \L^2\,T^{\ell-1})
\, \;\;\;\; \mbox{for} \; p^2= {\cal O}( \L^2)\, ,  \\
&&\nome{as5}
\frac {\partial}{\partial p_{\mu}}
\G_{2n,n-1}^{(\ell)}(p,-p,p_1,\cdots, p_{2n-2}; \L)
= {\cal O}( \L^{4-2n}T^{\ell-1} \; \frac {p_{\mu}
}{\L^2} )\, \;\;\;\; n \ge 1
\, ,
\eeeq
where $T=\ln(\L/\mu)$.
All these assumptions are satisfied for $\ell=0$ and $\ell=1$ (see
sect. 2.1.1). 
Anyway it is the  first equation that states the most important result
and it is our aim to show it holds even at loop $\ell+1$. This is the
reason why we need all other behaviours in the assumptions, and those
in turn must be satisfied at loop $\ell+1$.

\vskip 0.3 cm \noindent
(ii) {\it Auxiliary vertices at loop $\ell$}

\vskip 0.2 cm \noindent
(a) For the auxiliary vertices with just two soft momenta  the
sum in \re{auxv} is controlled, as $\l\to 0$, by the proper vertex,
giving
\beq\nome{ci6}
\bG^{(\ell)}_{2n+2,2}(q,p_1,\cdots,p_{2n},-q;\l)|_{q^2=\l^2}\sim
\G_{2n+2,1}^{(\ell)}(\l)\sim t^{\ell} \,.
\eeq
Powers of $t$ do not appear if the two soft momenta are inserted in
different vertices or in vertices without soft momenta of order $\l$.

\vskip 0.2 cm \noindent
(b) For an arbitrary even number  of soft momenta ($s>0$) we find
\beq\nome{ci10}
\bG_{2n+2,2s+2}^{(\ell)}(q,p_1,\cdots,p_{2n},-q;\l)|_{q^2=\l^2}\sim
\l^{-2s}t^{\ell}\,,\;\;\;\;\;\;\;
s=1,\cdots,n-1
\,.
\eeq
This behaviour is controlled by the largest number of internal
propagators with soft momentum of order $\l$, \ie by the contribution
of the graph depicted in fig.~6 in which the $2s$ soft momenta are
all emitted by the four-point vertices to the left (or right). 
\begin{figure}[htbp]
\epsfysize=1.6cm
\begin{center}
\epsfbox{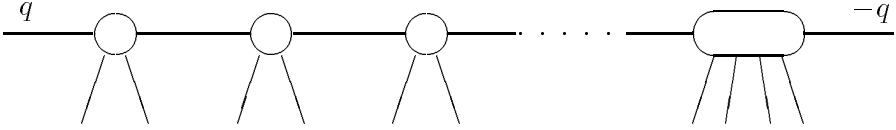}
\end{center}
\caption{\small{
The leading contribution for $\L \to 0$ of auxiliary
vertices in which the pairs of momenta in the four-point functions are
exceptional.
}}
\end{figure}
\noindent
In this way we have $s$ internal propagators with
momentum $q^2=\l^2$ producing  the factor $\l^{-2s}$ in \re{ci10}.
Loop corrections to the internal propagators results in non-leading
logarithmic powers.

In the same way we construct the auxiliary vertices for an odd number 
of soft momenta. The leading contribution is the same as in fig.~6 
since the presence of the last soft momentum is unimportant. We have
then
\beq\nome{lastmom}
\bG_{2n+2,2s+3}^{(\ell)}(\l) \sim
\bG_{2n+2,2s+2}^{(\ell)}(\l) \,.
\eeq

\vskip 0.2 cm \noindent
(c) Lastly, for the derivatives of the auxiliary vertices with all
exceptional momenta we have
\beq\nome{ci13}
\frac {\partial}{\partial p_{\mu}}
\bG_{2n+2,2n+1}^{(\ell)}(q,p,-p,p_1\cdots p_{2n-2},-q; \l)
={\cal O}( \l^{4-2n}t^{\ell}\; \frac {p_{\mu}
}{\l^4} )
\,,\;\;\;\;\;\;n>1\,.
\eeq
This behaviour can be read from the graphs of fig.~7
when the derivative acts on
an internal  propagator with momentum $P=p+\sum_{k=1}^a 
p_{i_{k}}$.
\begin{figure}[htbp]
\epsfysize=2cm
\begin{center}
\epsfbox{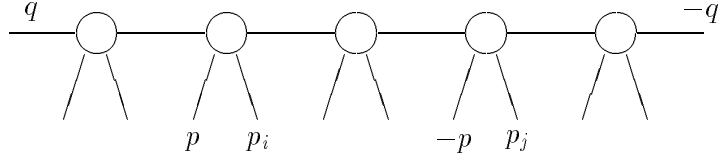}
\end{center}
\caption{\small{
The leading contribution for $\L \to 0$ of the derivative
with respect to $p$ of auxiliary vertices in which all momenta are soft. 
The derivative acts on the internal propagator of momentum $P=p+\sum_{k=1}^a 
p_{i_{k}}$.}}
\end{figure}
\newline
As noticed above, all the contributions with higher vertices, 
with loop corrections
in the propagators or with the derivatives acting on vertices lead to
lower powers of $t$.

\vskip 0.3 cm \noindent
(iii) {\it Iteration at loop $\ell+1$}

\vskip 0.2 cm \noindent
We now deduce the behaviour of the vertices at loop $\ell+1$ for \IR
by inserting the results for the auxiliary vertices at loop $\ell$ in
the integral equations.

\vskip 0.2 true cm \noindent
(a) For the two-point function we find
$$
\s_1^{(\ell+1)}(\L) = {\cal O}(\L^2 T^{\ell}) \to 0 \, ,
$$
$$
\s_2^{(\ell+1)}(\L)
= {\cal O}(\L^2 T^{\ell}) \to 0 \,,
$$
and, for a non-soft momentum $p$
$$
\S_2^{(\ell+1)}(p;\L) = {\cal O}(\L^{0}) \, .
$$
This follows from eqs. \re{as2} and \re{as3}.

\vskip 0.2 true cm \noindent
(b) For the four-point function at non-exceptional momenta
eq. \re{ci6} gives
\beq\nome{ci7}
\s_3^{(\ell+1)}(\L) = {\cal O}(\L^2 T^{\ell}) \to 0 \,,\;\;\;\;\;\;
\S_4^{(\ell+1)}(p_1,\cdots,p_4;\L)
= {\cal O}(\L^{0}) \,.
\eeq
Again loop corrections in $s(\l)$ do not contribute to the leading 
terms. 

\vskip 0.2 true cm \noindent
(c) Finally, for $\G_{2n,0}^{(\ell+1)}$ with $n>2$ we infer from \re{ci6}
\beq\nome{ci9}
\G_{2n,0}^{(\ell+1)}(p_1,\cdots,p_{2n};\L)={\cal O}(\L^{0})\,.
\eeq
Thus we conclude from \re{ci7}-\re{ci9} that all the physical vertices
at $\L=0$ and with non-exceptional momenta
are finite.
This verifies at loop $\ell+1$ the main assumption \re{as1}.

We then go on to test the additional assumptions.

\vskip 0.2 true cm \noindent
(d) To study the behaviour of $\S_{2,1}^{(\ell+1)}$ 
for $p^2={\cal O}(\L^2)$ we need the subtracted integral 
$\D\G_{4,3}^{(\ell)} (q,p,-p,-q;\l) $ with $q^2=\l^2$, $p^2 \sim 
\L^2$ and $\l > \L$. We appreciate that the last term in the 
definition \re{inte3} of $\D\G_4$ can be estimated from \re{as2}
and, since the $\l$-integration is finite, its contribution is 
proportional to $p^2 \sim \L^2$, without logarithmic powers. 
For the remaining terms, 
we use the Taylor expansion
$$
\G_{4,3}^{(\ell)}(q,p,-p,-q;\l) - \G_{4,3}^{(\ell)}(q,0,0,-q;\l)=
\int_0^1 dx (p \cdot \partial')
\G_{4,1}^{(\ell')}(q,p',-p',-q;\l)|_{\l^2=q^2,\; p'=xp}\,.
$$
Inserting this expansion and the result \re{ci13} in the integral 
equation \re{inte2}, we get
\beq\nome{ci51}
\S_{2}^{(\ell+1)}(p;\L) \sim \OO (p^2\,T^\ell)\,.
\eeq
This proves \re{as4} at loop $\ell+1$.

\vskip 0.2 true cm \noindent
(e) For the vertices with pairs of exceptional momenta
it is straightforward to prove \re{as2}-\re{as3} at  loop $\ell+1$
by using the result \re{ci10}.

\vskip 0.2 true cm \noindent
(f) For the vertices with all exceptional momenta
we show \re{as5} holds at $\ell+1$-loop order.
For $n=1$ this is simply obtained by taking the derivative of
eq.~\re{ci51} with respect to $p_{\mu}$.
For $n>1$ this can be done by writing
\beeqn
&&\frac {\partial}{\partial p_{\mu}}
\G_{2n,n-1}^{(\ell+1)}(p,-p,p_1,\cdots, p_{2n-2}; \L)
\\
&&
=\half\int_{\L} \d4 q \frac{s^{(\ell-\ell')}(\l)}{q^2}
\frac {\partial}{\partial p_{\mu}}
\bG_{2n+2,n}^{(\ell')}(q,p,-p,p_1,\cdots, p_{2n-2},-q; \l)|_{\l^2=q^2}
\eeeqn
and using \re{ci13}.

Summing up, in this chapter we have exploited the scalar theory as a toy
model to introduce the exact RG formulation and to show how in our language
perturbative renormalizability and infrared finiteness follow from
dimensional counting.
In the next chapter we will devote to the implementation  of
symmetries in this framework.

\chapter{The Quantum Action Principle}

Implementation of symmetries in the RG formulation is the main subject
of this thesis. We will realize throughout this section such a goal is
not really straightforward. As a matter of fact, in trying to apply
the RG method to a theory with local symmetry, we encounter the
problem that the division of momenta into large or small (according to
the scale $\Lambda$) is incompatible with gauge invariance. This is
easy to appreciate if we consider the homogeneous gauge transformation
$$
\Phi(x) \mapsto \Omega (x) \Phi (x)\,.
$$
Since in momentum space $\Phi(p)$ is mapped into a
convolution with the gauge transformation $\Omega$, any division of
momenta into high and low is seen not to be preserved by gauge
transformations.  In order to solve this problem, either we break the
gauge invariance in intermediate steps, aiming to recover it at the
physical point $\Lambda=0$ by imposition of some constraints or the RG
is generalized in such a way the symmetry is manifestly preserved.
Clearly, manifest preservation of gauge invariance would be
preferable.  Although an attempt in this direction has recently been
done in ref.~\cite{morrisfaro}, we will not discuss this option
further here, but instead give a detailed study of the former.

We will follow ref.~\cite{mt} and show an analogue of the
Quantum Action Principle (QAP) exists for the solution of the broken
Ward identities. 
When these are solved, the unbroken Ward identities are
guaranteed to hold once the cutoff is removed.
Indeed in this way, renormalized physical Green functions, 
with the correct gauge dependence, evaluated
at non-zero subtraction points (when necessary, \ie when massless particles
are present), may be
constructed  order by order in the couplings. 

In ref.~\cite{rw,newrw}  the background field method was used in the
RG formulation. This allows  to maintain background gauge invariance  
by replacing the division
of momenta into high or low by a division of eigenvalues of the 
background covariant Laplacian. 
However, we should note that
 background field invariance does not suffice to  ensure 
the quantum gauge invariance ---\ie BRS invariance in gauge
fixed systems.
For example, it is the latter that ensures that longitudinal modes are
properly 
cancelled
by ghosts in internal propagators, \ie that unitarity is maintained, not
the former. 

We will now address the problem of the broken Slavnov-Taylor (ST)
identities and  
discuss the so-called Quantum Action Principle.

\section{The Quantum Action Principle}
The Quantum Action Principle describes the 
response of a quantum field theory under a field transformation.
Thus it is a quite powerful  tool in the construction of field theories with
symmetry properties. Even if it was firstly established in the BPHZ
renormalization scheme \cite{qap1}, its validity was confirmed independently of
the renormalization scheme \cite{speer,qap2}.
Let us denote by  $\de\Phi_a$ an infinitesimal variation of the field
$\Phi_a$, by $\de \LL$ the corresponding infinitesimal variation of
the lagrangian. At the classical level we can write the trivial
identity
\beq\nome{qap1}
\int d^4 y \; \frac{\de \int_x \LL(x)}{\de \Phi_a(y)}
\, \de\Phi_a (y) =  \int d^4 x \, \de\LL (x)\,.
 \eeq
If we supply the classical Lagrangian $\LL(x)$ with an additional term
coupling the external source $\g_a(x)$ to the field variation,
eq. \re{qap1} reads
\beq\nome{qap2}
\int d^4 y \; \frac{\de\int_x \LL(x)}{\de\Phi_a(y)}\,
\frac{\de\int_x \LL(x)}{\de\g_a(y)}=\int d^4x \, \de \LL(x)\,.
\eeq
The QAP states this identity generalizes to all order in perturbation
theory under the form~\footnote{The presence of a regulator of UV
divergences  is assumed.}
\beq\nome{gammaqap}
\int d^4 x \; \frac{\de\G[\Phi,\g]}{\de\Phi_a(x)}\,
\frac{\de\G[\Phi,\g]}{\de\g_a(x)}=[\D\G]\,,
\eeq
where the insertion $\D$ is local and a normal product of degree $d$
$$
d= 4-\mbox{dimension}\,(\Phi_a) + \mbox{degree} \, (\de \Phi_a)\,.
$$
Identity \re{gammaqap} can be written for $Z[J,\g]$ 
$$
Z[J,\g]= \int {\mathcal D}\Phi \, \exp i \left\{
\sbrs[\Phi,\g]+S_{\mbox{\footnotesize{counterterms}}}[\Phi]+J_a\Phi_a
\right\}
\,,
$$
under the form
\beq\nome{qapz}
\int d^4x\, J_a (x)\, \frac{\de Z[J,\g]}{\de \g_a(x)}= \int
{\mathcal D}\Phi \, \D[\Phi,\g]\, \exp i \left\{
\sbrs[\Phi,\g]+S_{\mbox{\footnotesize{counterterms}}}[\Phi]+J_a\Phi_a
\right\}
\,.
\eeq
Then the response of the system is given by the insertion of a local
operator of dimension $d$ as above.
When removing the regulator (UV limit)  $\D$ is finite, at least in 
perturbation theory, and this ensures that also the insertion of
the operator $\Delta$ is finite in the UV limit.

In general one is interested in solving the equation $\D=0$.
On the other hand,  since in perturbation theory  
$[\D\G]=\D+{\mathcal O}(\hbar)$,
the insertion of $\D$ is also local at the first order in
which $\D$ itself is non-vanishing. Due to the existence of a finite number
of local operators of the correct dimension,  the equation $\D=0$
gives rise, order by order, to a finite number 
of conditions, which can eventually be satisfied by fine-tuning 
\cite{brs} the 
parameters in the action $\sbrs +
S_{\mbox{\footnotesize{counterterms}}}$. 

The QAP consists of the relations 
\re{gammaqap} and \re{qapz},
together with locality of $\D$, and must be understood as a general theorem of
renormalization theory to be used in any formalism
\cite{qap3}. 
However, it is not obvious  how the QAP can be obtained for an
effective theory. In fact, the procedure of integrating the high 
energy degrees of freedom  generates effective
non-local
interactions ---\ie a series of local interactions of arbitrarily high
numbers  of derivatives---
and also the field transformations  become non-local. 
We will see that QAP  is an extremely powerful theorem and
suffices to discuss ST identities or their generalizations describing
gauge invariance.

\section{Effective Slavnov-Taylor identities}

The gauge symmetry requires that the generating functional $Z[J,\g]$
satisfies the ST identity \cite{brs,becchi0}
\beq \nome{ST}
\SS_J Z[J,\g]=0 \,, 
\eeq
where $\SS_J$  is the usual ST operator
$$
\SS_J=\spint J_a(p)\,(-)^{\de_a}\dfud{}{\g_a(p)}\,.
$$
The ST identity can be
directly formulated for the Wilson effective action $\se$ at any $\L$.
Consider the generalized BRS transformation
\beq\nome{brseff}
\de\Phi_a(p)=\Ki(p)\,\eta\,\dfud{\se}{\g_a(-p)}\,,
\eeq
in which  $\eta$ is a Grassmann parameter. 
Performing such a change of variable in the
functional integral \re{Z'}, we  deduce the following identity
\beq \nome{steff}
\SS_J Z[J,\g]=N[J,\g;\L] \int \DD\Phi \exp{i\lgr
-\half(\Phi, \, D^{-1} \Phi)_{0\L}+(J,\Phi)_{0\L} 
+\se[\Phi;\L]\rgr} \, \De[\Phi,\g;\L]\, .
\eeq
The operator giving the ST identities violation at the effective level is
$$
\De[\Phi,\g;\L]=
i \spint  \Ki(p)\,\exp{(-i\se)}\lgr \dfud{}{\Phi_a(p)}
\,\dfud{}{\g_a(-p)}\rgr \exp{(i\se)}\qquad\qquad\qquad\;\;
$$
$$
+ i\spint  \Phi_a(p)\,D^{-1}_{ab}(p)\,
\dfud{\se}{\g_b(p)}
\,.
$$
Whereas the l.h.s of the identity \re{steff} arises from the variation of
the source term $(J,\Phi)_{0\L}$, the functional $\De$ originates from
the Jacobian of the transformation \re{brseff} and from
the variation of the rest of the exponent in \re{Z'}.
Restoration of symmetry, $\SS_J Z[J,\g]=0$, translates into 
$$
\De[\Phi,\g;\L]=0 \;\;\;\; {\mbox{for any}} \;\;\L\,.
$$
In order to get information about $\De$, in the following we will study 
in detail the properties of the flow of this operator. 
$\De$ satisfies a linear evolution equation 
(found by explicit derivation) \cite{b,bdm4,bdm5,mt}
\beq\nome{flowDe}
\LdL\De =
\int_p \, [\LdL K_{0\L}(p)] \, \bigl\{
L_1+\hbar L_2
\bigr\} \De\,,
\eeq
where the linear operators $L_1$ and $L_2$ are given by
\beeq\nonumber
&&
L_1
= - (-)^{\de_a} D_{ab}(p) \, \frac{\de\se}{\de\Phi_a(-p)}\,
\frac{\de}{\de\Phi_b(p)}\\
&& L_2=
\frac i 2 \, D_{ab}(p) \, \frac{\de^2}{\de\Phi_a(-p)\,\de\Phi_b(p)}
\,.\nome{L}
\eeeq
In eq.~\re{flowDe} we restored the powers of $\hbar$ in order to show
how in the r.h.s. of the flow equation for $\De$ there are terms at the
same loop order of the l.h.s..  

Since $\De$ satisfies a linear equation, the gauge symmetry condition 
$\De=0$ is verified for any $\L$ if we can set to zero the boundary 
conditions of \re{flowDe}. The main point is to fix to zero the ones 
for the relevant part $\Der$ of $\De$ for some value $\L_R$ of the IR 
cutoff. 
Normally $\Der(\L_R)=0$ is a set of constraints which overdeterminates 
the couplings in $\se(\L_R)$. 
If the boundary conditions are set at $\L_R\ne 0$, 
the number of independent constraints can be 
reduced by exploiting the so-called consistency conditions,
which are a set of algebraic identities 
coming 
 from the anticommutativity 
of the differential operator $\frac{\de}{\de\g}\frac{\de}{\de\Phi}$
\cite{b}. We can extract $\Der(\L_R)$ by expanding 
the vertices of $\De(\L_R)$ around vanishing momenta, even though we 
are considering massless particles. The result is that the consistency 
conditions constrain some couplings in $\Der(\L_R)$, so that the set 
$\Der(\L_R)=0$ can be fulfilled in some cases by tuning 
the parameters in $\ser$.  
Such an analysis was performed in ref.~\cite{b} for the pure gauge SU(2) model.

On the other hand, if the boundary conditions are set at
the physical point $\L_R=0$, the  consistency conditions
follow  from the nihilpotency
of the ST  operator.
In a theory with one or more massless particles, we 
have to introduce non-vanishing subtraction points in order to define 
$\Der(\L_R=0)$. This fact could spoil the power of the consistency 
conditions since they now involve also irrelevant vertices of $\De(0)$ 
evaluated at the subtraction points \cite{bdm4,bdm5}. 
Thus it seems that a case-by-case analysis based on a filtration of $\De$ 
is required in order to prove the locality of $\De$ so as to restore 
the usefulness of the consistency conditions.
However, these nasty irrelevant contributions are of 
the reducible type 
(see the form of $L_1$ in \re{L}), and we 
expect they will disappear when taking the Legendre transform,
similarly to what happened in passing from the Wilsonian effective action 
to the cutoff effective action.
Again, from a perturbative point of view, instead of studying $\De$ 
it is convenient to introduce 
\cite{bv,mt,ellw} its Legendre transform $\DG$, in which reducible
contributions are absent. 
Recalling \re{ZL} which relates $\se[\Phi,\g;\L]$ 
to $W[J,\g;\L]$, and using \re{Leg} we find 
\beeq
&&\DG[\Phi,\g;\L]=-\spint\lq \Kiu(p) 
\dfud{\G}{\Phi_a(-p)}\, \dfud{\G}{\g_a(p)}
+  
\frac{\Ki(p)}{\K(p)}\,D^{-1}_{ab}(p)\,
 \Phi_a(p) \dfud{\G}{\g_b(p)}\rq \no \\
&& \phantom{\DG[\Phi,\g;\L]}
-i\,\hbar \spqint \frac{\Ki(p)}{\K(p)}\,D^{-1}_{ab}(p)\,
(-)^{\de_a}
\frac{\de^2 W}{\de J_a(p) \de J_c(q)} \,
\frac{\de^2 \G}{\de\Phi_c(-q) \de \g_b(-p)} 
\,,
\eeeq 
where $\de^2W/\de J\de J$ is that functional of $\Phi$ and $\g$
appearing in the inversion eqs. \re{inversionw} and \re{wint}.
Finally, after performing such an inversion, 
the cutoff ST identity reads
\beq\nome{deltagamma}
\DG[\Phi,\g;\L]\equiv\DGb +\DGh=0\,,
\eeq 
with
\beq \nome{dgb'}
\DGb=
-\spint \Kiu(p) 
\dfud{\G}{\Phi_a(-p)}\dfud{\G}{\g_a(p)}
-
\spint \frac{\Ki(p)}{\K(p)}D^{-1}_{ab}(p)
\Phi_a(p) 
\dfud{\G}{\g_b(p)}
\eeq
and
\beeq\nome{deltagh}
&&\DGh = i \hbar \spqint \Ki(p)  
\Bigg\{
(-1)^{\de_c} 
\lp \G_2^{-1}(q;\L) \,\bG(-q,-p;\L)\rp_{bc} 
-\de_{bc}\, 
\de^4(p-q) \Bigg\}
 \nonumber \\
&&\phantom{\DGh = -i\spqint} \!\!\!
\times\lp \G_2^{-1}(p;\L) \, D_{\L\L_0}^{-1}(p) \rp_{ca} \,
\frac{\de^2\, \G}{\de\Phi_b(q)\, \de \g_a(p)} 
\,.
\eeeq
Notice that at $\L=0$ the cutoff ST identity reduces to $\DGb(0)=0$
and, in the UV limit, becomes the usual ST identity \re{ST}. Moreover we have 
inserted the factor $\hbar$ in \re{deltagh} to put into evidence that $\DGh$ 
vanishes at the tree level.

The expression of $\DG$ is simpler in terms of  $\Pi$,
defined in \re{defpi}.  This  functional
 differs from the cutoff effective action only in the tree-level 
two-point function, in which the IR cutoff has been removed. With
such a definition, in the \UV limit the tree-level contribution to $\Pi(\L)$
coincides with $\sbrs$, whereas at the tree level  $\G_2(\L)$ contains
the IR cutoff (see \re{gamma2}).
In terms of $\Pi$ the functional  $\DGb$ can be rewritten 
as  
\beq\nome{dgb}
\DGb[\Phi,\g;\L]=
-\spint \Kiu(p)\, 
\dfud{\Pi[\Phi,\g;\L]}{\Phi_a(-p)}\, \dfud{\Pi[\Phi,\g;\L]}{\g_a(p)}\,.
\eeq
Recalling the ST identity for the physical effective action
\beq\nome{lastqap}
\SS_{\G}\G[\Phi,\g]=0\,,
\eeq
where
\beq\nome{slavphys}
\SS_{\G}=\spint
\lp  \dfud{\G}{\Phi_a(-p)}\,
\dfud{}{\g_a(p)} +
\dfud{\G}{\g_a(p)}\,
\dfud{}{\Phi_a(-p)}\rp\,
\eeq
we see that in the \UV limit
\beq\nome{dgbl}
\DGb[\Phi,\g;\L]\to\SS_{\Pi(\L)}\Pi(\L) \quad\quad{\mbox{for}}\;\;
\L_0\to\infty
\eeq
at any $\L$. The existence of such a limit is guaranteed in
perturbation theory by the UV finiteness of the cutoff effective
action (perturbative renormalizability). In order to show this
property holds also for $\DGh$, it suffices to recognize that the
presence of cutoff functions having almost non-intersecting supports
forces the loop momenta in \re{deltagh} to be of the order of $\L$.
Henceforth we will take the \UV limit in $\DG$.  

\section{Perturbative solution of $\bom{\DG=0}$}

The proof of the ST identity \re{deltagamma} in the RG formalism, with
possible anomalies, is based on induction in the loop number and
is discussed in
\cite{bdm3,bdm4,bdm5,mt,bv,susy}.  
Once again the flow equation for the cutoff ST identity is found by explicit 
derivation \cite{mt}. Its expression looks quite involved
\beeq\nome{evST} 
&&\LdL \D_\G= -\half \int_{p,q,r}
[\LdL K_{\L\L_0}^{-1}(p)]  (-1)^{\de_a} \, D_{ab}^{-1}(p)
\G_{2\;be}^{-1}(-p)\, \bG_{ef}\,(-p,-r;\L) \no
\\ 
&& \phantom{\LdL \D_\G= -\half \int_{p,}}\times
\G_{2\;fd}^{-1}(-r)\,
\frac{\de^2 \DG}{\de\Phi_c(-q) \de\Phi_d(r)}
\G_{2\;cg}^{-1}(q)\, \bG_{gl}\,(q,p;\L)\,\G_{2\;la}^{-1}(p)\,,
\eeeq
but it is easy to grasp its meaning, that is the evolution of the
vertices of 
$\DG$ at the loop ${\ell}$ depends on vertices of $\DG$ itself at lower loop
order.

Therefore, if $\DG^{(\ell')}=0$ at any loop order $\ell'<\ell$,
then
\beq\nome{chie}
\LdL\DG^{(\ell)}=0\,.
\eeq
We can thus analyse $\DG$ at an arbitrary value of $\L$. 
There are two natural choices 
corresponding to $\L=0$ and $\L=\L_R$ much bigger than the subtraction
scale $\mu$, \ie $\L_R=\L_0$. With the former the gauge symmetry condition 
fixes the relevant part of the effective action 
in terms of the physical coupling $g(\mu)$ and provides the boundary
conditions of the RG flow, whereas with the latter 
the gauge symmetry condition determines  the cutoff dependent 
bare couplings.
With this choice the implementation of symmetry is simplified
due to the locality~\footnote{Here and in the following
a functional is said to be local when it contains only couplings 
with non-negative dimension.} of the 
functionals involved. Although the computation of physical 
vertices is generally cumbersome, this second possibility is more 
convenient in the computation of quantities which do not evolve with
the cutoff $\L$. To show how things work, in the following chapters we 
will compute ---starting
from $\DG$--- the beta function for the massless scalar theory and the
gauge anomaly.

We now discuss the vanishing of $\DG$.
Also  for this functional we define its relevant part, isolating all 
monomials in the fields, sources and their derivatives 
with  dimension three. The rest is included in $\DGi$.
At the UV scale $\L_0$ the functional $\DGh (\L_0)$ is schematically
represented in fig.~1.
\begin{figure}[htbp]\nome{figdelta}
\epsfysize=4cm
\begin{center}
\epsfbox{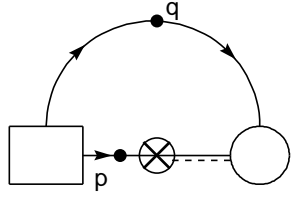}
\end{center}
\caption{\small{Graphical representation of $\DGh(\L_0)$. The box and 
the circle represent the functionals $\bG$ and
$\Pi$ respectively. The top line is the cutoff full propagator of the 
field $\Phi_b$;
the bottom full line represents the field $\Phi_a$ while the double
line is the corresponding source $\g_a$.
The cross denotes the product of the two functionals with the
insertion of the cutoff function $\Kiu(p)$.
Integration over the loop momentum is understood.}}
\end{figure}
\noindent
At this scale $\DG$ is local, or, more precisely, 
$\DGi(\L_0)={\cal O}(\frac1{\L_0})$,
so that the irrelevant contributions disappear in the \UV limit.  This
can be understood with the following argument.  From \re{dgbl},
$\;\DGb(\L_0)$ is manifestly relevant, since $\Pi(\L_0)=\Pir(\L_0)$,
while it is easy to convince oneself that
$\DGh(\L_0)=\DGhr(\L_0)+{\cal O}(\frac1{\L_0})$.  As a matter of fact,
from \re{deltagh} we notice that irrelevant terms may arise from
$\bG[\Phi,\g;\L_0]$ and the cutoff functions. At $\L_0$, $\bG$ is
given by either a relevant vertex or a sequence of relevant vertices
joint by propagators with a cutoff function $\Kin(q+P)$, where $P$ is
a combination of external momenta (see \re{gammab}).  Since the
integral is damped by these cutoff functions, only the contributions
with a restricted number of propagators survive in the
\UV limit. We can infer from power counting that they are of the 
relevant type. 
A similar argument holds for the possible irrelevant contributions 
coming from $\Kiu(p)$. Then \re{chie} ensures the locality of
$\DG(\L)$ at any $\L$.

\subsection{Solution of $\bom{\DG=0}$ at $\bom{\L=\L_0}$}

Once the locality of $\DG(\L)$ is shown, the solvability of the
equation $\DG(\L)=0$ can be proven using cohomological methods
\cite{brs,becchi0,piguet}.  This is a consequence of the $\L$-independence
of $\DG$ and the solvability of the same equation at $\L=0$, where the
cohomological problem reduces to the standard one.
Henceforth we will consider the first loop, the generalization to
higher loops being straightforward due to the iterative nature of the
solution.
Using \re{dgbl}, at  $\L=\L_0$ and at the first loop  \re{deltagamma} reads 
\beq \nome{fintun}
{\cal S}_{\Pi^{(0)}}\,\Pi^{(1)}(\L_0)
\,+\,\DGhr^{(1)}(\L_0)=0\,.
\eeq
This fine-tuning equation allows to fix some of the relevant 
couplings in $\Pi^{(1)}(\L_0)$. As a matter of fact 
the most general functional $\Pi^{(1)}(\L_0)$ can be cast into the form 
\re{pir} and  split into two contributions
\beq \nome{pigreco1}
\Pi^{(1)}(\L_0)=\Piinv^{(1)}(\L_0)+\Pit^{(1)}
(\L_0)\, ,
\eeq
where $\Piinv$ contains all the independent monomials which are
invariant, \ie ${\cal S}_{\Pi^{(0)}}\, \Piinv^{(1)}=0$. 
The remaining monomials contribute to $\Pit$.
Inserting \re{pigreco1} into \re{fintun}, we find
$$
{\cal S}_{\Pi^{(0)}}\, \Pit^{(1)}(\L_0)\,=-\, \DGh^{(1)}(\L_0)\,,
$$
which yields the couplings in $\Pit^{(1)}$ since $\DGh^{(1)}(\L_0)$
depends only on $\sbrs$.  An explicit calculation shows that the only
divergences are powers of $\L_0$ according to the dimension of the
relative vertex. In particular dimensionless couplings are finite, due
to the presence in \re{deltagh} of cutoff functions having almost
non-intersecting supports. In chapter 5 we will perform an  explicit
computation of some of these couplings in QCD. 
\chapter{The breaking of dilatation invariance: the Callan-Symanzik equation}

Let us go back to the massless $\phi^4$ theory, defined at the
classical level by the 
classical action
\beq\nome{sclsc}
\scl[\phi]= -\half \int \d4 p \, p^2  \, \phi(-p) \, 
\phi(p)   +\frac {g}{4!} \int_x \phi^4(x) \, .
\eeq
The classical theory is invariant under a dilatation transformation,
which, in the infinitesimal version, reads
\beq\nome{dilat}
\de \phi(x) = \a \,(1 + x_\m \p_\m ) \,\phi(x)\,.
\eeq 
Notice that this invariance is lost in presence of a mass term; in
that case the dilatation invariance is said to be softly broken.
The symmetry expressed in \re{dilat} is of the global type, the
infinitesimal parameter   $\a$ being constant.
Thus this invariance can be translated into a classical Ward identity
\beq\nome{wardclas}
W_D \scl \equiv \int_x (1 + x_\m \p_\m ) \, \phi(x) \frac{\de
\scl}{\de \phi(x)}=0\,.
\eeq
Let us consider the quantum extension of eq. \re{dilat}. We know the effective
action $\G[\Phi]$ has to fulfil the normalization conditions
\re{norm}.
Using the Quantum Action Principle eq. \re{gammaqap} for dilatation
invariance $W_D$ \re{wardclas}  we get
\beq\nome{wardq}
W_D \G=[\D\G]\,,
\eeq
where $\D$ is an even integrated insertion of dimension $4$. A basis
for $[\D\G]$ may be provided by the quantum extension of the terms
appearing in the classical action \re{sclsc} plus a mass term, or,
equivalently, by the three insertions
\beq\nome{insertion}
N\G\equiv \int_x \phi (x) \frac{\de\G}{\de\phi (x)}\,, \quad\quad
\p_g\G\,, \quad\quad  \D_m\G\equiv \int_x N_2[\phi^2](x)\,.
\eeq
The expansion of $[\D\G]$ in this basis yields the anomalous
dilatation Ward identity
\beq\nome{wanom}
W_D\G=(\g N-\b\p_g)\G+\a\,\D_m\G 
\eeq
where $\a$, $\b$, $\g$ are formal power series in $\hbar$ starting
from the order $\hbar$.
The term $\g N\G$ in the r.h.s. may be absorbed in the l.h.s. by
renormalizing the dilatation dimension of $\phi (x)$. As a consequence
the infinitesimal transformation \re{dilat} in eq. \re{wardq} has to
be replaced by 
\beq\nome{dilatren}
\de' \phi(x) = \a \, (1 -\g+ x_\m \p_\m ) \,\phi(x)\,,
\eeq 
calling $\g$ the anomalous dimension of the field $\phi$. However, 
even by renormalizing the dimension of the field, asymptotic invariance,
\ie invariance possibly broken by a soft mass term, can not be
maintained at the quantum level, since the second term in the
r.h.s. of eq. \re{wanom} is still present.

The anomalous Ward identity \re{wanom} can be rewritten in the form
of a parametric equation  with the help of the following identity
arising from  dimensional analysis  
\beq\nome{dimensional}
(\m \p_\m + W_D)\G=0\,,
\eeq
with $\m$ the normalization point introduced in
\re{norm}~\footnote{Recall the  introduction of a  non-vanishing
subtraction point, being
$\phi$ massless, is
required to avoid IR divergences.}.
Combining eqs. \re{wanom} and \re{dimensional} we derive the
Callan-Symanzik equation \cite{callan}
\beq\nome{callan}
(\m \p_\m + \b\p_g - \g N) \G = \a \, \D_m\G \,,
\eeq
which controls the asymptotic behaviour of the theory at large
euclidean momenta, when the soft mass insertion $\D_m\G$ becomes negligible.
Moreover, dimensional analysis enables us to identify $\b$ with the
Gell-Mann Low beta function for the coupling $g$ \cite{gell}
\beq\nome{betaf}
\b= \m \, \p_\m g
\eeq
and the anomalous dimension $\g$ with
\beq\nome{andim}
\g=-\half \mdm \log z\,,
\eeq
$z$ being the wave function renormalization of the field $\phi$.

\section{Dilatation invariance in the RG}

How symmetries can be implemented in the Wilson RG has been
extensively discussed in the previous chapter. We treated there the
general case of  gauge invariance, that is invariance of the classical
action under a local transformation of the fields. Anyway, it is
straightforward to adapt the derivation of the functional $\DG$
introduced in sect.~3.2 to dilatation transformations. As we are in
presence of a global symmetry, the ST identity can be replaced by a
Ward identity and we may think to the BRS source as a global,
anticommuting parameter.

We now want to formulate an effective dilatation Ward identity for the
cutoff effective action $\G(\L)$ (or its relative $\Pi$).
Let us start  from the anomalous Ward identity \re{wanom}. 
We expect  the presence of the scale $\L$ in the cutoff effective
action will make such a relation  unreliable. In fact what happens is
that by requiring the physical effective action to be independent of
the normalization point  $\m$ we force the functional $\Pi(\L)$ 
to obey a modified, effective identity, and this at any scale 
$\L$~\footnote{For an alternative derivation see \cite{prr}.}.
Since at the physical point $\L=0$ the functional  $\Pi$ coincides
with the physical effective action, we demand our modified identity to
reproduce eq. \re{callan}.  
We know from chapter 3 that in  the RG formulation the breaking of a
symmetry ---which in our case is the invariance of $\G$ under the
operator $\m \p_\m + W_D$--- is expressed by the functional $\DG$
introduced in sect.~3.2.
With the help of eq. \re{deltagamma}, the functional $\DG$ can be written as
\beq\nome{rgward}
(\mdm +W_D)\, \Pi(\L)+\DGh (\L)=\DG(\L)\,.
\eeq
On the other hand the functional $W_D\, \Pi(\L)$ can be expanded in the basis
\re{insertion}
\beq\nome{insereff}
W_D\G(\L)=(-\b(\L)\,\p_g +\g(\L)\, N)\,\Pi (\L)+\D_m\G(\L)\,
\eeq
where $\b(\L)$ and $\g(\L)$ are suitable coefficients. Taking into
account that $\Pi(\L=0)=\G$ and $\DGh(\L=0)=0$ (see \re{deltagh}), at the
physical point the functions $\b(\L)$ and $\g(\L)$ equal  the
corresponding physical ones
\beq\nome{bgphys}
\b(\L=0)= \b\,, \quad\quad \g(\L=0)=\g
\eeq
and the QAP tells us they are determined by physical vertices.

At the UV scale, instead, $\Pi(\L_0)$ is relevant and has the form
\re{intaction}, with the couplings $\s_i^{(B)}$ given by the relevant
couplings $\s_i(\L)$ evaluated at $\L=\L_0$. 
At this scale the symmetry of the physical effective action 
expressed by \re{callan} becomes
\beq\nome{diluv}
(\mdm +W_D) \left\{\half \int_p 
\phi(-p)\bigl[ (1+\s_1(\L_0)) p^2 + \s_2(\L_0)\bigr]
\phi(p)+\frac {\s_3(\L_0)}{4!} \int_x 
\phi^4(x) \right\}=-\DGh(\L_0)\,.
\eeq
We can extrapolate the values of the ``$\L$-beta function'' $\b(\L)$ and of the 
``$\L$-anomalous dimension'' $\g(\L)$ at the UV scale. 
Eq. \re{diluv} yields
$$
\b(\L=\L_0)=0\,, \quad\quad \g(\L=\L_0)=0\,.
$$
Thus in this limit it is only the mass insertion $\D_m\G(\L_0)$ which
contributes to $\DGh(\L_0)$.

Being allowed  to peruse the structure of the functional $\DG$ at an
arbitrary scale, we decide to set $\L=\L_0$ and then take the  \UV limit.
When eq. \re{diluv} is projected onto the basis  of the
monomials appearing in the classical action \re{sclsc} and in the bare
action \re{intaction}, it supplies the set of equations
\beeq
\nome{setcoupa}
\half \mdm \s_1(\L_0)
&=& \hD_{1}(\L_0)\,,\\
\nome{setcoupb}
\half \mdm \s_2(\L_0) + \s_2(\L_0)
&=&\hD_{2}(\L_0)\,,\\
\nome{setcoupc}
\mdm \s_3(\L_0)&=&\hD_{3}(\L_0)\,,
\eeeq
where $\hD_{1}$, $\hD_{2}$, $\hD_{3}$ are the coefficient of the
monomials $\half p^2\phi^2$, $\half \phi^2$, $\frac 1{4!}\phi^4$, 
respectively in
$\DGh(\L_0)$.
Eqs. \re{setcoupa}-\re{setcoupc}  can be thought of as the modified 
Ward identities for the cutoff effective action. 

It would be desirable to compute  the physical
beta function for the coupling $g$ and anomalous dimension for the
field $\phi$ in the RG formulation. Although the set of equations
\re{setcoupa}-\re{setcoupc}
enables us to get
closer to the objects we are searching, we still lack a piece of
information, that is we need to know how the functions $\b$ and $\g$
are related to the bare couplings $\s_i(\L_0)$. At this stage the RG
flow enters into the game.

It is acknowledged renormalization pertains to the UV property of the
theory. Thus we expect that $\m$-RG, governed by eq. \re{callan} which
expresses the requirement that physical observables are independent of
the specific value of $\m$, is connected to $\L$-RG, governed by
\re{eveq}, when $\L$ is in the UV region \cite{mich}. 
To find out such a connection, we rewrite eq. \re{intaction} as
\beq\nome{actionuv}
\Pi [\phi;\L_0]=\half \int_p 
\phi^{\mbox{\scriptsize{{UV}}}}(-p) \lp p^2 + \s_2^{\mbox{\scriptsize{{UV}}}} \rp
\phi(p)+\frac {g^{\mbox{\scriptsize{{UV}}}}}{4!} \int_x 
\phi^4(x) \,,
\eeq
where the UV fields and couplings are so defined
\beq\nome{baref}
\phi^{\mbox{\scriptsize{{UV}}}}=\sqrt z\,\phi\,, \quad\quad 
\s_2^{\mbox{\scriptsize{{UV}}}}=\frac{\s_2(\L_0)}{z}\,, \quad\quad 
g^{\mbox{\scriptsize{{UV}}}}=\frac{\s_3(\L_0)}{z^2}\,,\quad\quad  
z=\s_1(\L_0)+ 1\,.
\eeq
The beta function $\b(g)$ can now be inferred  either from the second
or the third of eqs. \re{baref} by demanding the
UV couplings $\s_2^{\mbox{\scriptsize{{UV}}}}$ and $g^{\mbox{\scriptsize{{UV}}}}$ are independent
of the normalization point $\m$. 
Choosing for instance the latter and recalling from dimensional
analysis both $\s_3$ and $z$ are
functions of $g(\m)$ and $\m/\L_0$, we get
\beq\nome{beta}
\b(g)=
\frac{\L_0\, \p_{\L_0}\, g^{\mbox{\scriptsize{{UV}}}}}{\p_g \, g^{\mbox{\scriptsize{{UV}}}}} =
\frac{
\L_0\, \p_{\L_0} \,\s_3  - 2\,\s_3 \,
\L_0\, \p_{\L_0}\,\log z }{
\p_g \,\s_3 -2 \, \s_3\,  \p_g \log z}\,.
\eeq
Since the  loop expansion for $z$ starts at the second order in
$\hbar$, at the first loop \re{beta} gives
\beq\nome{beta1l}
\b^{(1)}= \L_0\, \frac{\p \s_3^{(1)}}{\p \L_0}\,.
\eeq
Thus dimensional analysis and the Ward identity \re{setcoupc} provide
us with a recipe to compute the one loop beta function, since
\beq\nome{beta1calc}
\b^{(1)}(g)=-\hD_3^{(1)}(\L_0)\,.
\eeq
A thorough perusal of the procedure which has driven us so far should
make us realize that in this perspective the beta function appears as
the anomaly of the classical  dilatation invariance, the so-called
trace anomaly. However, such an anomaly proves to be harmless, since
the counterterms introduced by $\DGh$ are the same monomials the
classical action \re{sclsc} is made of.

In addition to the calculation of the beta function first order
coefficient from \re{setcoupc}, we will directly verify  eq. \re{setcoupa} 
leads to a
vanishing one-loop anomalous dimension and determine
the coefficient $\s_2^{(1)}(\L_0)$ through \re{setcoupb}. This will be
the topic of the following section.

\section{The one-loop beta function}

The one-loop coefficient of the  beta function is given, according to 
\re{beta1calc}, by an explicit calculation of the dilatation breaking
term  $\hD_3^{(1)}(\L_0)$, which can be built from
eq. \re{deltagh}. 
Since we are dealing with a Ward identity, the
term $\frac{\de^2\, \G}{\de\Phi(q)\, \de \g(p)}$ in \re{deltagh}  has
to be read as the functional derivative of the variation \re{dilat}
with respect to $\phi$ in the momentum space, which is simply the
field independent term $(-3-p_\m\, \p/\p p_\m )\;\de \,(p-q)$.
At the first order this is multiplied by the tree-level vertices of $\bG$.
The term
proportional to $\de_{bc}$ in \re{deltagh} gives rise to a 
contribution which, being field independent, will be
neglected. Moreover, the product of the full propagator $\G_2(-p;\L)$
by $D_{\L\L_0}$ is  $\de_{Li}$ at this order.
To extract the $\phi^4$ component of $\DGh$, in eq. \re{deltagh}
we have to insert the second term in the iterative expansion of the functional 
$\bG^{(0)}$ in vertices of $\G^{(0)}$  (see fig.~2.3).
Finally, the   $\phi^4$ component of $\DGh^{(1)}(\L_0)$  reads
\beeq\nome{3delta}
&&\int_{p_1\, p_2\, p_3} \hD_3^{(1)}(p_1,p_2,p_3,p_4;\L_0) 
\phi(p_1) \, \phi(p_2) \, \phi(p_3)\, \phi(p_4) \no \\
&&\phantom{\int\;\;}=
i \, g^2 \int_{p_1\,p_2\,p\,q\,k}  
\phi(p_1) \, \phi(p_2) \,\phi(-p_1-p-k)\no \\
&& \phantom{\int_{p_1\, p_2\, p_3}\;\;}
\times \de(p-q) \,\Kiu(p)\, \frac{\Kin(k)}{k^2}\,
\lp -3 -p_\m \, \frac{\p}{\p q_\m} \rp  \, 
\left[ \frac{\Kin(q)}{q^2}\,
\phi(-p_2+q+k) \right] \no \\
&&\phantom{\int_{p_1\, p_2\, p_3} \;\;}
+ \;\; \mbox{permutations} \,,
\eeeq 
where we have used $p_\m\; \p \,\de (p-q)/ \p \,p_\m = - p_\m\; 
\p \,\de (p-q)/\p \,q_\m$
and then  integrated by parts.
By performing a translation over  integration momenta, \re{3delta}
becomes
\beeqn
&&i\frac{g^2}{16\pi^2}\int_{p_1\, p_2\, p_3\,p_4} \de\lp\sum_i p_i\rp \,
\phi(p_1) \, \phi(p_2) \, \phi(p_4) \int d^4 p\, \Kiu(p)\,
 \frac{\Kin(p+p_1+p_4)}{(p+p_1+p_4)^2}\\
&& \phantom{i\frac{g^2}{16\pi^2}\int_{p_1\, p_2\, p_3} } \times
\lp -3 -p_\m \, \frac{\p}{\p p_{3\m}} \rp \left[ \phi(p_3) \, 
 \frac{\Kin(p+\sum_i p_i)}{(p+\sum_i p_i)^2}\right]\no \\
&& \phantom{i\frac{g^2}{16\pi^2}\int_{p_1\, p_2\, p_3} } 
+ \;\; \mbox{permutations} \,.
\eeeqn
This contribution to $ \hD_3^{(1)}$ is represented in fig.~1.
\begin{figure}[htbp]
\epsfysize=5cm
\begin{center}
\epsfbox{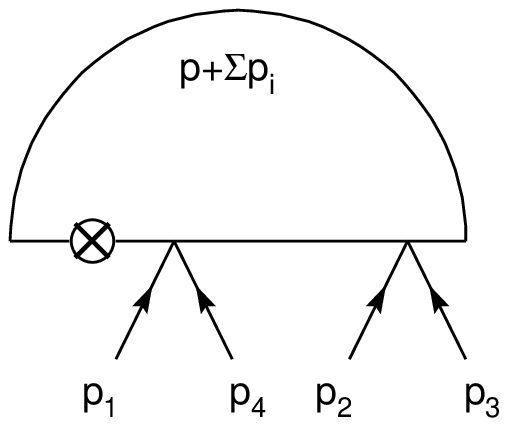}
\end{center}
\caption{\small{
 Graph contributing to $ \hD_3^{(1)}$.
}}
\end{figure}
\noindent
As we work in the limit \UV, the next step  consists in expanding into
the external momenta $p_i$ the cutoff functions. Afterwards, taking
into account the $K$'s are truly functions of $p^2/\L_0^2$ and 
exploiting the symmetry $p \to -p$, the previous expression can be
written as
\beeq\nome{inddep}
&& \frac{g^2}{16\pi^2} \int_{p_1\, p_2\, p_3} 
\phi(p_1) \, \phi(p_2) \, \phi(p_4) 
\no \\
&& \phantom{ \frac{g^2}{16\pi^2} \int_{p_1\, p_2\, p_3} }\times
\Biggl\{ \int_0^\infty \Kiu (x) \, \Kin (x) \frac{d
\Kin(x)}{d\,x} \,
\left[ \,2 + \half (p_1+p_4)_\m \, \frac{\p}{\p p_{3\m}}\right]
\,  \phi(p_3)\no \\
&& \phantom{ \frac{g^2}{16\pi^2} \int_{p_1\, p_2\, p_3} \quad}
-  \int_0^\infty \Kiu (x) \, \frac{\Kin^2 (x)}{x} \,
\left[ -1+ \half (p_1+p_4)_\m \, \frac{\p}{\p p_{3\m}}\right]
\,  \phi(p_3)
\no \\
&& \phantom{ \frac{g^2}{16\pi^2} \int_{p_1\, p_2\, p_3} \quad}
+ \OO(P^2/\L_0^2)\Biggr\}  + \;\; \mbox{permutations} \,,
\eeeq 
where $p_4=-p_1-p_2-p_3$,
$x=p^2/\L_0^2$, $P$ is  a combination of the momenta $p_i$.
Notice that whereas the second line of eq. \re{inddep} yields a cutoff
independent result, the integral over $x$ in the third line could
generate a cutoff dependent contribution.
Nevertheless, after adding the other five contributions from
permutations in the momenta $p_i$, so that $1 \to 6$ and $1/2 \,  (p_1+p_4)_\m 
\to -2\, p_{3\m}$, the ``would-be'' cutoff-dependent part disappears
due to classical dilatation invariance of the quartic term in the
fields
\beq\nome{qdil}
\int_{p_1\, p_2\, p_3} \, \phi(p_1) \, \phi(p_2) \, \phi(p_4) 
\lp -3 -p_\m \, \frac{\p}{\p p_{3\m}} \rp\, \phi(p_3) =0\,.
\eeq
At the same time, implementing \re{qdil} in \re{inddep} and taking the \UV 
limit, we
find~\footnote{The correct normalization factor $1/4!$, following
after symmetrization, is inserted.}
\beeq\nome{res3} 
&&\int_{p_1\, p_2\, p_3} \hD_3^{(1)}(p_1,p_2,p_3,p_4;\L_0) 
\phi(p_1) \, \phi(p_2) \, \phi(p_3)\, \phi(p_4)  \sim \no\\
&& \frac{g^2}{16\pi^2} \, \frac 1{4!}\,
\int_{p_1\, p_2\, p_3}  
\phi(p_1) \, \phi(p_2) \, \phi(p_3) \, \phi(p_4) \cdot 18 \cdot \lp
\frac 12 -\frac 13 \rp = \no \\
&&\frac{3\,g^2}{16 \, \pi^2}\, \frac 1{4!}\,
\int_x \phi^4 (x) \,.
\eeeq
As a matter of fact, the result of the integration over $x$ in the
second line of \re{inddep} is determined only by the values
$\Kin(0)=0$ and $\Kin(\infty)=1$ and therefore is independent of the
choice of the cutoff function. As a consequence, recalling \re{beta1calc},
in the UV limit we
recover the  one-loop beta function for the massless scalar
theory
\beq\nome{onebeta}
\b^{(1)}(g)=
\frac{3\,g^2}{16 \, \pi^2}\,.
\eeq

\section{Computation of $\bom{\g^{(1)}}$ and $\bom{\s_2^{(1)}(\L_0)}$ via 
$\bom{\hD^{(1)}_2(\L_0)}$} 

We now compute the $\phi^2$ one-loop coefficient of $\hD (\L_0)$. 
To extract such a term, we appreciate that it originates from the tree-level
irreducible vertex of $\bG$, \ie the four-point vertex of $\G$.
At the first order and in the UV limit, eq. \re{deltagh} gives
\beeq\nome{2delta}
&&\int_{p_1} \hD_2^{(1)}(p_1,p_2;\L_0) \,
\phi(p_1) \, \phi(p_2) =- \frac{ig}2 \int_{p_1\,p\,q} \de(p-q) \,
\Kiu (p) 
\, \phi(p_1)
\no \\
&&\phantom{\int_{p_1}\hD_2^{(1)}(p_1} 
\times \lp -3 -p_\m \, \frac{\p}{\p q_\m} \rp  
\left[ \, \phi(-p_1-p+q)\frac{\Kin(q)}{q^2}\right]\,
\eeeq
where again we have used $p_\m\; \p \,\de (p-q)/ \p \,p_\m = - p_\m\; 
\p \,\de (p-q)/\p \,q_\m$
and then  integrated by parts.
After a translation over  integration momenta, \re{2delta}
becomes
\beeq\nome{2deltaprimo}
&&- \frac{ig}2 \int_{p_1\,p_2\,p} \de\,(p_1+p_2)\, 
\phi(p_1) \, \Kiu(p)\\
&& \phantom{- \frac{ig}2 \int_{p_1\, p_2\,}}
\times\lp -3 -p_\m \, \frac{\p}{\p p_{2\m}} \rp \left[ \phi(p_2) \, 
 \frac{\Kin(p+p_1+p_2)}{(p+p_1+p_2)^2}\right]\no \,.
\eeeq
Such a contribution to $\hD_2^{(1)}$ is depicted in fig.~2.
\begin{figure}[htbp]
\epsfysize=5cm
\begin{center}
\epsfbox{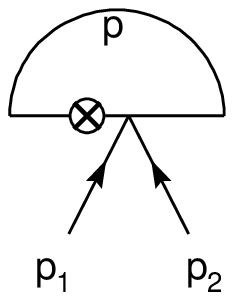}
\end{center}
\caption{\small{
Graph contributing to $ \hD_2^{(1)}$.
}}
\end{figure}
\noindent
Notice that in \re{2deltaprimo} when the derivative with respect to
$p_{2\m}$  acts on the
field, the $p$-integration vanishes due to Lorentz covariance. Thus
eq. \re{2deltaprimo} becomes 
\beq \nome{deltaprimobis}
- \frac{ig}2 \int_{p_1} 
\phi(p_1) \, \phi(-p_1) \int_p \Kiu(p)\,
\lp -3 -p_\m \, \frac{\p}{\p p_{\m}} \rp 
 \frac{\Kin(p)}{p^2}\,.
\eeq
We immediately recognize from the expression above that
$\hD^{(1)}(\L_0)=0$. Hence, by using the Ward identity \re{setcoupa}
and with the help of dimensional analysis we find out that the
anomalous dimension $\g$ vanishes at this order.
The  integral \re{deltaprimobis} produces a cutoff-dependent result which, 
together with
\re{setcoupb}, enables us to determine the coupling $\s_2(\L_0)$.
If a sharp cutoff is employed, we get
\beq 
\s_2(\L_0)=  - \frac{g}{32\pi^2} \, \L_0^2 
\eeq
to be compared with the one-loop two-point function found in
sect.~2.1.3 by solving the RG equation \re{eveq}.
\chapter{$\bom{SU(N)}$ Yang-Mills theory}

As widely discussed in chapter 3, the challenging problem with gauge 
theories  is  that local
gauge symmetry typically conflicts with the presence of a
momentum cutoff. 
What we have to show is that 
the Slavnov-Taylor  identities can be implemented perturbatively
by appropriately fixing the boundary conditions on the RG flow.
Thus the key issue is to constrain
the finite number of relevant parameters in $\Gr[\phi,\g;\L]$.
Actually it is at this stage we  implement  both the physical
parameters (such as masses, couplings, wave-function normalizations)
and the symmetry, \ie ST identities.

In ref.~\cite{b}  the exact RG 
is formulated for the $SU(2)$ Yang-Mills theory and 
the boundary conditions for
the relevant parameters in $\ser[\phi,\g;\L]$ are imposed at a non-physical
point $\L=\L'\ne 0$, so that the relevant parameters can be defined
by expanding the vertices around vanishing momenta.
Although with this choice the relevant parameters are not directly
related to the physical couplings in the effective action $\G[\phi,\g]$,
 the analysis of relevant parts of the ST identities becomes easy.
 
On the other hand, in ref.~\cite{bdm3,bdm4} the boundary conditions
for the same theory are given at the physical point $\L=0$.  In this
case the procedure of extracting the relevant parameters becomes rather
involved, since non-vanishing subtraction points appear when massless
particles are present.   The fine-tuning equation is
explicitly solved, that is the relevant parameters which 
are not fixed by the physical couplings and wave-function
normalizations are expressed in terms of physical vertices.

We propose here an alternative procedure, based on the solution of 
the fine-tuning equation at the ultraviolet scale. This will be the topic of 
the chapter.
 
\section{RG flow for $\bom{SU(N)}$ Yang-Mills theory}

In this section we introduce the cutoff effective action
for the $SU(N)$ Yang-Mills theory
and deduce the RG  flow equations. 
In the $SU(N)$ gauge theory the classical Lagrangian in the
Feynman gauge is
\beq \nome{YM}
S_{YM}=\int d^4x
\,
\left\{
-\frac 1 4 F^a_{\mu\nu}  \, F^{a\mu\nu}
- \frac 1 2 \left(\p^\mu A^a_\mu \right)^2
- \bc^a \p^\mu D_\mu^{ab} c^b
\right\}\,,
\eeq
where the gauge stress tensor and the covariant derivatives are
 given by
$F^a_{\mu\nu}=\p_\mu A^a_\nu -\p_\nu A^a_\mu
+ g \, f^{abc} A^b_\mu \,A^c_\nu$ and
$D^{ab}_\mu c^b =\p_\mu c^a + g \, f^{abc} A^b_\mu \, c^c$. Let  $\ta^a$ be
the SU(N) generator in the adjoint representation, such that 
$[\ta^{a}, \ta^{b}]=f^{abc} \ta^{c}$ and 
$\tr [\ta^{a} \ta^{b}]=\half\de^{ab}$. 
This action is invariant under the BRS transformations \cite{brs}
\beq\nome{brst}
\de A^a_\mu= \frac 1 g \eps \,D^{ab}_\mu c^b
\,,
\;\;\;\;\;\;\;
\de c^a= -\half\,\eps\, f^{abc} c^b \,c^c
\,,
\;\;\;\;\;\;\;
\de \bc^a= - \frac 1 g \,\eps\, \p^\mu A^a_\mu
\,,
\eeq
with $\eps$ a Grassmann parameter.
Introducing the sources $u^a_\mu$ and  $v^a$,   associated to 
the variations of $A^a_\mu$ and $c^a$, respectively,
we write the BRS action
\beq\nome{Stotym}
\sbrs [\Phi_I,\g_i]
=
S_{cl} +
\int d^4x
\left\{
\frac 1 g u^a_\mu  D^{ab}_\mu c^b - \half f^{abc} v^a\,c^b \,c^c
\right\}\,,
\eeq
where we have denoted by $\Phi_I$ and $\g_i$ the fields and the BRS 
sources
$$
\Phi_I=\{\,A^a_\mu, \,c^a, \,\bc^a\}
\,,
\;\;\;\;\;\;\;\;
\g_i=\{w_\mu^a,\, v^a\}
\,,
$$
and  $w_\mu^a=u_\mu^a+g\,\partial_\mu \bc^a$ (no source is introduced
for $\bc^a$). In fact $\sbrs$ depends on $u_\mu^a$ and $\bc^a$ only
through the combination $w_\mu^a$. We expect this property will hold
also for the cutoff effective action. 

We now come to the definition of the  cutoff effective action. 
We have  already appreciated that
in order to quantize the theory we need a regularization procedure of
the UV divergences. 
We regularize these divergences by assuming that in the path integral
only the fields with frequencies smaller than a given
UV cutoff $\L_0$ are integrated out. This procedure is equivalent 
to assume  the free
propagators vanish for $p^2 > \L_0^2$.
Then the physical theory is recovered in  the limit \UV.
The generating functional is defined in eq. \re{Z}
\beq\nome{WYM}
Z[J,\g]=e^{iW[J,\g]}=\int {\cal D}\Phi \, \exp{i\{-\half (\Phi, 
D^{-1}\Phi)_{0\L_0}+(J,\Phi)_{0\L_0}+\si[\Phi,\g;\L_0]\}}
\,,
\eeq
with  the cutoff scalar products between fields and sources  given by
\beq
\half (\Phi,\, D^{-1}\, \Phi)_{\L\L_0}\equiv
\int_p\, \, K^{-1}_{\L\L_0}(p)
\, p^2 \left[\half A^a_\mu(-p)\,A^a_\mu(p)
- \bc^a(-p)\, c^a(p)\right] \,,
\eeq
\beeq
&&(J,\Phi)_{\L\L_0}\equiv
\int_p \, K^{-1}_{\L\L_0}(p) 
\big\{j^a_\mu(-p)\, A^a_\mu(p) \no \\
&& \phantom{\half (\Phi,\, D^{-1}\, \Phi)_{\L\L_0}\equiv \int_p}
+ [\bchi^a(-p) - \frac i g p_\mu u^a_\mu(-p)] \, c^a(p) +
\bc^a(-p)\, \chi^a(p) 
\big\}\,.
\eeeq
The introduction of the cutoff function in the propagators breaks the 
gauge invariance, therefore the UV action $\si$ must contain 
all possible renormalizable interactions which are needed 
to compensate the symmetry breaking induced by the regularization. 
These are given by the monomials in the fields, BRS sources and
their derivatives which have dimension
not larger than four and are Lorentz and $SU(N)$ scalars, 
since Lorentz 
and global  gauge symmetries are  preserved by the cutoff 
regularization.
The independent monomials of the Yang-Mills sector are 
\beeq\nome{mon1}
&&
\tr(A_\mu \, A_\mu) \,,
\;\;\;\;\;\;\;
\tr(\partial_\nu A_\mu \, \partial_\nu A_\mu) \,,
\;\;\;\;\;\;\;
\tr(\partial_\mu A_\mu \, \partial_\nu A_\nu) \,,
\;\;\;\;\;\;\;
\tr(w_\mu \, \partial_\mu c )\,,\nonumber
\\
&&
\tr(A_\mu\, A_\mu \, \partial_\nu A_\nu) \,,
\;\;\;\;
\tr(A_\mu\, A_\nu \, \partial_\mu A_\nu) \,,
\;\;\;\;
\tr(w_\mu \, c \, A_\mu) \,,
\;\;\;
\tr(v \, c \, c )\,, 
\nonumber \\
&&
\tr(A_\mu \, A_\nu\, A_\mu \, A_\nu)\,,
\;\;\;\;\;\;
\tr(A_\mu \, A_\mu\, A_\nu \, A_\nu)\,,\nonumber
\\
&&
\tr(A_\mu \, A_\nu)\, \tr(A_\mu \, A_\nu)\,,
\;\;\;\;\;\;
\tr(A_\mu \, A_\mu)\, \tr(A_\nu \, A_\nu)\,,
\eeeq
where $A_\m=A^a_\m\, \ta^a$, $c=c^a\, \ta^a$, $w_\m=w^a_\m\, \ta^a$, 
$v=v^a\, \ta^a$ and the trace is over the adjoint representation. 
Notice that in the $SU(2)$ case there are  only one monomial 
with three
gauge fields and  two independent monomials with four gauge fields, 
while for $SU(3)$ only three monomials with four gauge fields are 
independent. 
 All these interactions appear in $\si$ with couplings of 
non-negative dimension (relevant parameters) which depend on $\L_0$.
In order to obtain the physical theory we have to show not only  that
these relevant parameters can be fixed in such a way that the \UV limit 
can be taken by fixing the physical parameters such
as the masses, the coupling $g$ and the wave function normalization 
constant at a subtraction point $\mu$, but also that
in the \UV limit the Slavnov-Taylor identities are satisfied.
Perturbative renormalizability ensures the first requirement can 
be fulfilled
\cite{p,b,bdm2,bdm3}.

According to Wilson, we then integrate the fields with frequencies
$\L^2<p^2<\L_0^2$ and define the Wilsonian effective action
$\se[\Phi,\g;\L]$.
The requirement that the generating functional \re{Z} is independent
of $\L$ gives rise to a flow equation  \re{erg1}, \re{eveq} for $\se$
or for its Legendre
transform  $\G[\Phi,\g;\L]$, respectively. All this procedure was
studied in detail in sect. 1.1, 1.2.

\subsection{Relevant parameters}

In order to set the
boundary conditions it is useful to separate relevant vertices from  
irrelevant ones. The relevant couplings  are involved in the 
$SU(N)$ singlets vertices with
$$
n_A+n_c+2n_w+2n_v \le 4\,,
$$
where $n_i$ is the number of fields of  type $i$.
In the case  we will mainly concentrate upon, that is $SU(2)$, the
most general form of the relevant part of the functional $\Pi$ is
contained in 
\beeq\nome{rel}
&&\Pi[\Phi,\g;\L] = \frac 1 2 \int_p \Pi_{\mu\nu}^{(AA)}(p;\L) A^a_\mu(-p)
A^a_\nu(p)+
\frac {1} {3!} \eps^{abc} \int_p \int_q \G_{\mu\nu\rho}^{(3A)}(p,q,r;\L)
A^a_\mu(p) A^b_\nu(q) A^c_\rho(r) \no \\
&&\phantom{\Pi[\Phi,\g;\L] }
+ \frac {1} {4!} \int_p \int_q \int_r
\G_{\mu\nu\r\s}^{(4A)abcd}(p,q,r,h;\L)
A^a_\mu(p) A^b_\nu(q) A^c_\r(r) A^d_\s(h) \no \\
&&\phantom{\Pi[\Phi,\g;\L] }
+ \int_p \G_{\mu}^{(wc)}(p;\L) w^a_\mu(-p) c^a(p)
+ \eps^{abc} \int_p \int_q  \G_{\mu\nu}^{(wcA)}(p,q,r;\L)
w^a_\mu(p) c^b(q) A_\nu^c(r) \no \\
&&
\phantom{\Pi[\Phi,\g;\L] }
+ 
\frac 1 2 \eps^{abc} \int_p \int_q
\G^{(vcc)}(p,q,r;\L) v^a(p) c^b(q) c^c(r) \,,
\eeeq
where $r=-p-q$, $h=-p-q-r$ (recall that the functionals $\Pi$ and $\G$
differ only in the tree-level two-point function). In \re{rel} the
relevant couplings are defined as follows.

\par\noindent
1) The vector propagator  has the structure
\beq\nome{propp}
\Pi_{\mu\nu}(p;\L)=
-g_{\mu\nu}\, p^2 +
g_{\mu\nu}\Pi_L(p;\L) + t_{\mu\nu}(p)\,\Pi_T(p;\L)\,,
\eeq
with
$$
t_{\mu\nu}(p)\equiv g_{\mu\nu}\,p^2\;-\;p_\mu p_\nu \,.
$$
The three relevant couplings are defined by
\beq\nome{sl}
\Pi_L(p;\L)=\s_{m}  (\L)+p^2 \s_\alpha(\L) + \Sigma_L(p;\L)\,,
\;\;\;\;\;
\Sigma_{L}(0;\L)=0\,,
\;\;\;\;\;
\frac{\partial \Sigma_{L}(p;\L)}{\partial p^2}|_{p^2=\mu^2}=0\,,
\eeq
and
\beq\nome{stran}
\Pi_T(p;\L)= \s_A(\L) + \Sigma_T(p;\L)\,,
\;\;\;\;
\Sigma_{T}(p;\L)\vert_{p^2=\m^2}=0\,,
\eeq
From these conditions we can factorize 
a dimensional function of $p$ in the vertices $\Sigma_{L,T}$. 
Thus they are irrelevant
and contribute to $\Gir(\L)$.

\par\noindent
2) The contribution to $\G(\L)$ due to three vectors can be written as
\beeq\nome{3A}
&&\G_{\mu\nu\r}^{(3A)}(p,q,r;\L)=
[(p-q)_\r g_{\mu\nu} +(q-r)_\mu g_{\nu\r} + (r-p)_\nu g_{\mu\r}]\;
[\s_{3A}+\S^{(3A)}(p,q,r)] \no \\ 
&& \phantom{\G_{\mu\nu\r}^{(3A)}(p,q,r;\L)\,}
+\tilde \G_{\mu\nu\r}^{(3A)}(p,q,r)\,.
\eeeq
In the last term all the three Lorentz indices are carried by external
momenta. Hence, after these have been factorized, the remnant is a
function of dimension $-2$. Then the vertex
$\tilde \G_{\mu\nu\r}^{(3A)}(p,q,r;\L)$ is irrelevant.
Also  $\S^{(3A)}(p,q,r;\L)$ is irrelevant, being defined by
$$
\S^{(3A)}(p,q,r;\L)|_{3SP}=0 \,.
$$

\par\noindent
3) The contribution of the four vectors to $\G(\L)$  is given by two
different $SU(2)$ scalars
\beeq\nome{4A}
&&\G_{\mu\nu\r\s}^{(4A)abcd}(p,q,r,h)
=t^{abcd}_{1;\mu\nu\r\s}[\s_{4A}+\S_1^{(4A)}(p,q,r,h)]
+t^{abcd}_{2;\mu\nu\r\s}[\s'_{4A}+\S_2^{(4A)}(p,q,r,h)]
\no \\ && \phantom{\G_{\mu\nu\r\s}^{(4A)abcd}(p,q,r,h)\,}
+\tilde \G_{\mu\nu\r\s}^{(4A)abcd}(p,q,r,h)\,,
\eeeq
where
$$
t^{a_1 \cdots a_4}_{1;\mu_1 \cdots \mu_4}
=\left(\eps^{a_1a_2c}\eps^{ca_3a_4}-\eps^{a_1a_4c}\eps^{ca_2a_3}
\right)g_{\mu_1\mu_3}g_{\mu_2\mu_4} + (2\leftrightarrow 3) +
(3\leftrightarrow 4)
$$
is the four-vector $SU(2)$ structure appearing in the BRS
action and
$$
t^{abcd}_{2;\mu\nu\r\s}=
(\de^{ab}\de^{cd}\,+\,\de^{ac}\de^{bd}\,+\,\de^{ad}\de^{bc})\;
(g_{\mu\nu} g_{\r\s}\, +\, g_{\mu\r} g_{\nu\s}\,+\,g_{\mu\s}
g_{\nu\r})\,.
$$
In the  vertex $\tilde \G_{\mu\nu\r\s}^{(4A)abcd}$ at least two 
Lorentz indices are carried by external
momenta and thus it is irrelevant.
The two relevant couplings $\s_{4A}(\L)$ and $\s'_{4A}(\L)$ are defined by
$$\S_1^{(4A)}(p,q,r,h)|_{4SP}=0\,, \;\;\;\;\;\;
\S_2^{(4A)}(p,q,r,h)|_{4SP}=0\,.
$$

\par\noindent
4) The $w$-$c$ vertex reads
\beq\nome{bcc}
\Pi_\mu^{(wc)}(p)=\frac{p_\mu}{g}[-i+\s_{wc}+\S^{(wc)}(p)]\,,
\eeq
with a relevant coupling  defined through
$$
\S^{(wc)}(p)|_{p^2=\mu^2}=0\,.
$$
As to the ghost propagator, it can be  expressed in terms of  the vertex 
$\Pi_\mu^{(wc)}(p)$
\beq\nome{ghprop}
\Pi^{(\bc c)}(p)=
p^2 + i p^2 [\s_{wc}+\S^{(wc)}(p)]\,.
\eeq

\par\noindent
5) The contribution of  $w$-$c$-$A$ to $\G(\L)$ is given by the
$SU(2)$ scalar
\beq\nome{wca}
\G_{\mu\nu}^{(wcA)}(p,q,r)=g_{\mu\nu}[\s_{wcA} +\S^{(wcA)}(p,q,r)]
+\tilde \G_{\mu\nu}^{(wcA)}(p,q,r)\,.
\eeq
The vertex $\tilde \G_{\mu\nu}^{(wcA)}(p,q,r;\L)$  is irrelevant since
the two Lorentz indices are carried by external momenta. 
Thus the relevant  coupling $\s_{wcA}(\L)$ is defined by
$$
\S^{(wcA)}(p,q,r;\L)|_{3SP}=0 \,.
$$
Due to the fact that $\G$ depends on $\bc$ and $u_\m$ only through the
combination $w_\mu$,  the coefficient of $\bc$-$c$-$A$ is
\beq\nome{bccA}
\G_{\nu}^{(\bc cA)}(p,q,r)=-ig\, p_\mu \, \G_{\mu\nu}^{(wcA)}(p,q,r)\,.
\eeq

\par\noindent
6) Finally, the only vertex involving the source $v$
and containing a relevant coupling is 
\beq\nome{vcc}
\G^{(vcc)}(p,q,r)=\s_{vcc}+ \S^{(vcc)}(p,q,r)\,,
\;\;\;\;
\S^{(vcc)}(p,q,r;\L)|_{3SP}=0\,.
\eeq
At the tree level $\Pi^{(0)}=\sbrs$, so that the relevant couplings
in the tree approximation have the following values
\begin{eqnarray}\nome{sigmatree}
&&
\s_{m}^{(0)}(\L)=\s_{\alpha}^{(0)}(\L)=
\s_{A}^{(0)}(\L)=\s_{wc}^{(0)}(\L)=\s_{4A}^{'(0)}(\L)=0\,,\no
\\ &&
\s_{3A}^{(0)}(\L)=-ig
\,,\;\;\;
\s_{4A}^{(0)}(\L)=-g^2
\,,\;\;\;
\s_{wcA}^{(0)}(\L)=\s_{vcc}^{(0)}(\L)=-1\,.
\end{eqnarray}

All the remaining vertices, being  coefficients of
monomials with dimension higher than four,  are irrelevant and
therefore belong to $\Gir(\L)$.

\subsection{Boundary conditions}

As discussed in chapter 1, for the irrelevant vertices
we assume the following boundary condition
\beq\nome{bcirr}
\Gir[\Phi,\g;\L_0] =0 \,,
\eeq
since, due to dimensional reasons, they must vanish at the UV scale to
ensure perturbative renormalizability.
Then at $\L=\L_0$ the cutoff effective action becomes  local and
corresponds to the bare action. 

As regards  the relevant part, we first have  to address the
fundamental issue of symmetry. That will be done in the next section.

\section{Effective ST identities} 
The gauge symmetry of the classical action \re{YM} translates into 
the ST identity for the effective action 
$\G[\Phi,\g]\equiv\G[\Phi,\g;\L=0]$, which reads 
$$
{\cal S}_{\G'} \G'[\Phi,\g]=0\,,
$$
where $\G'[\Phi,\g]=\G[\Phi,\g]-\int d^4x \, \frac 12 (\p^\m A^a_\m)^2$ and 
the  Slavnov operator \cite{becchi0} ${\cal S}_{\G'}$ was defined in 
\re{slavphys}.

In the context of RG, the ST identities are recovered by imposing the
constraint  \re{deltagamma}, which is exploited to set the boundary conditions
for the couplings 
in  the cutoff effective action (or $\Pi$).
For the  gauge theory we are considering, eqs. \re{dgb} and
\re{deltagh} specialize into
\beq\nome{dgbc}
\DGb[\Phi,\g;\L]=-\int d^4 p \; \Kiu (p)\,
\frac{\de\Pi'[\Phi,\g;\L]}{\de\Phi_i(-p)}\frac{\de\Pi'[\Phi,\g;\L]}{\de\g_i(p)}
\eeq
and 
\beeq\nome{dghc}
&&\DGh[\Phi,\g;\L]
=i\hbar\,
\int_{p,q} K_{0\L}(p)\left\{(-1)^{\de_L} \left(
\frac1{\G_2(q;\L)}\,\bG[-q,-p;\L]\right)_{JL} -\,\de_{JL}\, 
\de^4(p+q)\right\}
\nonumber\\&&
\;\;\;\;\;\;\;\;\;
\times 
\left(\frac1{\G_2(-p;\L)D_{\L \L_0} (-p)}\right)_{Li} \, 
\, \frac{\de^2}{\de\Phi_J(q)\de\g_i(p)}
\left(\Pi[\Phi,\g;\L]-\frac1g \int_{x}u_{\m}\p_{\m}c\right)
\,,
\eeeq
where $\Pi'$ is the expression obtained
by removing  the gauge fixing term in \re{defpi}.
We should observe that eq.  \re{dghc} differs from its analogue
\re{deltagh} in a twofold way.
First,  in the former the index $i$ does not run on the antighost
$\bc$ due to the fact we did not introduce a BRS source for it,
contrary to  summation over $J$ which includes $\bc$ (it propagates in
the loop). Second, the price we have to pay to eliminate the BRS
source for $\bc$ is the removal of the tree level  of $\Pi_\mu^{(wc)}$
in the argument of $\de^2 /\de\Phi\,\de\g$ in \re{dghc}.

Again we notice that at $\L=0$ and in the UV limit the gauge symmetry condition
\re{deltagamma} reduces to the usual ST identities, since $\Pi'$ becomes 
$\G'$ and $\DGh$ vanishes.

At this stage we are ready to discuss the solution of the effective ST
identity at the ultraviolet scale.

\section{Solution of the fine-tuning at $\bom{\L=\L_0}$} 

We have described in the previous section how the ST identities can been
directly formulated for the  cutoff effective action (or $\Pi$)
at any $\L$.
In this context the ST identities are recovered by imposing the
condition \re{deltagamma}, which, at the first loop and at $\L=\L_0$
becomes  eq. \re{fintun}.
In the following we discuss the solution of  this fine-tuning equation
at $\L=\L_0$
and at the first loop order.
Under these conditions, the functional
$\DGb^{(1)}$ is the
standard Slavnov operator applied to $\Pi$
$$
\DGb^{(1)}(\L_0)={\cal S}_{\Pi^{(0)}}\,\Pi^{(1)}(\L_0)\,.
$$
Let us analyse in detail the various vertices of $\DGb$, which can be
inferred from eq. \re{slavphys}.

\vskip0.2truecm
\par\noindent
(i) With two fields we have just one vertex
\beq\nome{dbac}
\BD^{(Ac)}_{\G,\mu}(p;\L_0)={\Pi'}_{\mu\nu}^{(AA)}(p;\L_0)\,  
\G_\nu^{(wc)}(p;\L_0)
\,.
\eeq
That \re{lastqap} holds at the tree level trivially follows from the
transversality of  ${\Pi'}_{\mu\nu}^{(AA)}(p)$.
At the first loop the term with $t_{\m\n}$ cancels out for the same
reason as above and what we get is
\beq\nome{dbac1loop} 
\BD^{(Ac,1)}_{\G,\mu}(p;\L_0)=
 -\frac i g \,p_\m\,[\s^{(1)}_m(\L_0)+\s^{(1)}_\a (\L_0)\,p^2]\,.
\eeq
This is no surprise, since we expected the breaking of gauge invariance.
On the other hand, we can match the values of the couplings
$\s^{(1)}_m(\L_0)/\L_0^2$ and $\s^{(1)}_\a (\L_0)$ with the finite coefficients of the
corresponding monomials in $\DGh$. This is the meaning of the fine-tuning.
We just have to prove the numbers coming out from $\DGh$ are
finite. This topic will be addressed in the following section.

\vskip0.2truecm
\par\noindent
There are two different vertices with three fields, 
$\BD^{(AAc)}_{\G,\mu\nu}$ and $\BD^{(wcc)}_{\G,\mu}$.
\vskip0.2truecm
\par\noindent
(ii) The former reads
\beeq\nome{dbacc}
&&\BD^{(AAc)}_{\G,\mu\nu}(p,q,k;\L_0)=\Pi_\r^{(wc)}(k;\L_0)\,
\G_{\mu\nu\r}^{(3A)}(p,q,k;\L_0)-
{\Pi'}_{\mu\r}^{(AA)}(p;\L_0) \, \G_{\r\nu}^{(wcA)}(p,k,q;\L_0)
\no \\ && \phantom{\BD^{(AAc)}_{\G,\mu\nu}(p,q,k;\L_0)}
+{\Pi'}_{\nu\r}^{(AA)}(q;\L_0) \, \G_{\r\mu}^{(wcA)}(q,k,p;\L_0).
\eeeq
At the tree level it vanishes since $\G_{\mu\nu\r}^{(3A)}$ saturated
with $k_\r$ is proportional to $t_{\m\n}(p)-t_{\m\n}(q)$.
At the first loop, exploiting the previous observation, we have
\beq\nome{dbacc1loop}
\BD^{(AAc,1)}_{\G,\mu\n}(p,q,k;\L_0)= \left[ i\lp \s^{(1)}_{wc}+
\frac{\s^{(1)}_{3A}}g\rp
+\s^{(1)}_A +\s^{(1)}_{wcA}\right](t_{\m\n}(p)-t_{\m\n}(q))+g_{\m\n}
\s^{(1)}_\a (p^2-q^2)\,.
\eeq
Due to the UV finiteness of $\DGh$  and $\s_\a$ at this order, the ST
identity can be
recovered only if the following relation among the divergent part of
the coupling constants at loop one ---which will be denoted by 
$\s_i^*(\L_0)$--- holds
\beq\nome{relat1}
i\s_{wc}^*+\frac i g \s_{3A}^* + \s_A^* +\s_{wcA}^*=0\,.
\eeq
The finite part of the functional, instead, enters the fine-tuning
equation, which allows to determine the finite parts of the  $\s^{(1)}_i$'s
at the UV scale.

\vskip0.2truecm
\par\noindent
(iii) The further contribution to $\DGb$ with three fields is given by
\beeq\nome{dbwcc}
&&\BD^{(wcc)}_{\G,\mu}(p,q,k;\L_0)=\Pi_\mu^{(wc)}(p;\L_0)\,
\G^{(vcc)}(p,q,k;\L_0)+
\Pi_\nu^{(wc)}(q;\L_0) \, \G_{\mu\nu}^{(wcA)}(p,k,q;\L_0)
\no \\ && \phantom{\BD^{(wcc)}_{\G,\mu}(p,q,k;\L_0)}
+\Pi_\nu^{(wc)}(k;\L_0) \, \G_{\mu\nu}^{(wcA)}(p,q,k;\L_0)\,.
\eeeq
At the tree level  momentum conservation tells us this vertex
vanishes.
On the contrary, at the first loop 
\beq\nome{dbwcc1loop}
\BD^{(wcc,1)}_{\G,\mu}(p,q,k;\L_0)= -\frac{ i}{g} p_\m (\s^{(1)}_{vcc}-\s^{(1)}_{wca})\,,
\eeq
so that gauge invariance requires
\beq\nome{relat2}
\s_{vcc}^*=\s^*_{wcA}\,.
\eeq

\vskip0.2truecm
\par\noindent
The functional $\DGb$ has the two four-point vertices
$\D^{(wAcc)abcd}_{\G,\mu\nu}$ and $\D^{(3Ac)abcd}_{\G,\mu\nu\r}$.
The former automatically vanishes at one loop once 
eq. \re{dbwcc1loop} is satisfied.
In fact we should observe there exists a  consistency condition 
relating $\D^{(wAcc)abcd}_{\G,\mu\nu}$ to  \re{dbwcc}. 
\vskip0.2truecm
\par\noindent
(iv) Finally, the last vertex  of $\DGb$ we have to consider is
\beeq\nome{dbaaac}
&&\D^{(3Ac)abcd}_{\G,\mu\nu\r}(p,q,k,h;\L_0)=
\Pi_\s^{(wc)}(h;\L_0)\,\G_{\mu\nu\r\s}^{(4A)abcd}(p,q,k,h;\L_0)
 \\
&&\phantom{\D^{(3Ac)abcd}_{\G,\mu\nu\r}(p,q,k,h;\L_0)\,}
+\eps^{eda}\eps^{ebc} \G_{\s\mu}^{(wcA)}(q+k,h,p;\L_0) \,
\G_{\s\nu\r}^{(3A)}(p+h,q,k;\L_0) \no \\
&&\phantom{\D^{(3Ac)abcd}_{\G,\mu\nu\r}(p,q,k,h;\L_0)\,}
+\eps^{edb}\eps^{eac} \G_{\s\nu}^{(wcA)}(p+k,h,q;\L_0) \,
\G_{\s\mu\r}^{(3A)}(q+h,p,k;\L_0) \no \\
&&\phantom{\D^{(3Ac)abcd}_{\G,\mu\nu\r}(p,q,k,h;\L_0)\,}
+\eps^{edc}\eps^{eba} \G_{\s\r}^{(wcA)}(q+p,h,k;\L_0) \,
\G_{\s\nu\mu}^{(3A)}(k+h,q,p;\L_0)\,. \no
\eeeq
Using \re{rel} and the definitions of the vertices of $\Pi$, we have
\beeq\nome{dbaaac1loop}
&&\D^{(3Ac)abcd}_{\G,\mu\nu\r}(p,q,k,h;\L_0)=
h_\m \, g_{\n\r} \left[ B_1 \lp \de^{ab}\de^{cd} + \de^{ac}\de^{bd}\rp
+ B_2 \, \de^{ad}\de^{bc}\right]\no\\
&&\phantom{\D^{(3Ac)abcd}_{\G,\mu\nu\r}(p,q,k,h;\L_0)\,} +
h_\n \, g_{\m\r} \left[ B_1 \lp \de^{ab}\de^{cd} + \de^{ad}\de^{bc}\rp
+ B_2 \, \de^{ac}\de^{cd}\right]\no\\
&&\phantom{\D^{(3Ac)abcd}_{\G,\mu\nu\r}(p,q,k,h;\L_0)\,} +
h_\r \, g_{\m\n} \left[ B_1 \lp \de^{ac}\de^{bd} + \de^{ad}\de^{bc}\rp
+ B_2 \, \de^{ab}\de^{cd}\right]\,,
\eeeq
where
\beeq\nome{prerelat34}
&&B_1 = \frac{\s'_{4A}}{g} (-i+\s_{wc}) - \frac{\s_{4A}}{g} (-i+\s_{wc}) 
+\s_{wcA}\s_{3A} \\
&&B_2 =  \frac{\s'_{4A}}{g} (-i+\s_{wc}) +2 \left[\frac{\s_{4A}}{g}
(-i+\s_{wc}) - \s_{wcA}\s_{3A} \right] \no
\eeeq
Recalling \re{sigmatree}, it is easy to verify $B_1=B_2=0$ at the tree
level, whereas at the first   loop order eq. \re{prerelat34} provides 
two independent relations among the couplings
\beq\nome{relat34}
\s^{'*}_{4A}=0\,,\;\;\;\;\;\;
-\frac ig \s_{4A}^*-g\, \s_{wc}^* =-\s_{3A}^*-ig \,\s_{wcA}^*\,.
\eeq
Hence from \re{relat34} we learn $\s'_{4A}$ is finite at this order
and we earn a further constraint to add to eqs. \re{relat1},
\re{relat2}.
In the whole the relevant part of the cutoff effective action for the
$SU(2)$
gauge theory contains nine couplings. The divergent contributions  of
six of those, namely  $\s_{wc}^*,$ $\s_{A}^*,$ $\s_{3A}^*,$
$\s_{wcA}^*,$ $\s_{vcc}^*$ and $\s_{4A}^*$, are related by three
equations, which express the BRS invariance of the divergent part of 
$\Pi (\L_0)$, whereas the counterterms necessary to restore gauge
symmetry are finite and non-invariant.
It follows only three
of the divergent couplings are independent, let us say   $\s_{A}^*,$ 
$\s_{wc}^*,$ and $\s_{wcA}^*$, and we ascribe them the role of wave
function renormalization for the vector field, $z_1$, for the ghost
field, $z_2$ and of the coupling $z_3\,g$.
Hence the functional  $\Pi(\L_0)$ can  be split into two contributions
\beq \nome{pigreco1ym}
\Pi^{(1)}(\L_0)=\Piinv^{(1)}(\L_0)+\Pit^{(1)}
(\L_0)\, ,
\eeq
where $\Piinv$ contains all the independent monomials which are
invariant, \ie ${\cal S}_{\Pi^{(0)}}\, \Piinv^{(1)}=0$. The explicit
form of $\Piinv$ is 
\beq \nome{piinv}
\Piinv^{(1)}(z_i(\L_0))=\int \mbox{d}^4 x
\left[ -\frac14\,z_1
\, 
{\cal F}^a_{\m\n} \, 
{\cal F}^{a\m\n} + z_2\,z_3 \,\left( \frac1{g\, z_3}\, w^a_\m\, 
{\cal D}^{ab}_\m c^b
-\frac12\, f^{abc}\, v^a\,c^b\,c^c \right)  \right]
\eeq
with ${\cal F}^a_{\mu\nu}=\p_\mu A^a_\nu -\p_\nu A^a_\mu
+ g \, z_3 \, f^{abc} A^b_\mu \,A^c_\nu$ and the covariant derivative
given by
${\cal D}^{ab}_\mu c^b =\p_\mu c^a + g \, z_3 \, f^{abc} A^b_\mu \, c^c$.
The key to pass from \re{rel} to \re{piinv} is
$$
z_1=1-\s_A^*\,, \quad \qquad z_2=1+i\s_{wc}^*\,,\quad \qquad z_3=
\frac{\s_{wcA}^*}{1+i\s_{wc}^*}\,.
$$
The remaining monomials contribute to $\Pit$ which contains the finite
couplings $\r_i = \s_i -  \s_i^*$ evaluated at $\L=\L_0$.
In the following section we will  determine the couplings  $\r_i(\L)$ 
via  fine-tuning. As to the couplings in \re{piinv}, they are not 
involved in the fine-tuning, so that they  are free and can be 
fixed at the physical point $\L=0$.

\subsection{Solution of the fine-tuning at the first loop}

We stated in the previous section we would demonstrate the finiteness
of the functional $\DGh$ at the first loop. In fact we will do more
than that, in the sense we will also explicitly compute some of the
finite parts of the relevant couplings, \ie $\r_i$,
through the fine-tuning eq. \re{fintun}. 
Let us now build up the vertices of $\DGh^{(1)}$. We start from
\re{dghc}, which  at the first loop and
in the \UV limit 
has the form
\beeq\nome{dgh1ym}
&&\DGh[\Phi,\g;\L_0]
=i\hbar\,
\int_{p,q} K_{0\L}(p)\, (-1)^{\de_i} \,\K(q)\,
D_{JL}(q) \, \frac{\de^2 \bG^{(0)}}{\de\Phi_i(-p)\de\Phi_L(-q)}
\no \\ && \phantom{\DGh[\Phi,\g;\L]\,}
\times \frac{\de^2}{\de\Phi_J(q)\de\g_i(p)}
\left(\Pi^{(0)}-\frac1g \int_{x}u_{\m}\p_{\m}c\right)\,.
\eeeq
To get such an expression we should notice  that in \re{dghc} the term 
proportional to $\de_{JL}$ does not 
contribute since in $\Pi[\Phi,\g;\L_0]$ 
diagonal interactions between a field and its own source are absent.
Moreover the product of the full propagator $\G_2(-p;\L)$ by
$D_{\L\L_0}$ in \re{dghc} is simply $\de_{Li}$ at this order.
As the second derivative of the functional $\Pi^{(0)}$ in \re{dgh1ym}
is concerned, we have just two possibilities, that is either
\beq\nome{poss1} 
\frac{\de^2 \lp \Pi^{(0)}-\frac1g \int_{x}u_{\m}\p_{\m}c \rp}{\de\Phi_J(q)\,
\de w^a_\m(p)}
\eeq
which in turn splits into
\beq \nome{subposs1}
\frac{\de^2 \lp \Pi^{(0)}-\frac1g \int_{x}u_{\m}\p_{\m}c \rp}{\de c^b(q)\,
\de w^a_\m(p)} \,,\;\;\;\;\;\; 
\frac{\de^2\Pi^{(0)}}{\de A_\n^b(q)\,
\de w^a_\m(p)}
\eeq
or we derive with respect to the source $v$
\beq\nome{poss2}
\frac{\de^2\Pi^{(0)}}{\de c^b(q)\, \de v^a(p)}\,.
\eeq
These three vertices are represented in fig.~5.1.
\begin{figure}[htbp]\nome{verticespi}
\epsfysize=3.4cm
\begin{center}
\epsfbox{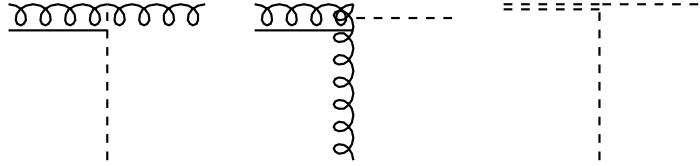}
\end{center}
\caption{\small{Vertices contributing to the second line of 
eq. \re{dgh1ym}.
The curly
and dashed  line denotes the gluon and the ghost field, 
respectively; the double lines represent the BRS source
associated to the field depicted by the top line. }}
\end{figure}
\newline
Going back to \re{dgh1ym}, we can sum up the three different
contributions and obtain 
\beeq\nome{dgh1ymspec}
&&\DGh[\Phi,\g;\L_0]
=i\hbar\,
\int_{p,q}  \eps^{abc}\,  \Kiu (p)\, \frac{ \Kin (q)}{q^2}\, 
\no \\
&&\phantom{\DGh[\Phi,\g;\L_0]=i\hbar\,\int}
  \left\{ 
- \frac{\de^2 \bG^{(0)}}{\de A_\m^a (-p)\, \de \bc^b(-q)}\,
\G_{\m\n}^{(wcA)}(p,q,r;\L_0)\, A_\n^c (r)\right. \no \\ 
&&\phantom{\DGh[\Phi,\g;\L_0]=i\hbar\,\int}
- \frac{\de^2 \bG^{(0)}}{\de A_\m^a (-p)\, \de A^b_\n(-q)}\,
\G_{\m\n}^{(wcA)}(p,r,q;\L_0)\, c^b (r)  \no \\
&&\left. \phantom{\DGh[\Phi,\g;\L_0]=i\hbar\,\int}
+\frac{\de^2 \bG^{(0)}}{\de c^a (-p)\, \de \bc^b_\n(-q)}\,
\G^{(vcc)}(p,q,r;\L_0)\, c^c (r) \right\}\,. 
\eeeq
Implementing the relation $\de/\de \bc^a (p) = -ig\,p_\m\, \de/\de
w_\m^a (p)$, which follows from the definition of $w_\m^a$, and
arresting at the first term in the expansion  \re{gammab} of $\bG$ in terms of
vertices of $\G$ ---so that $\bG$ coincides with $\G$ up to a sign, 
eq. \re{dgh1ymspec} supplies the $A$-$c$ component of $\DGh^{(1)}$
\beeq\nome{vertexac}
&&\DGh^{(Ac,1)}(\L_0)= \int_p \, A_\m^a (-p) \, c^a(p)\, 
\hD^{(Ac,1)}_{\G, \m} (p;\L_0)=
\int_p \, A_\m^a (-p) \, c^a(p)\no \\
&&\;\;\;\;\times \int_q  \Kiu (p-q)\, \frac{ \Kin (q)}{q^2}
2g \left\{ -q_\n \, \G^{(wcA)}_{\n\r} (q,p,-p-q)\, 
\G^{(wcA)}_{\r\m} (q+p,-q,-p) \right. \no \\ 
&& \phantom{\DGh^{(Ac,1)}(\L_0)= \int_p }
+\frac ig \, \G^{(wcA)}_{\r\n} (-p-q,p,q)\,
\G^{(3A)}_{\n\r\n}(-q,p+q,-p)
\no \\
&&\left. \phantom{\DGh^{(Ac,1)}(\L_0)= \int_p }
 -q_\n \, \G^{(wcA)}_{\n\m} (q,p-q,-p)\, 
\G^{(vcc)} (q-p,p,-q) \right\}\,,
\eeeq
where the vertices of $\G$ ---in which the dependence on the UV cutoff
$\L_0$ has been removed since we are at the tree level--- can be read
from eqs. \re{3A}, \re{wca}, \re{vcc} and \re{sigmatree}.
The contributions to $\DGh^{(Ac,1)}(\L_0)$ are depicted in fig.~2.
\begin{figure}[htbp]
\epsfysize=5cm
\begin{center}
\epsfbox{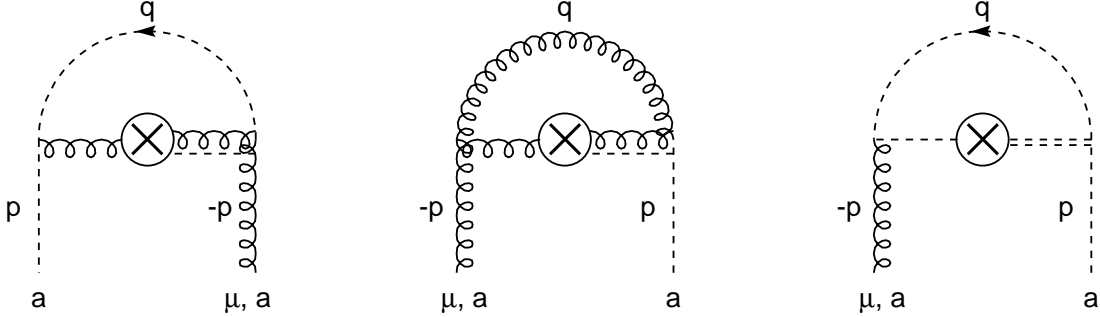}
\end{center}
\caption{\small{First loop contributions to $\hat{\D}_{\G,\m}^{(Ac)}$.
All momenta are incoming.}}
\end{figure}
\newline 
The final expression of the vertex  $\hD^{(Ac,1)}_{\G, \m} (p;\L_0)$
defined in \re{vertexac} is
\beq\nome{vertexacfinal}
\hD^{(Ac,1)}_{\G, \m} (p;\L_0)= 2\, g\, 
\int_q  \Kiu (p-q)\, \frac{ \Kin (q)}{q^2}\, (3p_\m-4q_\m)\,.
\eeq
When discussing the topic of symmetry, we stated this integration
produces a finite result, to be compared with  \re{dbac1loop}, so that
the requirement of gauge invariance forces us to fix some of the
relevant couplings. 
We can perform the $q$-integration  in
\re{vertexacfinal} with different choices of the cutoff function.
In any case we have to expand the integrand in powers of $p^2/\L_0^2$,
and we are allowed to do so since $p\ll\L_0$
\beq\nome{expand}
\hat{\D}_{\G\,\m}^{(Ac)}(p;\L_0)\,= \,p_\m\,[\hat\de_1\,\L_0^2+\hat\de_2\,p^2
\,+\, {\cal O}(p^4/\L_0^2)]\,.
\eeq

If we adopt a sharp cutoff, \ie $ \Kin (q)=\Theta (q^2 -\L_0^2)$,
only the divergent part can be calculated and we have~\footnote{We
should keep in mind there is a factor $i$ coming from integration over
$q$.}
$$
\hat\de_1 = -2i \frac g{16\pi^2}\,.
$$

On the contrary, if a smooth cutoff is used both $\hat\de_1$ and
$\hat\de_2$ 
can be determined. We are obviously led to pick up a cutoff which is
advantageous from the computational point of view. The possibility of
performing Feynman parametrization, for example, is definitely
appealing. With this in mind, we can employ the following cutoff
functions
\beq\nome{cutoffm} 
 \Kiu (q) = \frac{\L_0^4}{(q^2+\L_0^2)^2}\,, \quad\quad\quad
\Kin (q)=q^2 \frac{q^2+2\L_0^2}{(q^2+\L_0^2)^2}\,.
\eeq
The essential ingredients are  Feynman  parametrization and  the fundamental
integral
$$
\int_0^\infty \frac{t^{s-1}}{(t+1)^\n} \, dt =
\frac{\G(s)\,\G(\n-s)}{\G(\n)}\,.
$$
Finally, for the coefficients $\hat\de_i$ in \re{expand} we get
$$
\hat\de_1 = \frac 23  \frac {i\,g}{16\pi^2}\,, \quad \quad
\hat\de_2 = -\frac 7 {30}  \frac{i g}{16\pi^2}\,.
$$

Another manageable cutoff is the exponential function
\beq\nome{cutoffe} 
 \Kiu (q) = e^{-q^2/\L_0^2}\,,\quad\quad\quad
\Kin (q)= 1-e^{-q^2/\L_0^2}\,.
\eeq
In this case Feynman parametrization is replaced by the representation
$$
\frac 1{q^2+m^2} = \int_0^\infty \frac{d\a}{\L_0^2}\,e^{-\frac{\a}{\L_0^2}
(q^2+m^2)}
$$
and the integration is easily carried out by using
$$
\int \frac{d^{2\omega}q}{(2\pi)^{2\omega}} \exp{(-x \,q^2+2q\cdot b)}= \lp
\frac{\pi} x \rp ^\omega \frac{e^{b^2/x}}{(2\pi)^{2\omega}}\,, 
\quad\quad\quad x>0\,.
$$
At the end we find
$$
\hat\de_1 = 0\,, \quad \quad
\hat\de_2 = -\frac 1 {12} \frac{i g}{16\pi^2}\,.
$$

Hence, restoration of the ST identity for the physical effective
action at the first loop order implies, at $\L=\L_0$, the following 
constraints on two of the relevant coupling in $\Pi^{(AA)}_{\m\n}(\L_0)$
$$
\s_m^{(1)}(\L_0)=i\, g\,\L_0^2\,\hat\de_1 \,, \quad\quad\quad
\s_\a^{(1)}(\L_0)= i\,g\,\hat\de_2\,.
$$
In a following section  we will test the correctness of our results by
comparing the values of $\s_m^{(1)}(\L_0)$, $\s_\a^{(1)}(\L_0)$
determined via the fine-tuning with those obtained  \cite{bdm3} 
by solving the RG flow  \re{eveq}.

\subsection{Vertices of $\bom{\DGh}$ with more than two fields}

If we want to analyse the vertices of $\DGh^{(1)}$ with more than two fields
we have first to go back to eq. \re{dgh1ymspec} and then through the
cutoff function $\Kiu$ glue the vertices
\re{subposs1}, \re{poss2} with higher terms in the iterative expansion
of $\bG^{(0)}$ in vertices of $\G^{(0)}$.
We start from the second order of the expansion of $\bG^{(0)}$; a
graphical representation is given in fig.~3.
\begin{figure}[htbp]
\epsfysize=3cm
\begin{center}
\epsfbox{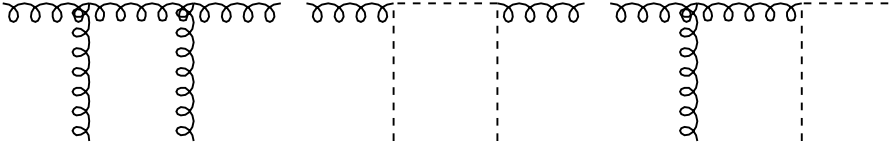}
\end{center}
\caption{\small{Graphical representation of the contribution to $\bG$
obtained in its expansion to the second order in vertices of $\G$.}}
\end{figure}
\newline
When we build all possible combinations of vertices of $\bG$ and $\Pi$
we obtain the contribution to $\DGh^{(1)}$ with three
fields, that is  $\hD^{(AAc)}_{\G,\mu\nu}$ and
$\hD^{(wcc)}_{\G,\mu}$, except one term which originates from the
irreducible four-vector vertex  of $\bG$. 
\begin{figure}[htbp]
\epsfysize=11cm
\begin{center}
\epsfbox{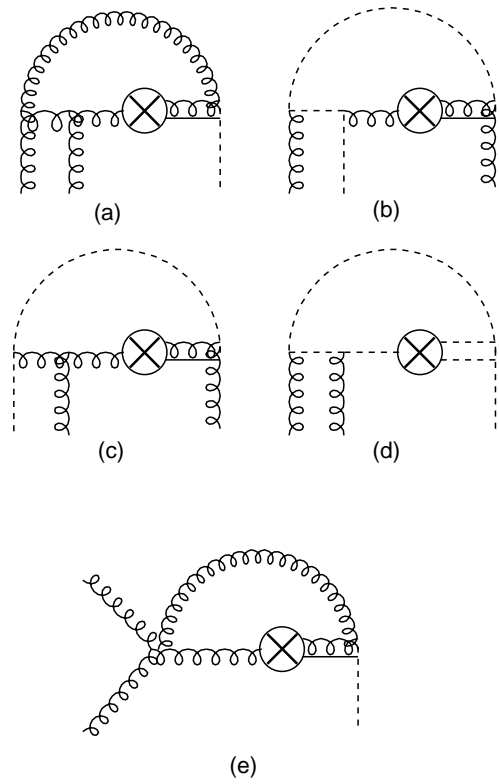}
\end{center}
\caption{\small{Graphs contributing to  $\hD^{(AAc)}_{\G,\mu\nu}$.
The graph (e) originates from the expansion of $\bG$ to the first
order in vertices of $G$.}}
\end{figure}
\newline 
\begin{figure}[htbp]
\epsfysize=8cm
\begin{center}
\epsfbox{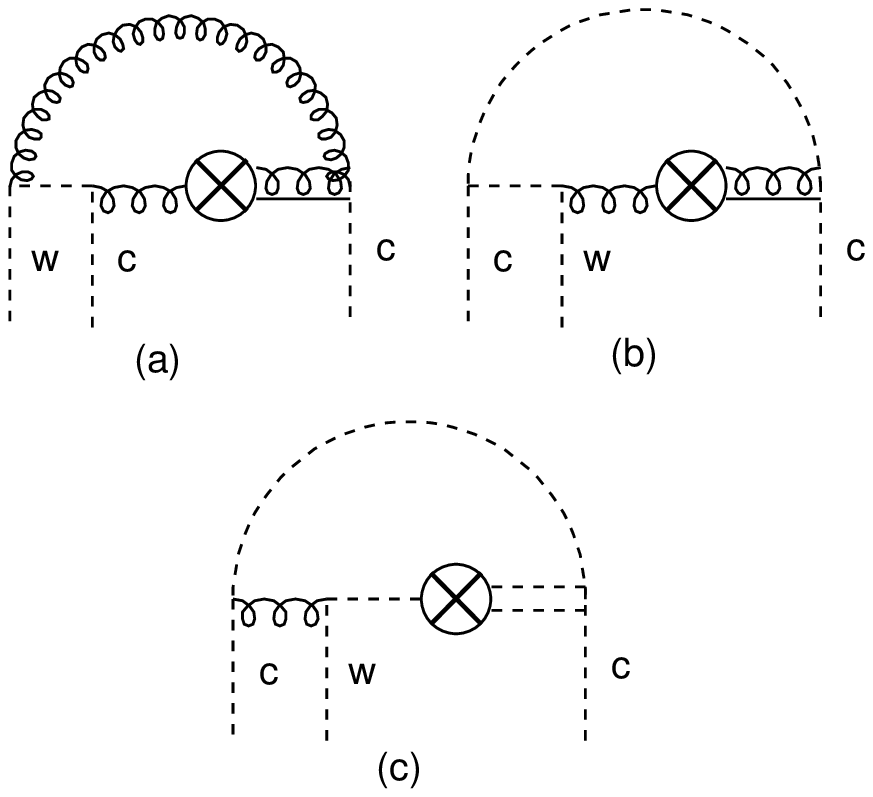}
\end{center}
\caption{\small{Graphs contributing to $\hD^{(wcc)}_{\G,\mu}$ }}
\end{figure}
\newline 
Fig.~4 and fig.~5 represent
the contributions to $\hD^{(AAc)}_{\G,\mu\nu}$ and
$\hD^{(wcc)}_{\G,\mu}$, respectively. Obviously some of the graphs
could also be excluded by Lorentz covariance and gauge
invariance. This is the case for example for the term mentioned above
and corresponding to the graph of fig.~4(e).
In order to clarify how things work, we compute ---up to a constant
factor--- one of the contribution to 
$\hD^{(wcc)}_{\G,\mu}$ given by the graph of fig.~5(a).
From eq. \re{dgh1ymspec} we deduce
\beeq\nome{onecontwcc}
&&\hD^{(wcc,1)}_{\G,\mu}(p,q,-p-q;\L_0) \sim \int_k D_{\L\L_0}(k)\, 
\G_\r^{(\bc c A)}(-k,q,k-q)\, D_{\L\L_0}(p+k)  \\
&&\phantom{\hD^{(wcc,1)}_{\G,\mu}(p,q,-p-q;\L_0) \sim \int_k } \times
\G_{\m\r}^{(w c A)}(p,k,-p-k) \, \G_{\r\n}^{(w c A)}(q-k,-p-q,p+k) \no
\\
&&\phantom{\hD^{(wcc,1)}_{\G,\mu}(p,q,-p-q;\L_0) }
\sim g \int_k \Kiu (q-k)\, \frac{\Kin(k)}{k^2}\,
\frac{\Kin(p+k)}{(p+k)^2}\,
k_\m \no \\
&&\phantom{\hD^{(wcc,1)}_{\G,\mu}(p,q,-p-q;\L_0) }
= i\, g \left[ \lp - \frac{11}{96} + \OO (P^2/\L_0^2) \rp p_\m + 
\lp \frac{37}{240} + \OO (P^2/\L_0^2) \rp q_\m \right]\,, \no
\eeeq
where $P$ is some combination of the momenta $p$ and $q$.
It should be clear at this point that the finiteness of the integrals 
\re{vertexacfinal}, \re{onecontwcc} is due to the presence of cutoff
functions having almost non intersecting supports.
 
When adding the contribution which arises from the graph of fig.~5(b)
what we obtain is a symmetric ---in a proper sense--- result, \ie
a term proportional to $p_\m$.
As we did for $\s_m$ and $\s_\a$, this result, together with all other
contributions coming from the graphs of fig.~5, can be used to fix the
combination $\r_{vcc}-\r_{wcA}$ in \re{dbwcc1loop}, via eq. \re{fintun}.

We can now carry on with the perusal of the functional $\DGh^{(1)}$.
Experience suggests us  our attention must be devoted to two kinds of
vertices, \ie    $\DGh^{((nA)c,1)}$ and $\DGh^{(w(nA)cc,1)}$, with $n$
the number of vector fields. Qualitatively, the behaviour of the
leading term of the two vertices can be inferred
from power counting.
\begin{figure}[htbp]
\epsfysize=8cm
\begin{center}
\epsfbox{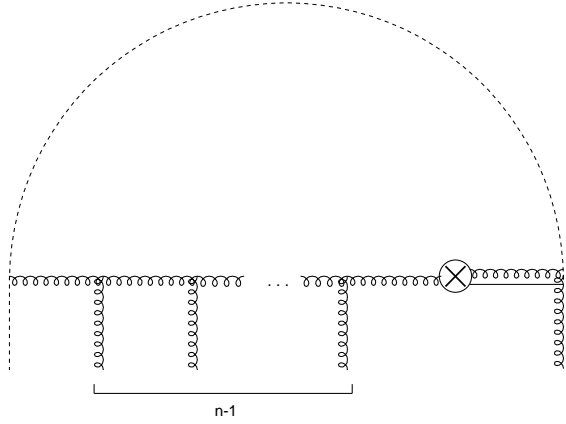}
\end{center}
\vskip-1truecm
\caption{\small{One of the prototype of graphs contributing to 
$\DGh^{((nA)c,1)}$ }}
\end{figure}
\newline 
The former can be either of the type represented in fig.~6 or of the
type of fig.~4(a) with the insertion of $n$ three-vector vertices. In
any case, 
the cutoff function $\Kiu$
---let us consider \re{cutoffm} just to fix
ideas--- brings a factor $\L_0^4/k^4$, $k$ being the
integration momentum; there are $n$ propagators producing a
factor $(1/k^2)^n$, $n$ three-vector vertices (or $n-1$ three-vector  
vertices and a $\bc$-$c$-$A$ vertex)
contributing at most
with $k^n$ and finally the integration measure goes like $k^4$. The
presence of the factor  $\L_0^4$ implies only with integrands behaving  at
least like $(1/k^4)$ we will end up with  finite, relevant  vertices
of $\DGh$. 
Hence, taking into account the powers of momentum, the possible value
of $n$ is restricted by 
$$
-4-2n+n+4 \ge -4 \quad\;\;\; \Longrightarrow  \quad\;\;\;  n\le 4\,.
$$
In fact  $n=4$ is excluded by Lorentz and gauge covariance.

On the other hand, the vertex   $\DGh^{(w(nA)cc,1)}$ is of the type of
fig.~5(a) with the insertion of $n$ three-vector vertices. Its
behaviour 
for large $\L_0$ is dictated
by the factor $\L_0^4/k^4$ brought by  $\Kiu$;  factors
$(1/k^2)^{(n+2)}$  and $k^{(n+1)}$ are  due to the propagators and to the
vertices, respectively and  the integration measure goes like $k^4$.
Therefore, in the \UV limit a non-vanishing result will be obtained
only if
$$
-4-2(n+2) +(n+1)+4 \ge -4 \quad\;\;\; \Longrightarrow  
\quad\;\;\; n\le 1\,.
$$

\section{Comparison with the fine-tuning at $\bom{\L=0}$}
So far our direction has been to discuss the symmetry at the
ultraviolet scale and determine the cutoff-dependent $\r_i(\L_0)$.
An alternative point of view is to set the boundary conditions for
these couplings at the physical point $\L=0$ \cite{bdm3,bdm4}.
In this way some of the relevant 
couplings (\ie the $z_i$'s) are related to physical couplings such as
the wave function normalizations and the three-vector coupling $g$ at
a subtraction point $\mu$. The remaining are fixed imposing the
symmetry at the physical point.

In order to show the equivalence of the two formulations, we have to
compute the couplings $\r_i^{(1)}(\L_0)$ starting 
from their boundary
conditions at $\L=0$ \cite{bdm3} and letting them evolve via eq. \re{eveq}
to the UV scale. Even though this will be done only for the couplings 
$\s_m(\L_0)$ and 
$\s_\a(\L_0)$, the argument can be generalized to all the couplings
$\s_i$. 
The fine-tuning at $\L=0$ provides for the two couplings we are examining 
the boundary conditions, which, due to gluon transversality, are simply
$$
\s_m(0)=\s_\a(0)=0\,.
$$
From eq. \re{eveq} we can now derive the loop expansion \cite{bdm3} and, at the 
first loop we have
$$
\s_m^{(1)}(\L_0) = \int_0^\L \frac{d\l}\l \, I_L(0;\l)
$$
$$
\s_\a^{(1)}(\L_0)  = \int_0^\L \frac{d\l}\l \, \frac{\p}{\p \bp^2} 
I_L(\bp;\l)\,,
$$
where
$$
I_{L}(p;\L)=
-i g^2 \LdL \left[
\int_q \frac{\K(q)\, \K(p+q)}{q^2 \, (p+q)^2}
(2 q^2+10 pq+3p^2+8\frac{(pq)^2}{p^2})
-6 \int_q \frac{K_{\L\L_0}(q)}{q^2} \right] \,.
$$
When we choose the sharp cutoff function only the first integral can be 
computed, whereas with the cutoffs \re{cutoffm}, \re{cutoffe} both 
$\s_m^{(1)}(\L_0)$ and $\s_\a^{(1)}(\L_0)$ can be found. 
In all cases the results coincides with those obtained in sect.~5.3.1.

\chapter{Chiral gauge theories and anomalies}

The problem of finding a consistent renormalization procedure of a 
chiral gauge theory (CGT) is still an active field of investigation,
both in the framework of perturbation theory \cite{tv}-\cite{chiral} 
and in lattice regularization \cite{roma,lattice}. In the presence of 
chiral fermions
no regularization is known to preserve  chiral symmetry.
This is not  a technical problem but it is related to the fact that
chiral symmetry is anomalous \cite{abj}. In dimensional regularization with 
minimal subtraction \cite{tv}-\cite{wdim} the
correct definition of the matrix $\g_5$ produces
chiral breaking terms \cite{gamma5}, although they formally vanish  for 
$d\to4$.
The appearance of these breaking terms is necessary to reproduce the
anomaly for external currents.  In order to ensure the
Slavnov-Taylor  identities of the local chiral symmetry for the
renormalized theory, the minimal subtraction procedure must be
supplemented by additional finite non-invariant counterterms 
\cite{tv}-\cite{rss}.
Similarly, in
lattice regularization one must introduce the Wilson term in
order to avoid the fermion doubling and reproduce the correct
anomaly.  This term explicitly breaks chiral invariance and one
adds all possible counterterms to the naive action to compensate
the explicit symmetry breaking induced by the Wilson term \cite{roma}.
We should notice that in both approaches the regularized Lagrangian
couples left and right fermions and therefore also global chiral
symmetry is broken.

Proving that it is possible to compensate the breaking due to
the regularization by a suitable choice of non-invariant counterterms
in the Lagrangian is an essential ingredient of the renormalization
procedure. If the fermionic content of the theory ensures the
cancellation of the one-loop chiral anomaly, this fine-tuning problem
has a solution.
Its solvability is independent of the regularization procedure since the 
classification of all possible anomalies is an algebraic problem
\cite{brs}-\cite{becchi0}.

We have appreciated in the previous chapters  that the Wilson renormalization
 group  formulation  can be used to deduce
the ST identities in the Yang-Mills theory. 
We know  that, by properly fixing the boundary
conditions of the RG equation, the ST identities for the pure 
Yang-Mills theory with gauge group $SU(N)$ can be satisfied when 
all cutoffs
are removed (at least in perturbation theory).  This has been shown
 both in terms of  the ``bare'' couplings of the effective
action at the ultraviolet scale \cite{b} and of the physical couplings
\cite{bdm3}-\cite{bdm5}. 
In this  case the possibility of solving the fine-tuning problem 
is guaranteed by the fact that, once the renormalization conditions 
are imposed and all cutoffs are removed, the renormalized Green 
functions are independent of the regularization procedure 
and, in particular, they coincide with those obtained via 
dimensional regularization.

In this chapter we address the extension of the RG formulation  to chiral gauge
theories.  As in this case there is no regularization which
preserves the symmetry, the RG formulation has no disadvantages
with respect to other procedures. As a matter of fact, in the RG
approach the space-time dimension is $d=4$ so that there is no
ambiguity in the definition of the matrix $\g_5$ and in the
regularized action left and right fermions are not coupled.
Therefore the solution of the fine-tuning procedure is
simpler than in the standard cases (\ie dimensional or lattice
regularization), since the possible counterterms must be
globally chiral invariant.  Although there is no left-right
coupling, we will show that we obtain the correct chiral
anomaly (if the matching conditions for the anomaly cancellation
are not fulfilled).
  
\section{Renormalization group flow and effective action}
Let us consider the $SU(N)$ chiral gauge theory described by the classical 
Lagrangian (in the Feynman gauge)
\beq\nome{YMF}
S_{cl}= S_{YM} + S_F \,,
\eeq
where
\beq\nome{ferm}
S_F=\int d^4x
\left(\bpsi_{L}\,i\, \ds{D}\, \psi_{L} + \bpsi_{R}\,i\, \ds{\p}\,
\psi_{R}\right)
\eeq
and $ S_{YM}$ was defined in eq. \re{YM}. The fermions $\psi_L$,
$\psi_R$ are in the fundamental representation of the gauge group $SU(N)$.
The action \re{YMF} is invariant under the BRS transformations
\cite{brs} eq. \re{brst} and
$$
\de \psi_{L} = - \eps\, c^a t^a\, \psi_{L}\,, 
\;\;\;\;\;\;\;
\de \bpsi_{L} = - \eps \,\bpsi_{L}\, c^a t^a\,, 
\;\;\;\;\;\;\;
\de \bpsi_{R} = \de \psi_{R} = 0
$$
with $\eps$ a Grassmann parameter.

Introducing the sources $\bl$ and $\l$  associated to 
the variations of $\psi$ and $\bpsi$, respectively 
we get the BRS action
\beq\nome{Stot}
S_{BRS}[\Phi_I,\g_i]
=
S_{cl} +
\int d^4x
\left\{
\frac 1 g u^a_\mu  D^{ab}_\mu c^b - \half f^{abc} v^a\,c^b \,c^c
+ \bl \,c^a\, t^a\, \psi_{L} +\bpsi_{L}\, c^a\, t^a\,\,  \l
\right\}
\eeq
where we have denoted by $\Phi_I$ and $\g_i$ the fields and the BRS 
sources
$$
\Phi_I=\{\,A^a_\mu, \,c^a, \,\bc^a, \,\bpsi,\,\psi\}
\,,
\;\;\;\;\;\;\;\;
\g_i=\{w_\mu^a,\, v^a, \,\bl, \, \l\}
\,.
$$

We now implement the RG method. As usual the generating functional is
given by \re{Z}
and the cutoff scalar products between fields and sources, introduced in
eq. \re{scalarprod}, read
\beeq&&
\half (\Phi,\, D^{-1}\, \Phi)_{\L\L_0}\equiv
\int_p\, \, K^{-1}_{\L\L_0}(p)
\big\{p^2 \left[\half A^a_\mu(-p)\,A^a_\mu(p)
- \bc^a(-p)\, c^a(p)\right] 
\nonumber
\\
&&
\;\;\;\;\;\;\;\;\;\;\;\;\;\;\;
\;\;\;\;\;\;\;\;\;\;\;\;\;\;\;\;\;\;\;\;\;\;\;\;\;\;\;\;
-\bpsi_L(-p)\,\ds{p}\,\psi_L(p)
-\bpsi_R(-p)\,\ds{p}\,\psi_R(p)
\big\}\,,
\eeeq
\beeq
&&(J,\Phi)_{\L\L_0}\equiv
\int_p \, K^{-1}_{\L\L_0}(p) 
\big\{j^a_\mu(-p)\, A^a_\mu(p) +
[\bchi^a(-p) - \frac i g p_\mu u^a_\mu(-p)] \, c^a(p) +
\bc^a(-p)\, \chi^a(p) 
\nonumber 
\\
&&
\;\;\;\;\;\;\;\;\;\;\;\;\;\;\;\;\;\;\;\;\;\;\;\;\;\;\;\;
+\bpsi(-p) \eta(p) 
+\bar\eta(-p) \psi(p) 
\big\}
\,.
\eeeq

Due to the breaking of gauge invariance caused by  the cutoff function
in the propagators,  the UV action $\si$ must contain all possible 
renormalizable interactions which are needed to compensate the symmetry 
breaking induced by the regularization. These are given by the monomials
in the fields, BRS sources and their derivatives which have dimension
not larger than four and are Lorentz and $SU(N)$ scalars, since Lorentz 
and global chiral gauge symmetries are  preserved by the cutoff 
regularization.
The independent monomials of the Yang-Mills sector have been given in
eq. \re{mon1}. Here we just recall that 
 in the $SU(2)$ case 
there are  only one monomial with three
gauge fields and  two independent monomials with four gauge fields, while
for $SU(3)$ only three monomials with four gauge fields are independent. 
In the fermionic sector one has the five monomials 
\beq\nome{mon2}
\bpsi_{L}\, \ds{\p}\, \psi_{L}\, , 
\;\;\;\;\;\;
\bpsi_{R} \,\ds{\p}\, \psi_{R}\, ,
\;\;\;\;\;\;
\bpsi_{L}\, \ds{A^a}t^a\, \psi_{L} \, ,
\;\;\;\;\;\;
\bl \,c\, \psi_{L} \, ,
\;\;\;\;\;\;
\bpsi_{L}\, c\,  \l\,.
\eeq
All these interactions appear in $\si$ with couplings of 
non-negative dimension (relevant parameters) which depend on $\L_0$.
In order to obtain the physical theory we have to show that
these relevant parameters can be fixed in such a way that:
\newline
(1) the \UV limit can be taken by fixing the physical parameters such
as the masses, the coupling $g$ and the wave function normalization 
constant at a subtraction point $\mu$.  Perturbative renormalizability 
ensures that this can be done \cite{p,b,bdm3} (for this reason the 
dependence on the ultraviolet cutoff $\L_0$ has been and will be
sometimes  understood).
\newline
(2) in the \UV limit the Slavnov-Taylor identities must be satisfied.
For a chiral gauge symmetry this requires that the gauge group is 
anomaly free or, more generally, constrains the fermionic content of 
the theory.
This is the crucial point we want to discuss in this chapter.

\subsection{Boundary conditions: physical parameters and symmetry}
The relevant part of the
cutoff effective action involves only monomials in the fields and
sources with dimension not larger than four
\beq\nome{pirc}
\Pir[\Phi,\g;\s_i(\L)]=\sum_i \s_i(\L)\,P_i[\Phi,\g]\,,
\eeq
where the sum is over the monomials $P_i(\Phi,\g)$  given in 
\re{mon1} and \re{mon2}.
The couplings $\s_i(\L)$ can be expressed in terms of the 
cutoff vertices at a given subtraction point.  Thus we need to extract
the relevant part from a given functional with a non-vanishing
subtraction point. In chapter 2 and 5 we faced
this problem for the scalar theory and 
the pure gauge case, respectively.  For the extraction of the relevant part 
of a given functional in a theory with fermions see \cite{anom}.

The remnant of the cutoff effective action 
is the irrelevant part.
Since we expect the theory to be renormalizable, the simplest boundary
condition for it is
$$
\Gir[\Phi,\g;\L=\L_0]=0\,.
$$
For $\L=\L_0$, then, the cutoff effective action becomes local and
corresponds to the bare action $\si$ in \re{Z}, with the bare
couplings given by  $\s_i(\L_0)$. 

In refs.~\cite{bdm3,bdm4} the boundary conditions for these couplings are set
at the physical point $\L=0$. In this way some of the relevant 
couplings are related to physical couplings such as
the wave function normalizations and the three-vector coupling $g$ at
a subtraction point $\mu$. The remaining are fixed imposing the
symmetry at the physical point. This procedure is highly not trivial
since one has to analyse  non-local functionals. 

Alternatively, we can discuss the symmetry at the ultraviolet scale
and determine the cutoff-dependent $\s_i(\L=\L_0)$'s.  This idea 
was described in details in
chapter 3. In this case the discussion is
simpler, since all functionals are relevant and we have to perform a
perturbative calculation (\ie to solve the RG equations)  to obtain
the physical couplings. 
Although  the couplings  $\s_i(\L)$ are determined at $\L=\L_0$, we
still set the wave function normalizations and the gauge coupling $g$
at a subtraction point $\mu$  at $\L=0$. As a matter of
fact  there are combinations of the monomials in \re{pirc} which 
are not involved in the fine-tuning, so that the corresponding 
couplings are free and can be fixed at the physical point $\L=0$.
Since we are already familiar  with the implementation of  the gauge
symmetry 
in the RG formulation, we jump to the solution 
 of the fine-tuning at the UV scale.

\section{Solution of $\bom{\DG=0}$ at $\bom{\L=\L_0}$}

In section 3.3 we showed that if $\DG^{(\ell')}=0$ at any loop order 
$\ell'<\ell$, then $\DG^{(\ell)}$ is constant and we can analyse it
at an arbitrary value of $\L$. 
We also chose to perform such an analysis at the UV point, where
the gauge symmetry condition 
determines  the cutoff-dependent bare couplings.
In fig.~3.1 we represented the functional $\DGh(\L_0)$  
at the  UV scale $\L_0$.
A prominent feature of the $SU(N)$ theory is that  in \re{dghc} the
term proportional to $\de_{JL}$ does not 
contribute since in $\Pi[\Phi,\g;\L_0]$ 
diagonal interactions between a field and its own source are absent.

We have seen before that the equation $\DG^{(1)}(\L_0)=0$ can
solved by tuning some of the
relevant couplings in $\Pi^{(1)}(\L_0)$ and we gave a practical
example in sect.~5.3.1.
As a matter of fact, the fine-tuning equation \re{fintun} allows 
to compute the couplings in
$\Pi^{(1)}(\L_0)$ since $\DGh^{(1)}(\L_0)$ depends only on $\Pi^{(0)}$.
At this loop order the line with the crossed circle in
fig.~3.1 contributes only through
a cutoff function $\Kiu$, since the additional full propagator
associated to this line cancels at this order (see \re{dgh1ym}).

As in the pure YM case, the most general functional $\Pi(\L_0)$
contains
 the relevant monomials
given in \re{mon1} and \re{mon2} and can be split into two contributions
\beq \nome{pigreco1c}
\Pi^{(1)}(\L_0)=\Piinv^{(1)}(\L_0)+\Pit^{(1)}
(\L_0)\, ,
\eeq
where $\Piinv$ contains all the independent monomials which are
invariant, \ie ${\cal S}_{\Pi^{(0)}}\, \Piinv^{(1)}=0$. The explicit
form of $\Piinv$ is 
\beeq \nome{pich}
\Piinv^{(1)}(z_i(\L_0))=\int \mbox{d}^4 x\, \Bigg\{\!\!\!&&\!\!\!\!\!\!\! -
\frac14\,z_1
\, 
{\cal F}^a_{\m\n} \, 
{\cal F}^{a\m\n} + z_2\,z_3 \,\left( \frac1{g\, z_3}\, w^a_\m\, 
{\cal D}^{ab}_\m c^b
-\frac12\, f^{abc}\, v^a\,c^b\,c^c \right)  \\ \nonumber
\!\!\!&&\!\!\!\!\!\!\!\!\!\!\! +\,z_4 \,\bpsi_L\,i\ds{\cal D}\,\psi_L\, +
\, z_5 \,\bpsi_R\,i\ds{\p}\,\psi_R 
\,+\, z_2\,z_3\,\left(\bl\,c\cdot t \, \psi_L\, + 
\,\bpsi_L\,c\cdot t \, \l \right) \Bigg\}\, ,
\eeeq
with ${\cal F}^a_{\mu\nu}=\p_\mu A^a_\nu -\p_\nu A^a_\mu
+ g \, z_3 \, f^{abc} A^b_\mu \,A^c_\nu$ and the covariant derivatives 
given by
${\cal D}^{ab}_\mu c^b =\p_\mu c^a + g \, z_3 \, f^{abc} A^b_\mu \, c^c$ 
and ${\cal D}_\m\,\psi_L=(\p_\m+z_3\, g\, A_\m^a\,\t,a)\,\psi_L$. 
The remaining monomials contribute to $\Pit$ which can be written as
\beeq \nome{pitilde}
\Pit^{(1)} [\Phi,\,\g;\,\s_i(\L_0)]\!\!\!& \equiv&\!\!\! \int\mbox{d}^4x\,
\bigg\{
\s_1\,\L_0^2\,\mbox{Tr}\,(A_\m\,A_\m)\,+\,
\s_2\,\mbox{Tr}\,(\p_\m A_\m\,\p_\n A_\n)\,
+\,\s_3\,\mbox{Tr}\,(\p_\m A_\m\,A_\n\,A_\n) \nonumber\\
&+&\!\!\!
\s_4\,\mbox{Tr}\,(A_\m \,A_\m\,A_\n\,A_\n)\,+\,
\s_5\,\mbox{Tr}\,(A_\m \,A_\n\,A_\m\,A_\n)\,+
\s_6\,\mbox{Tr}\,(A_\m \,A_\m)\,\mbox{Tr}\,(A_\n \,A_\n)\,
\nonumber \\
&+&\!\!\!
\s_7\,\mbox{Tr}\,(A_\m \,A_\n)\,\mbox{Tr}\,(A_\m \,A_\n)\,+\,
\s_8\,\mbox{Tr}\,(w_\m\,A_\m\,c)\,+\,
\s_9\, \mbox{Tr} \, (v\,c\,c)\,\nonumber\\
&+&\!\!\! 
\s_{10}\,\bpsi_L\,i\,\ds{A}^a\,\t,a\,\psi_L\,+ \,
\s_{11}\,\bl\,c\cdot t\,\psi_L\,+ \,
\s_{12}\,\bpsi_L\,c\cdot t\,\l\bigg\}\,.
\eeeq
However, for $SU(2)$ only nine and for $SU(3)$ only 11 of the monomials 
above are independent.    
Inserting \re{pigreco1c} into \re{fintun}, we find
\beq \nome{fintun1}
{\cal S}_{\Pi^{(0)}}\, \Pit^{(1)}(\L_0)\,=-\, \DGh^{(1)}(\L_0)\,.
\eeq
which fixes the $\s_i(\L_0)$'s whose finiteness is shown by explicit 
calculation in the next section.
On the contrary  the couplings $z_i(\L_0)$ 
are not fixed by the fine-tuning, so that we are allowed to set them equal 
to their physical values at $\L=0$, \ie $z_i(0)=1$. In the standard language
this corresponds to the renormalization prescriptions.

\subsection{ Explicit solution of $\bom{\DG^{(1)}(\L_0)=0}$}

In this section we solve the fine-tuning equations at the first loop
order and at the UV scale, $\L=\L_0$. In this case $\DGb$ contains
the UV couplings of $\Pit^{(1)}(\L_0)$, while the vertices of $\DGh$ 
are given by the product of the tree-level vertices of $\bG$ (obtained from
\re{gammab}) and those of $\Pi$. 

We first consider the $A$-$c$ vertex of $\DG$. 
From \re{dgbc} and \re{pitilde} it is easy to realize that 
$\bar{\D}_{\G\,\m}^{(Ac)}(p;\L_0)$ is given by 
\beq \nome{deltaac}
\bar{\D}_{\G\,\m}^{(Ac)}(p;\L_0)\,
=-\frac i g \,p_\m\,[\s_1(\L_0)\,\L_0^2+\s_2(\L_0)\,p^2]\,.
\eeq
On the other hand, the pure YM contribution to 
$\hat{\D}_{\G\,\m}^{(Ac)}(p;\L_0)$ was calculated in sect.~5.3.1 
and is given 
in \re{vertexacfinal}.
We now derive the total  fermionic (F) contribution, which is represented 
in fig.~1.
\begin{figure}[htbp]
\epsfysize=7cm
\begin{center}
\epsfbox{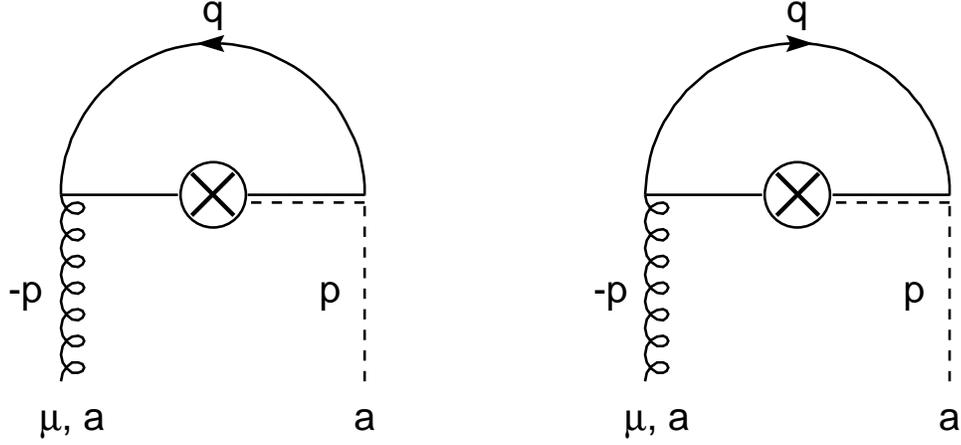}
\end{center}
\caption{\small{First loop contributions to $\hat{\D}_{\G\,\m}^{(Ac)}$ from the fermionic sector. The curly,
dashed and full line denotes the gluon, ghost and fermion field 
respectively; the double lines represent the BRS source
associated to the field depicted by the top line. All momenta are incoming.
}}
\end{figure}
\newline 
Using the vertices of $\sbrs$ we have
\beq \nome{grafacfer}
-2\,g\int_q \frac1{q^2} \,\Kiu(p-q)\,\Kin(q)\,\mbox{Tr}\,[\ds{q}\,
\g_\m\, (1-\gc)/2]\,.
\eeq
For  $p\ll\L_0$, we can write
\beq \nome{deltahac}
\hat{\D}_{\G\,\m}^{(Ac)}(p;\L_0)\,= \,p_\m\,[\hat\de_1\,\L_0^2+\hat\de_2\,p^2
\,+\, {\cal O}(p^4/\L_0^2)]\,.
\eeq
where the values of $\hat{\de}_1$ and $\hat{\de}_2$ can be obtained
from \re{vertexacfinal} and \re{grafacfer} once the cutoff function is 
specified. 
As in sect.~5.3.1, we have performed their calculation using three
different 
forms of $\Kiu$. 

We sum up the results  (in units of $\frac{ig}{16\,\pi^2}$) in the
following table
\begin{center}
\begin{tabular}{|l |r | r| r|r |} \hline
  &  &  &  & \\
$\Kiu(p)$ & $\hat{\de}_1^{\mbox{\tiny{(YM)}}}$ &$ 
\hat{\de}_1^{\mbox{\tiny{(F)}}}$& 
$\hat{\de}_2^{\mbox{\tiny{(YM)}}}$&
$\hat{\de}_2^{\mbox{\tiny{(F)}}}$ \\
 & & & & \\  \hline
 & & & & \\
$\Theta (1-p^2/\L_0^2)$ & $2$&1 & -  & -  \\
 & & & & \\
$\frac{\L_0^4}{(p^2+\L_0^2)^2} $&$ \frac23$ &$ \frac53$      
&  $-\frac7{30}$&
$-\frac13$ \\
 & & & & \\
$\exp{\left(-\frac{p^2}{\L_0^2}\right)} $ & 0 &$ \frac32$&$-\frac1{12}$ &
$-\frac13 $  \\

  &  &  &  &   \\
\hline
\end{tabular} 
\end{center}
The fine-tuning equation \re{fintun}, together with the results 
\re{deltaac} and \re{deltahac}, allows to find the values of
$\s_1(\L_0)$ and $\s_2(\L_0)$ which as a consequence depend on
the cutoff function. The finiteness of the results is due to the presence
in \re{vertexacfinal} and \re{grafacfer} of the two cutoff functions having 
almost non-intersecting supports (\ie $q^2 \gtrsim \L_0^2$, $(q-p)^2 
\lesssim \L_0^2$).
 
Also in this case, in order to check our calculation we have computed these 
relevant couplings at the physical point $\L=0$ using the evolution
equation \re{eveq} and the corresponding cutoff function. In the limit
\UV we find  $\s_1(0)=\s_2(0)=0$, as required by the fine-tuning equation
at the physical point $\L=0$ \cite{bdm3}-\cite{bdm5}.

The same analysis can be repeated for the other vertices 
of $\DG$ in order to fix the remaining couplings $\s_i(\L_0)$ in \re{pitilde}. 
Also these couplings are  finite, thanks to the same argument discussed
above.
However, we prefer to concentrate to the computation of the one-loop chiral 
anomaly, which yields a  cutoff-independent 
result. This will be the subject of the following section.

\section{The ABJ anomaly}
The ABJ anomaly represents the breaking of the classical chiral symmetry at 
the quantum level. One is obviously interested in theories where anomalies 
which affect currents coupled to propagating gauge fields
cancel. Nevertheless they have to be computed in order to test the
consistency of the regularization procedure. In this section we
concentrate upon a single left fermion (recall that in our
formulation right fermions are not coupled to the gauge field).
 
As the anomaly is absent at the tree-level,  the flow equation
(\ref{chie}) guarantees the $\L$-independence of $\DG$ at 
one loop.
Hence it may be convenient to compute the anomaly at $\L=\L_0$. 
There are  two relevant monomials of $\DG$,  
$\e\m\n\r\s \int \mbox{d}^4x\,
\mbox{Tr}\, \left[c\,\p_\m \,( A_\n\,\p_\r\,A_\s)\right]$ and 
$ \e\m\n\r\s \int \mbox{d}^4x\,
\mbox{Tr}\, \left[c\,\p_\m \,(A_\n\,A_\r\,A_\s)\right]$, 
which are absent 
in $\DGb$ but may be present in $\DGh$. This is due to the  locality of
$\Pi(\L_0)$, which in turn  implies that $\DGb$  is a trivial cocycle
of the cohomology of the BRS operator.
In other words a violation of the ST identity results in the impossibility of
fixing the relevant couplings $\s_i(\L_0)$ in $\Pi^{(1)}(\L_0)$ in such a
way the symmetry is restored, or, equivalently, some of the relevant parameters
in $\DG$ cannot be set to zero. Nevertheless, a consistency condition for them 
still holds (Wess-Zumino condition).

In the following  we compute the fermionic contribution to 
$\DGh$ at one-loop order which gives rise to the anomaly. 
Taking  the \UV limit in \re{dgh1ym} and setting  $\L=\L_0$, $\DGh$ 
becomes  
\beeq \nome{primodelta}
\DGh=\DGh^{YM}\!\!\!&+&\!\!
i\int_{p\,q} \frac{\Kin(q)}{q^2}\,\Biggl[\,\ds{q}_{\g \b}\, 
\frac{\de^2 \bG^{(0)}}{\de \psi_\a (-p) \de \bpsi_\b (-q)} \, 
\frac{\de^2 \Pi^{(0)}}{\de \psi_\g (q) \de \bl_\a (p)} \Kiu (p)\Biggr.
\\ \nonumber
& & \;\;\;\;\;\;\;\; + \Biggl.\,\psi\to\bpsi, \; \bl\to\l\;\Biggr]\, .
\eeeq 
In order to compute this functional we need only the tree-level
vertices of $\G$, \ie those of $S_{BRS}$, and in particular 
\beeq 
&&\G^{(\bpsi A \psi)}_{\m\,\a\b}(p,\,q,\,-p-q)=ig 
\left(\gm \frac {1-\gc} 2\right)_{\a\b}  \, ,  
\\  \nonumber
&&
\G^{(\bl c \psi)}_{\a \b}(p,\,q,\,-p-q) = \left(\frac{1-\gc}2
\right)_{\a\b} \, , \qquad
\G^{(\bpsi c \l)}_{\a \b}(p,\,q,\,-p-q)= \left(\frac{1+\gc}2
\right)_{\a\b} \,.
\eeeq
The fermionic contribution to the $c$-$A$-$A$ vertex of $\DGh$ is shown in
fig.~6.2. 
\begin{figure}[htbp]\nome{anomfig1}
\epsfysize=4.5cm
\begin{center}
\epsfbox{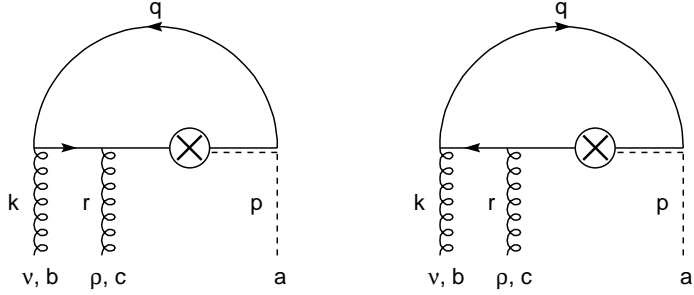}
\end{center}
\caption{\small{Fermionic contribution to the $c$-$A$-$A$ vertex of $\DGh$.
}}
\end{figure}
\newline
Notice that the line with the crossed circle in fig.~6.2 contributes 
only through the cutoff function $\Kiu$. 
From \re{primodelta} and \re{gammab}  the anomalous part (\ie  the part
giving rise to the antisymmetric tensor  $\e\m\n\r\s$) of this vertex reads 
\beq \nome{an1}
\int_{p\,k}c^a (p)\,A_\n^b (k)\,A_\r^c(r)
\left[\mbox{Tr}\,(\t,a\,\t,b\,\t,c) {\cal A'}^{(cAA)}_{\nu\r}(p,k,r)
+\mbox{Tr}\,(\t,a\,\t,c\,\t,b) {\cal A''}^{(cAA)}_{\nu\r}
(p,k,r)\right]\,,
\eeq
where $r=-p-k$ and 
\beeqn 
&&{\cal A'}^{(cAA)}_{\nu\,\r}(p,\,k,\,r)=
-\,{\cal A''}^{(cAA)}_{\r\,\n}(p,\,k,\,r)\\ \nonumber
&&= \frac{(ig)^2}2
\int_{q}\mbox{Tr}\, \left[\gc (-\ds{q})\,\gn\,(\ds{q}+
\ds{k})\,\gr\,\right] 
  \frac{\Kin (q)}{q^2} \,
\frac{\Kin (q+k)}{(q+k)^2}\,\Kiu (p-q)\,.
\\ \nonumber
\eeeqn
Performing the trace over Dirac matrices, one finds
\beq
{\cal A'}^{(cAA)}_{\nu\,\r}(p,\,k,\,-p-k)=
2\, i\,  g^2 \, \e\a\n\b\r  \,
\int_{q}(q_\a\,q_\b\,+\,q_\a\,k_\b)\, 
\frac{\Kin (q)}{q^2} \,
\frac{\Kin (q+k)}{(q+k)^2}\,\Kiu(p-q)\, .
\eeq
By expanding in the external momenta and taking into 
account the symmetry properties, we obtain 
\beq \nome{terpass}
{\cal A'}^{(cAA)}_{\n\r}\,(p,\,k,\,-p-k)\,=\,\frac{g^2}{16\pi^2}  
\,\e\a\n\b\r \,p_\a\,k_\b\,
\left[\int_0^\infty dx \, \Kin^2(x)\,\frac{\p \Kiu(x)}{\p x} +{\cal 
O}(P^2/\L_0^2)\,\right]\, ,
\eeq
with $x=q^2/\L_0^2$ and  $P$  some combination of the external momenta.
Notice that the result of the integral in \re{terpass} 
is determined only by the values  $\Kin(0)=0$ and $\Kin(\infty)=1$ and
therefore is independent of the choice of the cutoff function.
As a consequence, in the \UV limit, we recover the usual
contribution to the anomaly,
which is regularization independent. 
In the RG formulation this fact can be understood from the
$\L-$independence of $\DG^{(1)}$, so that the same
result is obtained if one computes the anomaly at 
the physical point $\L=0$. In this case the anomaly comes 
from $\DGb$ and is computed in terms of the physical vertices 
of $\G[\Phi,\g;\L=0]$, which are  regularization independent.

The fermionic contribution to the $c$-$A$-$A$-$A$ vertex of $\DGh$ is shown
in fig.~6.3.
\begin{figure}[htbp] \nome{anomfig2}
\epsfysize=4.5cm
\begin{center}
\epsfbox{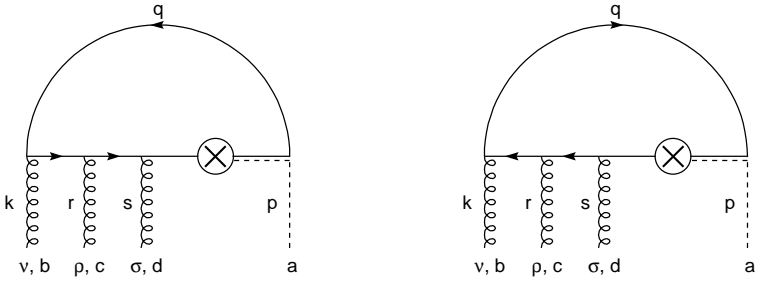}
\end{center}
\caption{\small{Fermionic contribution to the $c$-$A$-$A$-$A$ vertex of 
$\DGh$.
}}
\end{figure}
\newline
According to \re{primodelta} and \re{gammab} the  anomalous part of this vertex
reads 
\beeq \nome{an2}
\int_{p\,k\,r}c^a (p)\,A_\n^b (k)\,A_\r^c(r)\,A_\s^d(s)
\!\!\!&&\!\!\!\left[\mbox{Tr}\,(\t,a\,\t,b\,\t,c\,\t,d) {\cal A'}^{(cA
AA)}_{\nu\,\r\,\s}(p,\,k,\,r,\,s)\right.\\ \nonumber
\!\!\!&&\!\!\!+\left.\mbox{Tr}\,(\t,a\,\t,d\,\t,c\,\t,b) {\cal A''}^{(cA
AA)}_{\nu\,\r\,\s}(p,\,k,\,r,\,s)\right]\, ,
\eeeq
where $s=-p-k-r$ and 
\beeq 
{\cal A'}^{(cAAA)}_{\nu\,\r\,\s}(p,\,k,\,r,\,s)\!\!\!&=&\!\!\!
-\,{\cal A''}^{(cAAA)}_{\s\,\r\,\n}(p,\,-(s+p),\,r,\,-(k+p))
\\ \nonumber
\!\!\!&=&\!\!\!\frac{(ig)^3}2
\int_{q}\mbox{Tr}\, \left[\gc (-\ds{q})\,\gn\,(\ds{q}+
\ds{k})\,\gr\,(\ds{q}+\ds{k}+\ds{r})\,\gs \right] \\ \nonumber
\!\!\!&&\qquad \quad \times \frac{\Kin (q)}{q^2} \,
\frac{\Kin (q+k)}{(q+k)^2}\,\frac{\Kin (q+k+r)}{(q+k+r)^2}\,\Kiu (p-q)
\,.
\eeeq
Performing the trace over the Dirac matrices, we get
\beeq\nome{grafcaaa}
{\cal A'}^{(cAAA)}_{\nu\,\r\,\s}(p,\,k,\,r,\,s)\!\!\!&=&\!\!\!
g^3\,\e\n\r\a\s \,\int_q q^2\, (2\,q_\a\,+\,2\,k_\a\,+r_\a)\\
\!\!\!&\times&\!\!\!\frac{\Kin (q)}{q^2} \,
\frac{\Kin (q+k)}{(q+k)^2}\,\frac{\Kin (q+k+r)}{(q+k+r)^2}\,
\Kiu (p-q)\, . \no
\eeeq
Again, exploiting  symmetry properties and expanding into external 
momenta, we have
\beeq
&&{\cal A'}^{(cAAA)}_{\nu\,\r\,\s}(p,\,k,\,r,\,s)= \\ \nonumber
&&\frac{i\,g^3}{16\pi^2} \, \e\n\r\a\s\,\int_0^\infty dx 
\left[\frac13\, (2\,k\,+\,r)_\a\,
\frac{\p \Kin^3(x)}{\p x}\,\Kiu(x)\,-\,p_\a\,
\frac{\p \Kiu(x)}{\p x}\,\Kin^3(x)\right]\, ,
\eeeq
where $x=q^2/\L_0^2$ (terms of order ${\cal O}(P^2/\L_0^2)$ are
omitted).
As previously discussed, the integral over $x$ is independent of the
specific cutoff function and in the \UV limit is easily proven to give 
\beq
{\cal A'}^{(cAAA)}_{\nu\,\r\,\s}(p,\,k,\,r,\,s)=
\frac{i}{192\,\pi^2}
\, g^3 \, \e\n\r\a\s\,(2\,k\,+\,r\,+\,3\,p)_\a\,.
\eeq
Combining the two contributions as in (\ref{an2}), 
the $c$-$A$-$A$-$A$ part of the anomaly reads
\beq
\frac{i}{48\,\pi^2}\,g^3\,\e\m\n\r\s \, \int_{p\,k\,r} 
c^a (p)\,A_\n^b (k)\,A_\r^c(r)\,A_\s^d(s) \,
\mbox{Tr}\,(\t,a\,\t,b\,\t,c\,\t,d)\,p_\m\,.
\eeq

There could be in principle a $c$-$A$-$A$-$A$-$A$ vertex in $\DGh$ 
(this monomial is 
also relevant), but it is straightforward to show that the graphs which could 
give rise to such a term vanish because of the $\g$-trace.
Finally, in the coordinate space the anomaly 
has the well-known form \cite{abj}
\beq
{\cal A}=\frac{g^2}{24\,\pi^2}\, \e\m\n\r\s \int \mbox{d}^4x\,
\mbox{Tr}\, \left[c\,\p_\m(A_\n\,\p_\r\,A_\s\,+\,\frac{g}2\,A_\n\,A_\r\,A_\s
)\right]\,.
\eeq

To sum up, in this  chapter we have shown that, despite the loss of
chiral gauge invariance induced by the cutoff,  the Slavnov-Taylor
identities can be perturbatively recovered~\footnote{As long as the
theory is anomaly-free!} by solving, at the UV
scale, a fine-tuning equation which fixes all the couplings in the
bare action except the five $z_i$'s.
Then we could exploit such a freedom by fixing these undetermined couplings
at the physical point $\L=0$. 
In order to recover the complete UV action, in addition to the
fine-tuning procedure we have explicitly computed some of the couplings which,
apart from the $z_i$'s, have turned out to be finite. 
This is similar to what happens in dimensional regularization
where the definition of the matrix $\gc$ explicitly breaks the 
chiral symmetry by a term which lives in $\eps$ dimensions 
(the so-called evanescent vertex). Once this vertex is inserted
in primitively divergent Feynman diagrams (\ie generating poles in
$\eps$) it produces a finite contribution as $\eps\to 0$. Thus additional
finite counterterms must be introduced in order to restore the 
ST identities. This  fine-tuning involves all possible relevant 
interactions, since the matrix $\g_5$ couples left to right fermions.
On the contrary in our formulation the regularization breaks local 
gauge invariance but preserves global chiral symmetry, so that only 
globally invariant monomials are involved in the 
fine-tuning, which, though  unavoidable,  is thus simplified (the 
situation is even worse with lattice regularization due to
the additional breaking of Lorentz invariance). 

Finally, with a simple one-loop calculation, we have reproduced the
chiral anomaly, which has proven  to be universal, \ie independent
of the choice of the cutoff function, as it should.

\chapter{The Wess-Zumino model}
The aim of this chapter is the extension of the RG formulation to 
supersymmetric theories. Our basic model will be the simplest
supersymmetric theory that can be conceived, namely the Wess-Zumino model
\cite{wz}.

As in non-supersymmetric theories, perturbative calculations in field
 theories require a regularization
procedure to deal with ultraviolet divergences.  A powerful and simple
method is that of dimensional regularization which has the remarkable
feature of preserving gauge symmetry.  However, it is clear that this
regularization breaks supersymmetry, since fermionic and bosonic
degrees of freedom match only in fixed dimensions.  The only
modification of dimensional regularization compatible with
supersymmetry (the so-called dimensional reduction \cite{dred}) turns
out to be inconsistent \cite{incon}.  
The lack of a consistent
regularization scheme which manifestly preserves supersymmetry implies,
in particular, that superspace formalism can be used only with some
care since naive manipulations may lead to ambiguities
\cite{ambi}. Although dimensional reduction does not seem to cause any
practical difficulty and is extensively used \cite{jones} to perform
perturbative calculations, it is worthwhile to look for manifestly
supersymmetric regularizations which take advantage of the superspace
technique \cite{gates} and are free of ambiguities.

Now that the RG method has been successfully applied in a large
variety of non-supersymmetric models, our goal  will be its
generalization to  supersymmetric
theories, implementing the regularization in such a way that
supersymmetry is preserved. That this program is  practicable is
easily understood recalling
how the cutoffs are introduced in the RG formulation: we  split the
classical action into two parts, the quadratic and the interacting
one, and then multiplies the former by a cutoff function
$\K(p)$. Hence, in  the supersymmetric case it suffices to write
the classical action in terms of superfields and follow the same
procedure (in components this corresponds to use the same cutoff
function for all fields).  As our formulation works in $d=4$,
supersymmetry is maintained and, from the very beginning, we can
exploit the superspace technique, which simplifies perturbative
calculations and is now unambiguous.

We begin with the set up the RG
formalism for the massless  WZ model  and then, as an example
of how to perform perturbative calculations, we compute the one-loop
two-point function and discuss the boundary
conditions for the flow equation.
Our conventions are given in Appendix A.

\section{The RG flow for the Wess-Zumino model}

The massless WZ model is described by the classical Lagrangian \cite{wz}
$$
S_{cl}= S_2 + \si^{(0)}
$$
where 
\beeq &&
S_2=\frac{1}{16}\zint \bphi\; \phi\,, \qquad\qquad
\;\;\;\;\;\;\; \zint=\zzint\,,
\nonumber\\ &&
\si^{(0)}=
\frac{\l}{48}\,\cint  \phi^3 + \mbox{h.c.}
\eeeq
and $\phi$ ($\bphi$) is a chiral (anti-chiral) superfield satisfying 
$\bD_\ad\phi=0$ ($D^\a\bphi=0$).

As we have done a number of times so far, 
we regularize ultraviolet divergences by integrating out 
 the fields with frequencies smaller than a given
UV cutoff $\L_0$ in the path integral or, equivalently, 
we  modify the free
action to make the free
propagators vanish for $p^2 > \L_0^2$.

The generating functional of the theory is
\beq\nome{WWZ}
Z[J]=e^{iW[J]}=\int {\cal D}\Phi  \, \exp{i\lgr\half (\Phi, \,
\DD^{-1} \Phi)_{0\L_0}+(J,\Phi)_{0\L_0}+\si[\Phi;\L_0]\rgr}
\,,
\eeq
where we have collected the fields and the sources in $\Phi_i=(\phi,\,\bphi)$
and $J_i=(J,\,\bJ)$ respectively,
and introduced the general  cutoff scalar products between 
fields and sources
\beeq&&
\half (\Phi,\,\DD^{-1}\Phi)_{\L\L_0}\equiv\frac{1}{16}
\spint\, K^{-1}_{\L\L_0}(p)\,
\bphi(-p,\,\th)\,\phi(p,\,\th)\,,
\,\;\;\;\;
\spint \equiv \int \frac{d^4p}{(2\pi)^4} \,d^2\th \,
d^2\bt
\nome{prop}
\\&&
\nonumber
\\&&
(J,\Phi)_{\L\L_0}\equiv\frac{1}{16}
\spint\, K^{-1}_{\L\L_0}(p) \,
\lgr J(-p,\,\th)\,\frac{D^2}{p^2}\,\phi(p,\,\th)\,
+\bJ(-p,\,\th)\,\frac{\bD^2}{p^2}\,\bphi(p,\,\th)
\rgr
\,,
\eeeq
with $K_{\L\L_0}(p)$ a cutoff function which is one for $\L^2 < p^2 <
\L_0^2$ and rapidly vanishes outside~\footnote{The factors
$D^2/(16p^2)$, $\bD^2/(16p^2)$ are needed to write the chiral and
anti-chiral superspace integral respectively, as integrals over the
full superspace (see the appendix A
).}.  The introduction of such a cutoff
function in \re{prop} yields a regularized propagator which preserves
supersymmetry, this being a global, linearly realized
transformation. Hence the UV action $\si[\Phi;\L_0]$ in \re{WWZ}
contains all possible renormalizable supersymmetric interactions, \ie
superspace integrals of superfields and their covariant derivatives
which are local in $\th$.  Dimensional analysis tells us that they are
given by the monomials $\phi\bphi$, $\phi\,,\phi^2\,, \phi^3$,
$\bphi\,,\bphi^2\,, \bphi^3$, properly integrated.

Following  Wilson, we then   integrate over the fields with frequencies
$\L^2<p^2<\L_0^2$ and rewrite the generating functional in terms of
the Wilsonian effective action $\se[\Phi;\L]$
\beq\nome{Z'WZ}
e^{iW[J]}=N[J;\L]\int {\cal D}\Phi \, 
\exp{i\lgr \half(\Phi, \,\DD^{-1} \Phi)_{0\L}+(J,\Phi)_{0\L}
+\se[\Phi;\L]\rgr}
\,,
\eeq
where  $N[J;\L]$ contributes to the
quadratic part of $W[J]$. As usual, here and in the following
we explicitly write only the dependence on the cutoff $\L$, since the
theory is renormalizable and we are interested in the limit \UV.
As the regularization preserves supersymmetry, the functional $\se$
contains all possible supersymmetric interactions.  We are well aware $\se$
can be interpreted  as the generating functional of the connected amputated
cutoff Green functions --- except the tree-level two-point function
--- in which the free propagators contain $\L$ as an infrared cutoff
\cite{bdm}.  That is to say, the functional
\beq\nome{WLWZ}
W[J';\L]= 
\se[\Phi;\L]+\half (\Phi,\, \DD^{-1}\Phi)_{\L\L_0} \,,
\eeq   
with the sources $J'$ given by 
\beq\nome{J'WZ}
J_i'(-p,\th)=\K^{-1}(p)\,D^{2\ej}(\th)\Phi_j(-p,\th)\, \DD^{-1}_{ji}(p)\,,
\eeq
is the generator of the cutoff connected Green functions.  The matrix
$\DD^{-1}_{ij}$ is defined through \re{prop} and its entries are
$\DD^{-1}_{ij}=1/16$ if $i \ne j$ and zero otherwise. Moreover, in
order to keep formulas more compact, we have introduced the
two-component vector $\eps_k = \lp 1, -1\rp$ and the  shortened notation
$D^{-2}\equiv\bD^2$ which allow to treat simultaneously chiral and
anti-chiral fields.

\subsection{Evolution equation}
From a conceptual point of view there is nothing to add to the
procedure outlined in sections 1.1, 1.2 to derive the RG flow for a
supersymmetric model. However, if on one side the introduction of 
covariant derivatives, projectors, etc. 
enables us to have compact equations ---which will be directly exported
in supersymmetric gauge theories, on the other side some care must be
taken in writing them.
The requirement that the generating functional \re{Z'WZ} is independent 
of the IR cutoff $\L$ gives rise to a differential equation for the
Wilsonian effective action, the well known exact RG equation
\cite{p,b}, which can be translated into an equation for $W[J;\L]$ 
\beq\nome{eveqWWZ}
\L\p_{\L} W[J;\L]=\half\spint \L\p_{\L} \K^{-1}(p)\,\DD^{-1}_{ij}(p)
\lp
\frac{\de W}{\de J_i(-p,\th)}\frac{\de W}{\de J_j(p,\th)}
-i \,\frac{\de^2 W}{\de J_i(-p,\th) \de J_j(p,\th)}
\rp
\,.
\eeq

The following step consists in  introducing the cutoff effective
action which is given by the Legendre transform of $W[J;\L]$
\beq\nome{LegWZ}
\G[\Phi;\L]=W[J;\L]-\cint J\phi -\acint \bJ\bphi\,.
\eeq
This functional generates the cutoff vertex functions in which the 
internal propagators have frequencies in the range $\L^2<p^2<\L_0^2$ 
and reduces to
the physical quantum effective action in the limits $\L\to 0$ and \UV
\cite{bdm,mo,we}.

The evolution equation for the functional $\G[\Phi;\L]$ can be derived
from \re{eveqWWZ} by using \re{LegWZ} and inverting the functional
$\frac{\de^2 W}{\de J\de J}$.  As in \re{inversionw}, \re{wint}, this 
inversion can be performed
isolating the full two-point contributions $\G_2$ in the
functional $\G[\Phi;\L]$
\beeqn
&&(2\pi)^8 \frac{\de^2\G}{\de\Phi_j(p,\th_1)\,\de\Phi_k(k,\th)}
= (2\pi)^4 \G_{2\;kj}(k;\L) \, D^{-2\ek}(\th)\,  
D^{-2\ej}(\th_1)\,
\sde (k+p) \\ &&
\phantom{(2\pi)^8
\frac{\de^2\G}{\de\Phi_j(p,\th_1)\,\de\Phi_k(k,\th)}
}
+ \Gi_{kj}[\Phi;k,p;\L] 
\eeeqn
and  $W_2$ in $W[J;\L]$
\beeq\nome{inversionwWZ}
&&(2\pi)^8 \frac{\de^2 W}{\de J_k(-k,\th)\,\de J_{i}(q,\th_2)}
= (2\pi)^4 W_{2\;ik}(k;\L) \, D^{-2\ei}(\th_2)\,  
D^{-2\ek}(\th)\,
\sde (q-k)  \no \\
&& \phantom{
(2\pi)^8 \frac{\de^2 W}{\de J_k(-k,\th)\,\de J_{i}(q,\th_2)}
} + \Wi_{ik}[J;q,-k;\L]\,,
\eeeq
where the dependence on Grassmann variables in $\Gi$ and $\Wi$ is
understood. Henceforth we will prefer writing all integrals in the
full superspace, so that we have to cope with factors like
$\frac{D^2(\th)}{16 k^2}$ and $\frac{\bD^2(\th)}{16 k^2}$ originating
from chiral and anti-chiral projectors, respectively.  
These two factors can be simultaneously treated  with the help of the
vector $\ek$ and identifying $\lp\frac{D^2(\th)}{16k^2}\rp^{-1}$ with
$\frac{\bD^2(\th)}{16 k^2}$.

Then making use of the identity 
\beeq
\dfud{\Phi_i(-q,\,\th_2)}{\Phi_j(p,\,\th_1)} &=& D^{-2\ei}(\th_1)\, 
\sde (q+p) \, \de_{ij}
\nonumber \\
&=& (2\pi)^8 \skint 
\frac{\de^2 W}{\de J_k(-k,\th) \de J_{i}(q,\th_2)}
\lp \frac{D^2(\th)}{16 k^2}\rp^{\ek} 
\frac{\de^2\G}{\de\Phi_j(p,\th_1) \de\Phi_k(k,\th)}\no
\eeeq
we can express $\Wi_{ij}$ in \re{inversionwWZ} as a functional of 
$\Phi$ obtaining
\beq \nome{wintWZ}
\Wi_{ij}[J(\Phi);q,\,p;\L]= -\G_{2\;lj}^{-1}(p;\L)\lp 
\frac{D^2(\th_2)}{16\,q^2}\rp^{\ek}
\,\lp\frac{D^2(\th_1)}{16\,p^2}\rp^{\el}\,\bG_{kl}[\Phi;q,\,p;\L]\,
\G_{2\;ik}^{-1}(q;\L)
\,,
\eeq
where the auxiliary functional $\bG$ satisfies the recursive equation
\beq \nome{gammabWZ}
\bG_{ij}[\Phi;q,p;\L]= (-)^{\de_j}\Gi_{ij}[\Phi;q,p;\L]-\skint 
\lp\frac{1}{16k^2}\rp^{\abs{\ek}}
\,\Gi_{kj}[\Phi;k,p;\L]\,\G_{2\;lk}^{-1}(k;\L)\,\bG_{il}[\Phi;q,-k;\L]
\eeq
which gives $\bG$ in terms of the proper vertices of $\G$.  The
grassmannian parity $\de_j$ is zero for the (anti)chiral superfield
and the factor $(-)^{\de_j}$  has been introduced to take into 
account the possible anti-commuting nature of the field (it will be 
needed in SYM).

Finally, inserting \re{inversionwWZ}  in \re{eveqWWZ} and using \re{wintWZ},
we obtain the evolution equation for the functional $\G[\Phi;\L]$ 
\beeq \nome{eveqWZ}
&&\L\p_{\L}\lq \G[\Phi;\L]-
\half \,\spint \K^{-1}(p)\, \Phi_i(-p,\th)
\,\DD^{-1}_{ij}(p)\,\Phi_j(p,\th)
\rq
=-\frac i 2 \sqint \L\p_{\L} \K^{-1}(q)
\nonumber \\ 
&&\;\;\;\;\;\;\;\;\;\;\;\times\,
\G_{2\;lj}^{-1}(q;\L)\;
\DD^{-1}_{ji}\;
\G_{2\;ik}^{-1}(q\;\L)
\lp\frac{D^2(\th)}{16\,q^2}\rp^{\ek}
\lp\frac{D^2(\th)}{16\,q^2}\rp^{\el}
\bG_{kl}[\Phi;q,\,-q;\L]\,.
\eeeq
This equation, together with a set of suitable boundary conditions,
can be thought as an alternative definition of the theory which in
principle is non-perturbative. As far as we are concerned with its
perturbative solution, the usual loop expansion is recovered by
solving iteratively \re{eveqWZ}. We have already pointed out  such a
solution is possible since the
l.h.s. of \re{eveqWZ} at a given loop order depends only on lower loop
vertices. The proof of 
 perturbative renormalizability, \ie that the
\UV limit can be taken,  is a straightforward generalization of 
that given in sec.~2.2   for non-supersymmetric theories
\cite{p,b,bdm} .

\section{Relevant couplings and boundary conditions}

We are familiar enough to relevant couplings to appreciate via
dimensional analysis  they originate from the monomials 
$\phi\bphi$, $\phi\,,\phi^2\,, \phi^3$, $\bphi\,,\bphi^2\,, \bphi^3$,
properly integrated. 

The massless chiral multiplet two-point function 
(\ie the $\phi\bphi$-coefficient of the cutoff effective action)
\beq \nome{gamma2WZ}
\G_{2\;ij}(p;\L)=\DD^{-1}_{ij}\K^{-1}(p)+\S_{2\;ij}(p;\L)
\eeq
contains the relevant coupling 
$$
Z_{ij}(\L)=\S_{2\;ij}(p;\L) 
\left. \right|_{p^2=\mu^2}\,,
$$
where $\mu$ is some non-vanishing subtraction point, whose
introduction, being $\phi$ a massless field, is required to avoid 
the IR divergences.
Additional relevant couplings are contained in the
$\phi^3$, $\bphi^3$-coefficients of the cutoff effective action, \ie
$\G_{3\phi}(p,q,-p-q;\L)$ and $\G_{3\bphi}(p,q,-p-q;\L)$, 
and are defined by
$$
\s_3(\L)=\G_{3\phi}(p,q,-p-q;\L)\left. 
\right|_{3\mbox{\scriptsize{SP}}}\,,\quad\quad\quad
\bar\s_3(\L)=\G_{3\bphi}(p,q,-p-q;\L)\left. 
\right|_{3\mbox{\scriptsize{SP}}}\,.
$$
We need not define the remaining relevant couplings since the
corresponding monomials are not generated in perturbation theory.

All the vertices appearing  
with a number of $\phi\bphi$ larger than one are irrelevant.   
Further contributions to the irrelevant part of $\G$ comes from 
the two-point and three-point functions, and are given by 
$$
\S^{\mbox{\scriptsize{irr}}}_{2\;ij}(p;\L)\equiv\S_{2\;ij}(p;\L)-Z_{ij}(\L)\,,
$$
$$
\G_{3\phi}^{\mbox{\scriptsize{irr}}}(p,q,-p-q;\L)\equiv
\G_{3\phi}(p,q,-p-q;\L)-\s_3(\L)
$$
and
$$
\G_{3\bphi}^{\mbox{\scriptsize{irr}}}(p,q,-p-q;\L)\equiv
\G_{3\bphi}(p,q,-p-q;\L)-\bar\s_3(\L)\,.
$$

We assume that at the UV scale $\L=\L_0$ all irrelevant vertices
vanish. As a matter of fact $\G[\Phi;\L=\L_0]$ reduces to the bare
action, which must contain only renormalizable interactions  
in order to guarantee perturbative renormalizability. 

As to the relevant couplings, they  are fixed at the physical
point $\L=0$ in terms of the physical couplings, such as the wave 
function normalization, the three-point coupling and the mass. 
Hence the boundary conditions to be imposed  on the relevant couplings 
are 
\beq\nome{bc}
Z_{ij}(\L=0)=0 \,,\;\;\;\;\;\;\;\;
\s_3(\L=0)=\bar\s_3(\L=0)=\l\,.
\eeq

\subsection{Loop expansion}

\noindent{\it (i) Tree level}

\noindent
The starting point of the iteration is the tree-level 
interaction
\beq\nome{gammaintWZ}
\Gio_{ij}[\Phi;q,\,p;\L]=\frac{\l}{8}\, \de_{ij} \,\sppint
\de^4(\th_1-\th') D(\th')^{-2\ei}
\de^4(\th_2-\th')\,\Phi_j(p')\,\de^4(p+q+p')
\eeq
together with the tree-level two-point function 
$\G_{2\;ij}^{(0)}(p;\L)=\DD^{-1}_{ij} \K^{-1}(p)$. Inserting these
expressions in \re{gammabWZ} one obtains the tree-level functional 
$\bG_{ij}^{(0)}[\Phi]$.

\noindent{\it (ii) One-loop calculations}

\noindent
The evolution equation for the functional $\G[\Phi]$ at one-loop order can
be derived by writing the r.h.s of \re{eveqWZ}  in terms of the known objects
$\bG_{ij}^{(0)}[\Phi]$ and $\G_{2\;ij}^{(0)}$. One immediately realizes that
only the vertices with an equal number of $\phi$ and $\bphi$ are 
generated at this order. 

As an example we compute the one-loop two-point function.
The evolution equation for this vertex 
is determined by the $\phi\bphi$-coefficient in \re{eveqWZ} which, at the
tree level, originates only from the second term in the r.h.s. of 
\re{gammabWZ}, \ie 
$$
-\skint 
\,\Gio_{ml}[\Phi;k,\,q;\L]\,\frac{\K(k)}{16k^2}\DD_{nm}(k)\,
\Gio_{kn}[\Phi;-q,\,-k;\L]
\,.
$$ 
Next, substituting \re{gammaintWZ} in the expression above and
carrying out some standard $D$-algebra manipulations (reported in the
appendix), we find
\beeq\nome{gamma21}
&& \spint \bphi (-p,\,\th)\, \L \p_\L \S_2^{(1)} (p;\, \L)
 \, \phi  (p,\,\th)\,=
\frac i{64}\l^2\,
\spqint
\frac {\K(p+q)\L \p_\L\K(q)}{q^2(p+q)^2}\, \nonumber \\
&&\;\;\;\;\;\;\;\;\;\;\;\;\;\times\,
\bphi (-p,\,\th_1)\, \phi  (p,\,\th_2)\, \de^{4} (\th_1-\th_2)\,
\bD^2 D^2(q,\,\th_2)  \, \de^{4} (\th_1-\th_2)\,.
\eeeq
Notice that eq.~\re{eveqWZ} describes only the evolution of the interacting
part of $\G$, since the tree level in \re{gamma2WZ} cancels out.

Recalling the property
\beq\nome{susydelta}
 \de^{4} (\th_1-\th_2)\,\bD^2\,D^2  \, \de^{(4)} (\th_1-\th_2)=
 \de^{4} (\th_1-\th_2)\,,
\eeq
we get
\beq\nome{sigma2WZ}
\L \p_\L  \S_2^{(1)} (p;\, \L) =\frac i{128} \,\l^2\,
\int \frac{d^4 q}{(2\pi)^4}
 \; \frac {\L \p_\L(\K(q)\,\K(p+q))}{q^2\,(p+q)^2}\,.
\eeq
Implementing  the  boundary conditions \re{bc}, the solution of 
\re{sigma2WZ} at the physical point $\L=0$ and in the \UV limit is 
$$
\S_2^{(1)} (p;\, \L=0) =\frac i{128} \,\l^2\,
\int \frac{d^4 q}{(2\pi)^4}
 \; \lp\frac {1}{q^2\,(p+q)^2}-\left.\frac
{1}{q^2\,(p+q)^2}\right|_{p^2=\mu^2}\rp
\,.
$$
Notice the crucial role of the boundary condition for $Z_{ij}$, \ie 
$Z^{(1)}_{ij}(0)=0$,
which naturally provides the necessary 
subtraction to make the vertex function $\S_{2\;ij}$ finite for \UV. 
Conversely  we can see from power counting that the remaining 
irrelevant vertices (\ie  the coefficients of
$(\phi\bphi)^n$ with $n>1$) are finite,
and no subtraction is needed.
This property holds at any order in perturbation theory \cite{bdm}.

Before switching to the analysis of the $N=1$ super Yang-Mills theory,
we should comment on non-renormalization theorem for
the WZ model.
In our framework it  can be derived with no substantial modification with 
respect to the
standard proof \cite{grisaru}.
At the first loop it is straightforward recognizing that chiral
superfield interactions of the type $\cint \lp \z\,\phi +
m\,\phi^2+\l\,\phi^3
\rp $ do not
receive any finite or infinite perturbative contributions.   
As a matter of fact, inserting \re{gammabWZ} at the tree level in
\re{eveqWZ} we can  see that only vertices with an equal number of
chiral and anti-chiral fields acquire one-loop corrections.  However
in the massless case there are violations to this theorem.  In
ref. \cite{jjw} it was explicitly shown that the chiral interaction
$\phi^3$ receives a finite contribution at the two-loop level.  In
fact elementary power counting tells that this vertex stays finite at
any loop order.  The same result can be derived in our formalism.

\chapter{ Supersymmetric Gauge Theories and Gauge Anomalies}

The previous chapter was devoted to the implementation of the RG
formalism in supersymmetric theories. The WZ model was our laboratory
and we were able to regularize the theory in such a way supersymmetry
is preserved.
This holds also
for a supersymmetric gauge theory, but in this case
gauge symmetry is explicitly broken by the regularization.  As for
non-supersymmetric gauge theories, we will show that by properly
fixing the boundary conditions of the RG flow the ST identity
associated to the gauge symmetry is recovered, when the matter
representation is anomaly free. However, if the matching conditions
for the anomaly cancellation are not fulfilled, we will be able to
reproduce the chiral anomaly.

\section{$\bom{N=1}$ Super Yang-Mills}

The super Yang-Mills (SYM) action reads~\cite{sym-orig} (the
conventions are those of~\cite{piglect})
$$
S_{\mbox{\scriptsize{SYM}}}
= -\frac{1}{128g^2}\tr\cint \WW^\a \WW_\a\,, \;\;\;\;\;\;\;\;
\WW_\a = \bD^2\lp e^{-gV}D_\a e^{gV}\rp\,,
$$
where $V(x,\th)$ is the $N=1$ vector supermultiplet which belongs to
the adjoint representation of the gauge group $G$. In the matrix
notation $V=V^a\tau_a$, 
with  the matrices $\tau_a$ satisfying 
$[\tau_a,\tau_b]=if_{abc}\tau_c\,$,$\tr\tau_a\tau_b=\de_{ab}$.
The classical action is invariant under the gauge transformation
\beq
e^{gV'} = e^{-i\bchi}e^{gV} e^{i\chi}\, ,\quad\quad\quad  \bD_\ad\chi=0\,,
\quad  D^\a\bchi=0\,,
\eeq
where $\chi=\chi^a\tau_a$.

In order to quantize the theory we have to fix the gauge and choose a
regularization procedure. From what we have seen so far it should be
manifest that the introduction of the cutoff
does not spoil global symmetries as long as they are linearly
realized. If this is not the case the transformation of the quadratic
part of the action mixes with the transformation of the rest (recall that the
cutoff function multiplies only the quadratic part of the classical action).
Therefore, we shall choose a supersymmetric gauge fixing
instead of the familiar Wess-Zumino one in which the
supersymmetry transformation is not linear.

As described in ref.~\cite{gates}, we add  to the action a 
gauge fixing term which is a supersymmetric extension of the
Lorentz gauge and the corresponding Faddeev-Popov term
\beeq
&&\sgf=-\frac{1}{128\a}\tr\zint D^2V\bD^2V
\nonumber\\
&&\sfp= -\frac{1}{8}\tr\zint\lp\cm +\bcm\rp \lq
\half L_{gV}(\cp+\bcp)+ \half 
\lp L_{gV}{\rm coth}(L_{gV}/2)\rp\lp\cp-\bcp\rp \rq
\nonumber
\\&&\phantom{\sfp}=
 -\frac{1}{8}\tr\zint\lp\cm +\bcm\rp\lq 
\cp -\bcp +\half g\,[V,\cp+\bcp]+\cdots \rq\,,
\eeeq 
where  the ghost $\cp$ and the anti-ghost $\cm$ are chiral fields,
like the gauge  parameter $\chi$, and $L_{gV}\,\cdot=[gV,\cdot]$.
The classical action
$$
\scl=S_{\mbox{\scriptsize{SYM}}}
+\sgf+\sfp
$$
is invariant under the BRS transformation 
\beeq
&&
\de V = \eta \lq
\half L_{gV}(\cp+\bcp)+ \half 
\lp L_{gV}{\rm coth}(L_{gV}/2)\rp\lp\cp-\bcp\rp \rq \,,
\nonumber \\[1mm]&&
\de\cp= -\eta\, \cp^2\,, \;\;\;\;\;\;\;\;\;\;\;\;\;\;\;\;\;\;\;\;
\de\bcp=-\eta \,\bcp^2\,,
\nonumber \\&&
\de\cm= - \eta \frac1{16\a}\bD^2 D^2V \,, \;\;\;\;\;
\de\bcm= - \eta \frac1{16\a} D^2\bD^2 V\no
\eeeq
with $\eta$ a Grassmann parameter. Introducing the sources 
$\g_i=(\gv$, $\g_{\cp}$,
$\g_{\bcp})$, 
associated to the BRS variations of
the respective superfields, the BRS action in the Fermi-Feynman gauge
($\a=1$) reads 
\beeq \nome{brsacsym}
&&\sbrs =\scl +\zint 
\gv \lq \half L_{gV}(\cp+\bcp)+ \half 
\lp L_{gV}{\rm coth}(L_{gV}/2)\rp\lp\cp-\bcp\rp \rq  
\nonumber \\&&
\phantom{\sbrs =\scl}
-\cint \g_{\cp} \cp^2 - \acint \g_{\bcp}\bcp^2 
\nonumber \\&& 
\phantom{\sbrs}
= S_2 +\si^{(0)}
\eeeq
with 
$$
S_2=\zint\lq\frac{1}{16}V\partial^2V +\frac{1}{8}\lp\cm\bcp -\bcm\cp\rp\rq\,.
$$
Notice that in \re{brsacsym} we did not introduce the BRS 
sources for $\cm$ and $\bcm$
since one can show that the effective action depends on these fields
and the source $\gv$ only through the combination
$$
\tgv=\gv-\frac18\lp \cm+\bcm\rp\,.
$$

As described in the previous section for the WZ model ---and for
non-supersymmetric theories, we
regularize the UV divergences multiplying the free propagators 
by a cutoff function $\K$, so that the generating functional 
$Z[J,\g]$ can be written as in \re{WWZ}
with 
$$
\Phi_i= (V,\, \cp,\,         \bcm,    \,  \cm,   \, \bcp)\,,\;\;\;\;\;
J_i=(J_V, \, \xi_{-}+\bD^2\gv,\, -\bxi_{+},\, -\xi_{+},\, \bxi_{-}-
D^2\gv)
$$
and the cutoff scalar product between fields and sources given by
\beeq\nome{phiphi}
&&
(\Phi,\,\DD^{-1}\Phi)_{\L\L_0}=\spint \K^{-1}(p)
\lgr-\frac{1}{16}V(-p,\th)\,p^2V(p,\th) \right.
\nonumber\\&&
\phantom{(\Phi,\,\DD^{-1}\Phi)_{\L\L_0}=\spint \K^{-1}(p)}\left.
+\frac{1}{8}\lq\cm(-p,\th)\bcp(p,\th)
-\bcm(-p,\th)\cp(p,\th)\rq\rgr
\eeeq
and
\beeq\nome{jphi}
&&(J,\Phi)_{\L\L_0}=\spint \K^{-1}(p)\lgr J_V(-p,\th)\,V(p,\th)+
\frac1{16}\lq
\lp\xi_{-}+\bD^2\gv\rp(-p,\th)\,\frac{D^2}{p^2}\cp(p,\th) 
\nonumber \right. \right. \\ && 
\phantom{(J,\Phi)_{\L\L_0}=\spint \K^{-1}(p)}
+\frac{\bD^2}{p^2}\bcm(-p,\th)\,\bxi_{+}(p,\th)+\frac{D^2}{p^2}\cm(-p,\th)
\,\xi_{+}(p,\th)
\nonumber \\ && 
\phantom{(J,\Phi)_{\L\L_0}=\spint \K^{-1}(p)}
+\left. \left.
\lp\bxi_{-}-D^2\gv\rp(-p,\th)\,\frac{\bD^2}{p^2}\bcp(p,\th)\rq \rgr\,. 
\eeeq
The UV action 
$\si[\Phi,\g;\L_0]$ contains all possible relevant interactions 
written in  terms of $\Phi_i$, $\g_i$ and superspace derivatives,
which are invariant under Lorentz and global gauge transformations. 
Notice that at the tree level all quadratic contributions in the
fields and sources are gathered in \re{phiphi} and \re{jphi}. 

Afterwards we integrate over the fields with frequencies
$\L^2<p^2<\L_0^2$ and the result is the  analogue of \re{Z'WZ}
where the Wilsonian effective action $\se[\Phi,\g;\L]$ depends also on
the BRS sources.
The generating functional of the cutoff connected Green functions
$W[J,\g;\L]$ is given by \re{WLWZ} and \re{J'WZ} with 
$\ek$ the five-component vector $\ek=(0,1,-1,1,-1)$ and 
the matrix $\DD^{-1}_{ij}$ defined through \re{phiphi}.  
This matrix turns out to be block-diagonal and its entries are
$1/8(-p^2,\, \eps_{\mbox{\tiny{AB}}},\,  
\eps_{\mbox{\tiny{AB}}})$, ${\mbox{\footnotesize{A}}}=(+,-)$, 
with  $\eps_{\mbox{\tiny{AB}}}=-\eps_{\mbox{\tiny{BA}}}$ and 
$\eps_{+-}=1$.
The derivation of the evolution equation for the functional $W$
exactly follows that of the WZ model presented in sect.~7.1.1. 
Finally the cutoff effective action $\G$ 
\beeq \nome{Legsym}
&&\G[\Phi,\g;\L]=W[J,\g;\L]-\zint J_V V -\cint \lp \xi_-\cp + \cm
\xi_+ \rp  
\nonumber \\ &&
\phantom{\G[\Phi,\g;\L]=W[J,\g;\L]}
-\acint  \lp \bxi_-\bcp + \bcm \bxi_+ \rp 
\eeeq
evolves according to \re{eveqWZ} with the appropriate vertices, $\DD_{ij}$ 
and $\ek$. 

\subsection{Matter fields}
When adding matter fields to the pure  super Yang-Mills action one
gets SQCD, the supersymmetric generalization of QCD.
Matter is described by a set of chiral superfields 
$\phi^{\mbox{\tiny{I}}}(x,\th)$  which
belong to some representation $R$ of the gauge group. 
Their BRS transformation reads
$$
\de \phi^{\mbox{\tiny{I}}}= -\eta\,\cp^a \,
T_a{}^{\mbox{\tiny{I}}}{}_{\mbox{\tiny{J}}} \,\phi^{\mbox{\tiny{J}}} \equiv
-\eta(\cp\phi)^{\mbox{\tiny{I}}}  
\,,\quad 
\de \bphi_{\mbox{\tiny{I}}} =  \eta\,\bphi_{\mbox{\tiny{J}}} 
\,T_a{}^{\mbox{\tiny{J}}}{}_{\mbox{\tiny{I}}}\,\bcp^a 
\equiv \eta(\bphi\, \bcp)_{\mbox{\tiny{I}}} \, ,
$$
where the hermitian matrices $T_a$ are the generators of the gauge
group in the representation $R$. 

The BRS action for the matter fields is
\beq \nome{matter-action}
S_{\rm matter} = \frac{1}{16} \zint  \bphi\, e^{gV^a T_a}\phi - 
\cint \g_\phi\,  \cp\,\phi + \acint \g_{\bphi}\, \phi\,\bcp  
\eeq
plus a possible superpotential $W$ having the general form 
$W(\phi) = \frac{1}{8}m_{({\mbox{\tiny{IJ}}})}\phi^{\mbox{\tiny{I}}} 
\phi^{\mbox{\tiny{J}}} + \l_{({\mbox{\tiny{IJK}}})}\phi^{\mbox{\tiny{I}}} 
\phi^{\mbox{\tiny{J}}}
\phi^{\mbox{\tiny{K}}}$,
the mass matrix $m_{{\mbox{\tiny{IJ}}}}$ and the Yukawa coupling constants 
$\l_{{\mbox{\tiny{IJK}}}}$
being invariant symmetric tensors in the representation $R$.

Developing the RG formalism in presence of matter
fields is straightforward once we have replaced the sets of fields and sources
with 
\beeq 
&&
\Psi_i= (V,\, \cp,\,         \bcm,    \,  \cm,   \, \bcp\,\phi,\,\bphi)
\,,\;\;\;\;\; \g_i=(\gv\,,\g_{\cp}\,,\g_{\bcp}\,,\g_\phi\,,\g_{\bphi})\,,
\nonumber \\ 
&&
J_i=    (J_V, \, \xi_{-}+\bD^2\gv,\, -\bxi_{+},\, -\xi_{+},\, \bxi_{-}-
D^2\gv,\,J,\,\bJ)\,.
\eeeq
The evolution equation for the effective action 
has the usual form \re{eveqWZ}, with a natural redefinition of $\eps_k$
and $\DD^{-1}_{ij}$  to take into account matter fields
(\eg $\,\eps_k = \lp 0,\,1,\, 1,\, -1,\, -1,\, 1, -1 \rp$).

\subsection{Boundary conditions} 
As discussed in subsect.~7.2 we first distinguish between relevant and 
irrelevant vertices.  The relevant part of the
cutoff effective action involves full superspace integrals 
of monomials in the fields, sources and derivatives local in $\th$ and
with dimension not larger than two 
\beq\nome{gammarelsym}
\Gr[\Psi,\g;\s_i(\L)]=\sum_i \s_i(\L)\,P_i[\Psi,\g]\,,
\eeq
where the sum is over the monomials $P_i[\Psi,\g]$ invariant under
Lorentz and global gauge transformations. Due to the dimensionless nature of
the field $V$ this sum contains infinite terms which can be classified
according to the number of gauge fields.
The couplings $\s_i(\L)$ can be expressed in terms of the 
cutoff vertices at a given subtraction point, generalizing the
procedure used in  subsect.~7.2 to define the coupling $Z_{ij}(\L)$    
(see  also \cite{bdm3}-\cite{bdm5} for the technique  of
extracting the relevant part from a given functional with a
non-vanishing subtraction point in the non-supersymmetric Yang-Mills case).

As usual, the  boundary condition we impose on  the 
irrelevant part of the cutoff effective action, \ie the remnant,
is that it vanishes at $\L=\L_0$.
For $\L=\L_0$, then, the cutoff effective action becomes ``local'',
\ie an infinite sum of local terms,  and
corresponds to the UV action $\si[\Psi,\g;\L_0]$, with the bare
couplings given by  $\s_i(\L_0)$. 

The way in which the boundary conditions for the relevant couplings
$\s_i(\L)$ are determined is not straightforward and closely follows
the procedure introduced in chapter 6 for chiral gauge theories. In
sect.~7.2 we fixed
them at the physical point $\L=0$ in terms of the value of the
physical couplings (such us the normalization of the chiral field).
We have seen that in the case of a gauge theory, as the one we are 
considering, there
are interactions in \re{gammarelsym} which are not present in $\sbrs$, so
that only some of the relevant couplings are connected to the physical
couplings (such as the wave function normalizations and the
three-vector coupling $g$ at a subtraction point $\mu$).  For instance
the contribution to \re{gammarelsym} with two gauge fields consists of
three independent monomials 
$$
\zint \tr \lq \s_1\, V\,V 
+ \s_2\,V\,D^\a\,\bD^2\,D_\a\, V 
+ \s_3\,V\,D^2\,\bD^2\,V\rq
$$
instead of the two in $\sbrs$.
Therefore, in order to fix the boundary conditions for all the
relevant couplings, we need the additional  fine-tuning procedure which
implements  the gauge symmetry at the physical point.
However, this analysis involves non-local functionals and 
is highly not trivial.
Alternatively we can discuss the symmetry at the ultraviolet scale
and determine the cutoff-dependent $\s_i(\L=\L_0)$'s.  In this case
the discussion is
simpler, since all functionals are relevant, but  we have to perform a
perturbative calculation (\ie to solve the RG equations)  to obtain
the physical couplings. 

As we did for the chiral gauge theory, we consider the second possibility,
although  the wave function normalizations and the gauge coupling $g$
at a subtraction point $\mu$ are still set at $\L=0$. As a matter of
fact  there are combinations of the monomials in \re{gammarelsym} which 
are not involved in the fine-tuning, so that the corresponding 
couplings are free and can be fixed at the physical point $\L=0$.
Before explaining the details of the fine-tuning procedure we recall how 
to implement the gauge symmetry in the RG formulation.

\section{Effective ST identity}

The gauge symmetry requires that the physical effective action
satisfies the ST identity \cite{brs,becchi0}
\beq\nome{ST0sym}
\SS_{\G'}\G'[\Psi,\g]=0\,,
\eeq
where
$\G'[\Psi,\g]=\G[\Psi,\g]+\frac1{128} \tr \zint D^2 V\bD^2 V$
and~\footnote{From now on the sum over the fields in $\Psi$ will not include
$\cm$ and $\bcm$.}
\beq \nome{STsym}
\SS_{\G'}=\spint
\lq\lp \frac{D^2}{16p^2} \rp^{\ei}
 \dfud{\G'}{\Psi_i(-p)}\,
\dfud{}{\g_i(p)} +
\lp \frac{D^2}{16p^2} \rp^{\ei} 
\dfud{\G'}{\g_i(p)}\,
\dfud{}{\Psi_i(-p)}\rq
\eeq
is the Slavnov operator. In sect.~3.2 we showed the ST identity can be
directly formulated for the Wilson effective action $\se$ at any
$\L$. We give here a sketchy derivation for the specific case of $N=1$
SYM.
Consider the generalized BRS transformation
\beq\nome{susybrs}
\de\Psi_i(p)=\Ki(p)\,\eta\,\dfud{\st}{\g_i(-p)}\,, \;\;\;\;\;
\de\cm= - \eta \frac1{16}\bD^2 D^2V \,, \;\;\;\;\;
\de\bcm= - \eta \frac1{16} D^2\bD^2 V\,,
\eeq
where $\eta$ is a Grassmann parameter and $\st$ is the total action
(\ie $\se$ plus the source and the quadratic terms in \re{Z'WZ}). 
Performing such a change of variable in the
functional integral \re{Z'WZ}, one deduces the following identity
\beq \nome{STeffsym}
\SS_J Z[J,\g]=N[J,\g;\L] \int \DD\Psi \exp{i\lgr
\half(\Psi,\DD^{-1} \Psi)_{0\L}+(J,\Psi)_{0\L} 
+\se[\Psi;\L]\rgr} \, \De[\Psi,\g;\L]\, ,
\eeq
where $\SS_J$  is the usual ST operator
$$
\SS_J=\spint J_i(p)\,(-)^{\de_i}\dfud{}{\g_i(p)}+
\frac1{16} \spint \lq D^2 \xi_+ (p)+\bD^2 \bxi_+ (p)\rq
\dfud{}{J_V(p)}
$$
with ${\de_i}$  the source ghost number, and the functional $\De$ reads:
$$
\De[\Psi,\g;\L]=
i \spint  \Ki(p)\,\exp{(-i\se)}\lgr \dfud{}{\Psi_i(p)}
\,\dfud{}{\g_i(-p)}\rgr \exp{(i\se)}\qquad\qquad\qquad\;\;
$$
$$
-i\spint  \lq \Psi_i(p)\,\DD^{-1}_{ij}(p)\,
\dfud{}{\g_j(p)}+(\cp-\bcp)(p)\dfud{}{V(p)}
-\frac1{16} V(p)\Bigl(D^2 \dfud{}{\cm(p)}+\bD^2  \dfud{}{\bcm(p)}\Bigr)
\rq \se
\,.
$$
Whereas the l.h.s of the identity \re{STeffsym} arises from the variation of
the source term $(J,\Psi)_{0\L}$, the functional $\De$ originates from
the Jacobian of the transformation \re{susybrs} and from
the variation of the rest of $\st$.
Restoration of symmetry, $\SS_J Z[J,\g]=0$, translates into 
$$
\De[\Psi,\g;\L]=0 \;\;\;\; {\mbox{for any}} \;\;\L\,.
$$
However,  an analogous condition can be formulated in terms of $\DG$, the
Legendre transform of $\se$, in which reducible
contributions are absent. 
Recalling \re{WLWZ} and \re{J'WZ} which relate $\se[\Psi,\g;\L]$ 
to $W[J,\g;\L]$, and using \re{LegWZ}, \re{Legsym} we find 
$$
\DG[\Psi,\g;\L]=-\spint\lq \Kiu(p) \lp \frac{D^2(\th_1)}{16\,p^2}
\rp^{\ei} 
\dfud{\G'}{\Psi_i(-p)}\, \dfud{\G'}{\g_i(p)}
-\frac{\Ki(p)}{\K(p)}\,\DD^{-1}_{ij}(p)\,
 \Psi_i(p) \dfud{\G'}{\g_i(p)}\rq
$$
\beeq
&&
-i\,\hbar \spqint \frac{\Ki(p)}{\K(p)}\,\DD^{-1}_{ij}(p)
\lp \frac{D^2(\th_2)}{16 q^2}
\rp^{\ek} (-)^{\de_i}
\frac{\de^2 W}{\de J_i(p) \de J_k(q)} \qquad\qquad
\no\\&&
\phantom{-i\hbar \spqint \frac{\Ki(p)}{\K(p)} }
\times\frac{\de^2}{\de\Psi_k(-q) \de \g_j(-p)} 
\lp \G -\zint \gv (\cp-\bcp) \rp \,,
\eeeq 
where $\de^2W/\de J\de J$ is that functional of $\Psi$ and $\g$
appearing in the inversion \re{inversionwWZ} and \re{wintWZ}.
Finally, after performing such an inversion, 
the cutoff ST identity reads
\beq\nome{DGsym}
\DG[\Psi,\g;\L]\equiv\DGb +\DGh=0\,,
\eeq 
with
\beq \nome{dgb'sym}
\DGb=
-\spint \Kiu(p) 
\lp \frac{D^2(\th_1)}{16p^2} \rp^{\ei} 
\dfud{\G'}{\Psi_i(-p)}\dfud{\G'}{\g_i(p)}
+\spint \frac{\Ki(p)}{\K(p)}\DD^{-1}_{ij}(p)
\Psi_i(p) 
\dfud{\G'}{\g_i(p)}
\eeq
and
\beeq\nome{dghsym}
&&\DGh = i \hbar \spqint \Ki(p)  
\lp \frac{D^2(\th_1)}{16p^2}\rp^{\el}
\Bigg\{
\frac{(-1)^{\de_l} }{\lp 16q^2 \rp^{\abs{\ej}}} 
\lp \G_2^{-1}(q;\L) \,\bG(-q,-p;\L)\rp_{jl} 
-\de_{jl}\, 
\sde(p-q) \Bigg\}
 \nonumber \\
&&\phantom{\DGh = i\hbar} 
\times\lp \G_2^{-1}(p;\L) \, \DD^{-1}(p) \K^{-1}(p)\rp_{li} \,
\frac{\de^2}{\de\Psi_j(q)\, \de \g_i(p)} \lp \G -\zint \gv (\cp-\bcp) \rp\,.
\eeeq
Notice that at $\L=0$ the cutoff ST identity reduces to $\DGb(0)=0$
and, in the UV limit, becomes the usual ST identity
\re{ST0sym}. Moreover 
we have 
inserted the factor $\hbar$ in \re{dghsym} to put into evidence that $\DGh$ 
vanishes at the tree level.

In terms of the already familiar functional $\Pi$, expressed by
\re{defpi},
the functional  $\DGb$ can be rewritten as  
$$
\DGb[\Psi,\g;\L]=
-\spint \Kiu(p)\, \lp \frac{D^2(\th_1)}{16\,p^2}
\rp^{\ei} 
\dfud{\Pi'[\Psi,\g;\L]}{\Psi_i(-p)}\, \dfud{\Pi'[\Psi,\g;\L]}{\g_i(p)}\,,
$$
where 
$\Pi'$ is the expression obtained by removing the gauge fixing term in
$\Pi$. Thus, in the \UV limit, with the help of \re{slavphys} we have
\beq\nome{dgbsym}
\DGb[\Psi,\g;\L]\to\SS_{\Pi'(\L)}\Pi'(\L) \quad\quad{\mbox{for}}\;\;
\L_0\to\infty
\eeq
at any $\L$. The existence of such a limit is guaranteed in
perturbation theory by the UV finiteness of the cutoff effective
action (perturbative renormalizability). In order to show this
property holds also for $\DGh$, it suffices to recognize that the
presence of cutoff functions having almost non-intersecting supports
forces the loop momenta in \re{dghsym} to be of the order of $\L$.
Henceforth we will take the \UV limit in $\DG$.  

\subsection{Perturbative solution of $\DG=0$}
The proof of the ST identity \re{DGsym} in the RG formalism, with
possible anomalies, is based on induction in the loop number and
closely follows that of non-supersymmetric gauge theories discussed in
chapters 3, 5, 6 \cite{bv,susy}.  For the sake of completeness we resume
here the key issues.

We have  shown  that the evolution of the vertices of $\DG$ at
the loop ${\ell}$ depends on vertices of $\DG$ itself at lower loop
order \cite{mt}, so that if $\DG^{(\ell')}=0$ at any loop order $\ell'<\ell$,
then
\beq\nome{chiesym}
\LdL\DG^{(\ell)}=0\,.
\eeq
Thus we can analyse $\DG$ at an arbitrary value of $\L$. 
There are two natural choices 
corresponding to $\L=0$ and $\L=\L_R$ much bigger than the subtraction
scale $\mu$, \ie $\L_R=\L_0$. With the former the gauge symmetry condition 
fixes the relevant part of the effective action 
in terms of the physical coupling $g(\mu)$ and provides the boundary
conditions of the RG flow, whereas with the latter 
the gauge symmetry condition determines  the cutoff dependent 
bare couplings.
With this choice the implementation of symmetry is simplified
due to the locality~\footnote{Here and in the following
locality means that each term in the expansion of the functionals in
the gauge field $V$ contains only couplings with non-negative
dimension.} of the 
functionals involved. Although the computation of physical 
vertices is generally cumbersome, this second possibility is more 
convenient in the computation of quantities which do not evolve with
the cutoff $\L$, such as the gauge anomaly.
This is the reason why we will adopt the second possibility.

We now discuss the vanishing of $\DG$.
Also  for this functional we define its relevant part, isolating all 
supersymmetric monomials in the fields, sources and their derivatives 
with ghost number one and dimension three. The rest is included in $\DGi$.

At the UV scale $\DG$ is local, or, more precisely, 
$\DGi(\L_0)={\cal O}(\frac1{\L_0})$,
so that the irrelevant contributions disappear in the \UV limit.  This
can be understood with the  same argument we gave in the non-supersymmetric
case (see sec.~3.3). Then \re{chiesym} ensures the locality of
$\DG(\L)$ at any $\L$.

Once the locality of $\DG(\L)$ is shown, the solvability of the
equation $\DG(\L)=0$ can be proven using cohomological methods
\cite{brs,becchi0,piguet}.  This is a consequence of the $\L$-independence
of $\DG$ and the solvability of the same equation at $\L=0$, where the
cohomological problem reduces to the standard one.

Henceforth we will consider the first loop, the generalization to
higher loops being straightforward due to the iterative nature of the
solution.
Using \re{dgbsym}, at  $\L=\L_0$ and at the first loop  \re{DGsym} reads 
\beq \nome{symfintun}
{\cal S}_{\Pi^{(0)}}\,\Pi^{(1)}(\L_0)
\,+\,\DGhr^{(1)}(\L_0)=0\,.
\eeq
This fine-tuning equation allows to fix some of the relevant 
couplings in $\Pi^{(1)}(\L_0)$. As a matter of fact 
the most general functional $\Pi^{(1)}(\L_0)$ can be cast into the form 
\re{gammarelsym} and  split into two contributions
\beq \nome{pigreco1sym}
\Pi^{(1)}(\L_0)=\Piinv^{(1)}(\L_0)+\Pit^{(1)}
(\L_0)\, ,
\eeq
where $\Piinv$ contains all the independent monomials which are
invariant, \ie ${\cal S}_{\Pi^{(0)}}\, \Piinv^{(1)}=0$. The explicit
form of $\Piinv^{(1)}$ is obtained from $\sbrs$ in \re{brsacsym} and 
\re{matter-action} with the replacement 
$$
(V\,,\,\g_i\,,\,\cp\,,\,\bcp\,, \,g\,,\,\phi\,,\,\bphi)\to 
(\sqrt{z_1}\,V\,,\, \sqrt{z_2}\,\g_i\,,\, \sqrt{z_2\,}\cp\,,\,
\sqrt{z_2}\,\bcp\,,\, z_3 g\,,\, \sqrt{z_4}\,\phi\,, \,\sqrt{z_4}\,\bphi)\,.
$$
The remaining monomials contribute to $\Pit$.
Inserting \re{pigreco1sym} into \re{symfintun}, we find
$$
{\cal S}_{\Pi^{(0)}}\, \Pit^{(1)}(\L_0)\,=-\, \DGh^{(1)}(\L_0)\,,
$$
which yields the couplings in $\Pit^{(1)}$ since $\DGh^{(1)}(\L_0)$
depends only on $\sbrs$.  An explicit calculation shows that the only
divergences are powers of $\L_0$ according to the dimension of the
relative vertex. In particular dimensionless couplings are finite, due
to the presence in \re{dghsym} of cutoff functions having almost
non-intersecting supports~\footnote{See sec.~5.3.1 for the explicit
computation of some of these couplings in non-supersymmetric QCD.}.

As to the couplings $z_i(\L_0)$, which are not involved in the
fine-tuning, we are allowed to set them equal to their physical values
at $\L=0$, \ie $z_i(0)=1$. In the standard language this corresponds
to the renormalization prescriptions.

Instead of solving the fine-tuning equation and determine the 
(cutoff-dependent) couplings of the UV action, in the next section
we will deal with the computation of the
gauge anomaly,  which well illustrates how the method works and
meanwhile is a cutoff independent result. At one loop such 
independence is guaranteed by the absence of the anomaly at the tree level
and by the evolution equation \re{chiesym}.

\section{Gauge anomaly}

For N=1 SYM within the superspace approach it has been demonstrated
\cite{piguet} that the only possible anomaly is the
supersymmetric extension of the standard Adler-Bardeen anomaly
\cite{abj} and its explicit form is given in ref.~\cite{tonin,GGS}.
As well known, its structure is non-polynomial \cite{tonin,FGPS} and
can be expressed as an infinite series in the gauge field $V$. 
In the following we restrict ourselves to the first term
of this expansion, since higher order polynomials can be inferred \cite{FGPS}
using the consistency condition \cite{brs,becchi0} which, at this order, 
forces the one-loop anomaly $\AA^{(1)}$ to obey 
${\cal S}_{\Pi^{(0)}}\AA^{(1)}=0$.

In our framework a violation of the ST identity results in the
impossibility of fixing the relevant couplings $\s_i(\L_0)$ in
$\Pi^{(1)}(\L_0)$ in such a way that \re{symfintun} is satisfied.  In
other words, this happens when there are relevant monomials in $\DGh$
which are not trivial cocycles of the cohomology of the BRS operator.

As a first step we write $\DGh$ at one loop order. Performing the \UV limit
in \re{dghsym} and setting $\L=\L_0$, we have
\beeq&&
\DGh^{(1)}= i \spqint   \Kiu(p) 
\lq\lp \frac{1}{16q^2} \rp^{\abs{\ej}}
\Kin(q) (-)^{\de_i}
\DD_{jk}(q)
\bG_{ki}^{(0)}(-q,-p;\L) - \de_{ij}\de^8(p-q)\rq
\no\\
&& \ph{\DGh^{(1)}= i \spqint   \Kiu(p) }
\times\lp \frac{D^2(\th_1)}{16p^2}\rp^{\ei}
\frac{\de^2}{\de\Psi_j(q)\, \de \g_i(p)} 
\lp \sbrs -\zint \gv (\cp-\bcp) \rp\,.
\eeeq
Then we isolate the matter contribution in $\DGh^{(1)}$ which, depending on
the representation of the matter fields, can  possibly give rise to the 
anomaly
\beeq \nome{sprimod}
\DGh^{(1)}=\DGh^{\mbox{\scriptsize{SYM}}\,(1)}\!\!\!&+&\!\!
i\spqint \Kiu (p) \frac{\Kin(q)}{q^2}\,
\Biggl[
\frac{\de^2 \bG^{(0)}}{\de \phi(-p) \de \bphi(-q)} \, 
\frac{D^2(\th_1)}{16p^2}
\frac{\de^2 \sbrs}{\de \phi(q) \de \g_\phi(p)} \Biggr. 
\no \\ 
& & \;\;\;\;\;\;\;\; + \Biggl.\,D\to\bD,\; \phi\to\bphi, \;
\g_\phi\to\g_{\bphi}\;\Biggr]
\,. 
\eeeq 
\begin{figure}[htbp]
\epsfysize=6.5cm
\begin{center}
\epsfbox{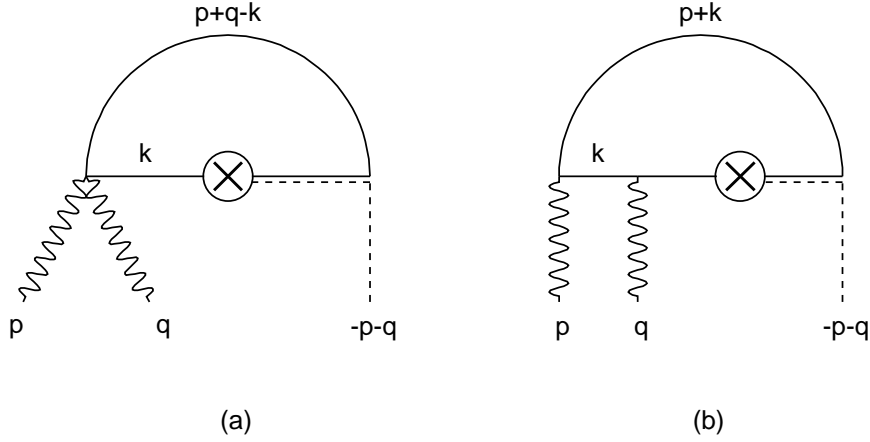}
\end{center}
\caption{\small{Matter contribution to the $\cp$-$V$-$V$ vertex of $\DGh$.
The wavy, dashed and full line denotes the vector, ghost and matter
fields respectively; the double line represents the BRS source
associated to the matter field. The cross denotes
the insertion of the cutoff function $\Kiu$ in the product of the
$\cp$-$\phi$-$\g_\phi$ vertex of $\sbrs$ with: (a) the irreducible
$\bphi$-$V$-$V$-$\phi$ vertex of $\bG$; (b) the reducible
$\bphi$-$V$-$V$-$\phi$ vertex of $\bG$.  All external momenta are
incoming and integration over the loop momentum is understood.  }}
\end{figure}
\newline
Inserting \re{gammab} in \re{sprimod} and extracting the tree-level
vertices of $\bG$ from $\sbrs$, we  see that the matter
contribution to the $\cp$-$V$-$V$ vertex of $\DGh$ is made of two
pieces, as shown in fig.~1.  
The first, originating from the irreducible
part of the $\bphi$-$V$-$V$-$\phi$ vertex of $\bG$, 
is given by
\beeq\nome{cvva}&&
-\frac{ig^2}{32}
\spqint \tr[\cp (-p-q,\th_1) V(p,\th_2) V(q,\th_2)]
\int\frac{d^4k}{(2\pi)^4}
\frac{\Kiu(k)\Kin(p+q-k)}{(p+q-k)^2}\, 
\no\\&&
\phantom{-\frac{ig^2}{32}
\spqint \tr} 
\times \de^4 (\th_1-\th_2) \,\bD^2 D^2 \de^4 (\th_1-\th_2)
\eeeq
and, as suggested from the graph depicted in fig.~1a which is not
typically triangle-shaped, does not contribute to the anomaly.  As a
matter of fact, by restricting to the Yang-Mills sector, we
immediately recognize that the anti-symmetric tensor $\eps_{\m\n\r\s}$
can not be generated from such a term.  Indeed using \re{susydelta}
and performing the loop integration, the expression in \re{cvva}
becomes
$$
g^2
\int\frac{d^4p\,d^4p\,d^4\th}{(2\pi)^8}\lgr(a_1\,\L_0^2+ a_2\,(p+q)^2)
\tr\lq\cp (-p-q,\th) V(p,\th) V(q,\th)\rq 
+\OO ((p+q)^4/\L_0^2)\rgr 
$$ 
where the $a_i$'s are finite
cutoff-dependent numbers which can be explicitly computed once the
cutoff function is specified.  The finiteness of such coefficients is
due to the presence of cutoff functions having almost non-intersecting
supports, \ie $k^2\lesssim \L_0^2$ and $(p+q-k)^2\gtrsim \L_0^2$.
These two monomials belong to the trivial cohomology of $\SS_\G$ and
their coefficients, together with those stemming from analogous
monomials of $\DGh^{\mbox{\scriptsize{SYM}}}$, fix the parameters in
$\Pit^{(1)}$ via \re{symfintun}.

We turn now to the contribution associated to the 
graph represented in fig.~1b, which originates from the second term in
the iterative expansion of $\bG$ in vertices of $\G$.
It reads
\beeq\nome{cvvb}&&
i\frac{g^2}{256}\,
\spqint \int\frac{d^4k}{(2\pi)^4}
\tr[\cp (-p-q,\th_2) V(p,\th_1) V(q,\th_2)]\,
\frac{\Kiu(k-q)\Kin(p+k)\Kin(k)}{k^2(k+p)^2} 
\no\\&&
\phantom{i\frac{g^2}{256}\,
\spqint \int\frac{d^4k}{(2\pi)^4}\tr}
\times
\bD^2 D^2(k,\th_1)\, \de^4 (\th_1-\th_2) \,
D^2 \bD^2(k+p,\th_1)\, \de^4 (\th_1-\th_2)\,.
\eeeq
After integrating the $\bD^2 D^2$ derivatives by parts and using the 
algebra of covariant derivatives (reported in  Appendix A)  
and \re{susydelta}, we find that 
the only non-vanishing terms in  \re{cvvb} are
\beeq\nome{cvvb'}
&&
i\frac{g^2}{256}\,
\int\frac{d^4p\, d^4q \,d^4\th}{(2\pi)^8}\int\frac{d^4k}{(2\pi)^4}\;
\frac{\Kiu(k-q)\Kin(k)\Kin(p+k)}{k^2(k+p)^2}
\\&&\phantom{i\frac{g^2}{256}\,}
\times
\tr \lq\cp (-p-q,\th)
\lp\lp  \bD^2 D^2+ 8  k_{\a\ad}\bD^\ad D^\a+16 k^2\rp  V(p,\th)\rp\, 
V(q,\th)\rq\,.\no
\eeeq
By performing the loop integration we find out that
the first and the third term in the trace  generate
only monomials which belong to the trivial
cohomology of $\SS_\G$, \ie
\beeq&&
g^2\int\frac{d^4p\,d^4q\,d^4\th}{(2\pi)^8}\lgr a_3
\tr\lq\cp (-p-q,\th) \lp\bD^2\,D^2 V(p,\th) \rp\,V(q,\th)\rq\right.
\no\\&&
\phantom{ g^2\int\frac{d^4p\,d^4q\,d^4\th}{(2\pi)^8} } \left.
+ (a_4\,\L_0^2+ a_5\, P^2 )\tr\lq\cp (-p-q,\th) V(p,\th)V(q,\th)\rq
+\OO (P^4/\L_0^2)\rgr\no
\eeeq
where $P$ is some combination of the momenta $p$ and $q$ and the
$a_i$'s are finite cutoff-dependent numbers.
We are now left with the second term in the trace in \re{cvvb'}.
Exploiting symmetry properties and expanding into external momenta we obtain
\beq \nome{anomaly1}
\frac{g^2}{1024\,\pi^2}
\int\frac{d^4p\,d^4q\,d^4\th}{(2\pi)^8}
\tr[ \cp (-p-q,\th) \lp\bD^\ad D^\a V(p,\th)\rp\,V(q,\th)]\,
(q_{\a\ad}\,I_1+  p_{\a\ad}\,I_2)
\eeq
where
\beeq&&
I_1=
\int_0^{\infty} dx \,\Kin^2 (x)\dpad{\Kiu(x)}{x} +\OO(P^2/\L_0^2)\no\\&&
I_2=\int_0^{\infty} dx \lq
\Kin^2(x)\dpad{\Kiu(x)}{x} + \frac{\Kin^2(x)}{x}\Kiu(x)\rq
+\OO(P^2/\L_0^2)\no
\eeeq
with $x=k^2/\L_0^2$ and $P$ as above.  Notice that in the \UV limit
$I_1$ yields a cutoff independent number, \ie $-1/3$, since it is
determined only by the values $\Kin(0)=0$ and $\Kin(\infty)=1$. On the
contrary $I_2$ depends on the choice of the cutoff function.

In \re{anomaly1} the structure proportional to $p_{\a\ad}$ does not
contribute to the anomaly, basically because, in the coordinate space,
all derivatives act on the same superfield.  On the other hand, had it
played a role in determining the anomaly, like all other contributions
analyzed above, our method would have led to an inconsistent result,
as $I_2$ and the $a_i$'s depend on the cutoff function.  Hence, only
the term with $q_{\a\ad}$ can generate a genuine anomaly.  By setting
$I_1=-1/3$ in \re{anomaly1} we get
\beq \nome{anomaly2}
\frac{g^2}{3072\,\pi^2}
\int\frac{d^4p\, d^4q \,d^4\th}{(2\pi)^8}
\tr[\cp (-p-q,\th) \,
\lp\bD^\ad D^\a V(p,\th)\rp\,q_{\a\ad}\,V(q,\th)]
\eeq
which has the true structure of the anomaly.  

The $\bcp$-$V$-$V$ vertex of $\DGh$ can be derived repeating the
steps described above. Also in this case we can identify the anomalous
contribution by isolating its cutoff independent part, which turns out to be
\beq \nome{anomaly3}
-\frac{g^2}{3072\,\pi^2}
\int\frac{d^4p\, d^4q \,d^4\th}{(2\pi)^8}
\tr[\bcp (-p-q,\th) \,
\lp D^\a \bD^\ad V(p,\th)\rp q_{\a\ad}\,V(q,\th)]\,.
\eeq
Finally, summing up  \re{anomaly2} and \re{anomaly3}, and switching to 
the coordinate space, the anomaly has the well-known form
\beq \nome{anomaly}
\AA=\frac{g^2}{6144\,\pi^2}
\zint\lp
\tr[\cp \,\bD^\ad D^\a V\,\{D_\a,\bD_{\ad}\}V]
-\tr[\bcp\,D^\a \bD^\ad V \,\{D_\a,\bD_{\ad}\}V]\rp\,.
\eeq
As a remark, we notice that in order to reproduce the standard abelian
anomaly in non-supersymmetric QCD we should perform the integration
over the grassmannian variables, identify the ghost $c$ with
$\cp+\bcp$ and replace $g$ with $2g$ to recover the usual
gluon-fermion coupling (see eq.~\re{matter-action}).  Then we would find
that the coefficient of the monomial $\eps^{\m\n\r\s}\tr[\p_\m c\, \p_\r
A_\n A_\s]$ is exactly $g^2/(24\pi^2)$.

In this chapter we have considered supersymmetric (gauge) theories within the
RG approach. Although we restricted to the WZ model and N=1 SYM, the
formalism is developed in such a way it can be applied to any
supersymmetric theory with an arbitrary field content and with
extended supersymmetry.
An advantage of the RG formulation is that the regularization is
implemented by introducing a cutoff in the loop momenta which makes all
the Green functions UV finite. This means one need not analytically
continue the Feynman integrals in the space-time dimension $d$, which
is kept fixed (in our case $d=4$). Therefore both the equality of
bosonic and fermionic degrees of freedom is safe --a necessary
condition for supersymmetry-- and the superspace technique presents no
ambiguity, for instance in handling the algebra of covariant
derivatives, traces of $\s$ matrices and using Fierz identities.

Unfortunately, in the RG approach the presence of the cutoff explicitly
breaks gauge symmetry. This is an unavoidable consequence of the
absence of a regularization scheme that manifestly preserves both
supersymmetry and BRS invariance, which in turn is intimately related
to the existence of the chiral anomaly.
However, we showed that the Slavnov-Taylor identity for the
physical effective action of an anomaly-free theory
is perturbatively recovered by solving the fine-tuning equation
\re{DGsym} at the UV scale. 
Such a procedure was sketched in subsect.~8.2.1.
On the other hand, in case of unfulfilled matching conditions for the anomaly
cancellation,
we have reproduced the supersymmetric 
chiral anomaly by a simple one-loop
calculation.
We performed a one-loop analysis, but the procedure systematically
generalizes to higher order. 

As well known, in the superspace formulation of SYM 
one has to face with the problem of infrared singularities,
due to the appearance of the pseudoscalar field $C(x)$, the $
\theta = 0 $ component of the gauge superfield (this difficulty is 
obviously circumvented in the Wess-Zumino gauge
\cite{wzgauge}, where the field $C$ is absent).
To avoid this problem one can assume \cite{piglect} that all fields are
made massive by adding suitable supersymmetric mass terms in the
action.  Since these masses break BRS invariance, the corresponding
Slavnov-Taylor identity holds only in the asymptotic region of
momentum space.

A prominent feature of  our formulation is the presence of the IR
cutoff $\L$ which
naturally makes
all cutoff vertices IR finite for $\L\ne 0$. Furthermore, for a
non-supersymmetric massless theory we have proven in sec.~2.3, by
induction in
the number of loops \cite{IR}, that the vertex functions without
exceptional momenta are finite for $\L\to 0$, once the relevant
couplings are defined in terms of cutoff vertices evaluated at some
non-vanishing subtraction points. In this proof the convergence of
loop integrals for $\L\to 0$ is simply controlled by the number of
soft momenta in the vertices which appear in the iterative solution of
the RG equation \re{eveqWZ}. Therefore we believe its generalization to
the supersymmetric case presents no difficulty.

Finally, though we restricted our analysis to the perturbative regime,
the RG formulation is in principle non-perturbative and provides a
natural context in which to clarify the connection between exact
results and those obtained in perturbation theory. In particular, it
would be interesting to consider issues such as the anomaly puzzle and
the violation of holomorphicity \cite{shifman} and reinterpret the recent
results on these topics \cite{ahm} in the RG approach.
\begin{appendix}
\chapter{Supersymmetric conventions}

The notations and conventions are those of~\cite{piglect}. 
Given a  Weyl spinor  $\psi_\a$,  $\a =1,2$,  indices can be raised and lowered
as follows
$$
\psi^\a=\eps^{\a\b}\psi_\b\, ,\quad \psi_\a=\eps_{\a\b}\psi^\b\, ,
$$
with  
$$
\eps^{\a\b}=-\eps^{\b\a}\, ,\quad\eps^{12}=1\, ,\quad
  \eps_{\a\b}=-\eps^{a\b}\, ,\quad \eps^{\a\b}\eps_{\b\g}=\de^\a_\g\, ,
$$
(the same for dotted indices). The 
summation convention is $\psi\chi=\psi^\a\chi_\a$ and 
$\bpsi\bchi=\bpsi_\ad\bchi^\ad$.

The matrices $\s^\m$ with lower indices are
$$
\s^\m_{\a\bd}=(\identity, \s^i)_{\a\bd}\,,
$$ 
where the $\s^i$'s are the Pauli matrices, whereas those with 
upper indices are
$$
\bs_\m^{\ad\b}=\s_\m^{\b\ad}
    =\eps^{\b\a}\eps^{\ad\bd}\s_{\m\,\a\bd}\,.
$$
A vector superfield $V(x,\th,\bt)$ has the following expansion
\beeq&&
V(x,\th,\bt) = C(x) + \th\chi(x) + \bt\bar\chi(x) + \half\th^2M(x)
+ \half\bt^2\bar M(x)  \no\\&&
\phantom{V(x,\th,\bt)}
+ \th\s^\m\bt A_\m(x) + \half\bt^2\th\l(x)
+ \half\th^2\bt\bar\l + \frac 1 4  \th^2\bt^2D(x)\, ,
\eeeq
where the components are ordinary space-time fields. 
A chiral (anti-chiral) superfield $\phi$  ($\bphi$)
expanded in component fields is
\beeq&&
\phi(x,\th,\bt) = 
  e^{-i\th\s^\m\bt\p_\m}\lp \phi(x) + \th\psi(x) + \th^2 F(x)\rp\,
\no\\&&
\bphi(x,\th,\bt) =
e^{i\th\s^\m\bt\p_\m}\lp \bar \phi(x) + \bt\bpsi(x) + \bt^2 \bar F(x)\rp\,.
\eeeq
The components of a vector superfield
transform under supersymmetry as
\beq\begin{array}{ll}
\de_\a C = \chi                
            &\bar\de_\ad C = \bar\chi \\[2mm]
\de_\a \chi^\b = \de_\a^\b M   
            &\bar\de_\ad {\bar\chi}^\bd=-\de_\ad^\bd\bar M\\[2mm]
\de_\a\bar\xi_\ad = \smuaad (A_\m+i\p_\m C) 
            &\bar\de_\ad\xi_\a = -\smuaad (A_\m-i\p_\m C) \\[2mm]
\de_\a M = 0  
            &\bar\de_\ad \bar M = 0 \\[2mm]
\de_\a \bar M = \l_\a-i(\s^\m\p_\m\bar\chi)_\a
            &\bar\de_\ad  M = \bar\l_\ad+i(\p_\m\chi\s^\m)_\ad \\[2mm]
\de_\a A_\m = \half(\s_\m\bar\l)_\a - \frac{i}{2}(\s^\n\bs_\m\p_\n\chi)_\a\quad
            &\bar\de_\ad A_\m = \half(\l\s_\m)_\ad 
                     + \frac{i}{2}(\p_\n\bar\chi\bs_\m\s^\n)_\ad \\[2mm]
\de_\a\l^\b = \de_\a^\b D + i(\s^\n\bs^\m)_\a{}^\b\p_\n A_\m
            &\bar\de_\ad{\bar\l}^\bd = -\de_\ad^\bd D 
                     + i(\bs^\m\s^\n)^\bd{}_\ad \p_\n A_\m \\[2mm]
\de_\a\bar\l_\ad = i\smuaad\p_\m M  
            &\bar\de_\ad\l_\a = i\smuaad\p_\m\bar M  \\[2mm]
\de_\a D = -i(\s^\m\p_\m\bar\l)_\a
            &\bar\de_\ad D = i(\p_\m\l\s^\m)_\ad\,.
\end{array}
\eeq
For the components of the chiral and anti-chiral superfields one has
\beq\begin{array}{ll}
\de_\a \phi = \psi_\a          &\bar\de_\ad \bphi = \bpsi_\ad   \\[2mm]
\de_\a\psi^\b = 2\de_\a^\b F   &\bar\de_\ad\bpsi^\bd = -2\de_\ad^\bd\bar F \\[2mm]
\de_\a F = 0                 &\bar\de_\ad\bar F = 0  \\[2mm]
\de_\a\bphi = 0              &\bar\de_\ad \phi= 0  \\[2mm]
\de_\a\bpsi_\ad = 2i\smuaad\p_\m\bphi \quad
                             &\bar\de_\ad\p_\a = 2i\smuaad\p_\m \phi  \\[2mm]
\de_\a\bar F = -i(\s^\m\p_\m\bpsi)_\a \quad
                             &\bar\de_\ad F = i(\p_\m\p\s^\m)_\ad \,.
\end{array}
\eeq
The covariant derivatives,  defined 
such as to anti-commute with the supersymmetry
transformation rules,  are given by
\beq
D_\a  =  \dpad{}{\th^\a} - i\smuaad\bt^\ad\p_\m \, ,\quad
\bD_\ad =  -\dpad{}{\bt^\ad} + i\th^\a\smuaad\p_\m \, .
\eeq
They obey the algebra
\beq
\lgr D_\a,\bD_\ad \rgr = 2i\smuaad\p_\m
\eeq
(the other anti-commutators vanish). Useful relations these covariant
derivatives satisfy are
\beeq&&
[D_\a,\bD^2]=4i(\s^\m\bD)_\a\p_\m\, ,\quad
[\bD_\ad,D^2]=-4i(D\s^\m)_\ad\p_\m \no\\&&
[D^2,\bD^2]=8iD\s^\m\bD\p_\m+16\p^2 = 
           -8i\bD\bs^\m D\p_\m-16\p^2\no\\&&
D\bD^2D=\bD D^2\bD\no\\&&
D\bD_\ad D = -\half\bD_\ad D^2-\half  D^2\bD_\ad\, , \quad
\bD D_\a \bD = -\half D_\a \bD^2-\half  \bD^2D_\a\,.
\eeeq

The superspace integral
of a superfield $V$, or of a (anti)chiral superfield 
$\phi$ ($\bar \phi$) is given by
\beq
\zint V=\int d^4x\, D^2\,\bD^2 V\,,\quad
\cint  \phi=\int   d^4x\, D^2 \phi\,,\quad
\acint \bar \phi=\int d^4x\, \bD^2\bphi\,,
\eeq
the integral with respect to the Grassmann variable $\th$
being defined by the derivative $\p/\p\th$.

The following  operators
\beq
P^{\rm T} = \frac{D\bD^2D}{8\p^2}\, , \quad
P^{\rm L} = -\frac{D^2\bD^2+\bD^2D^2}{16\p^2} 
\eeq
are projectors. In particular, $P^{\rm L}$ can be used to write
integrals of chiral (or anti-chiral) superfields as integrals over the
full superspace measure (recall that only for this measure
the integration by parts holds). For instance 
$\int\frac{d^4p\,d^2\th}{(2\pi)^4}\, \phi= \spint \frac{D^2}{16p^2}\,\phi$.

The delta function is defined by
$$
\de^8(z_1-z_2) = \de^4(\th_1-\th_2) \,\de^4(x_1-x_2) =
\frac{1}{16}(\th_1-\th_2)^2\,(\bt_1-\bt_2)^2
\,\de^4(x_1-x_2)\,.
$$
The functional derivatives are 
\beq
\dfud{V(z_1)}{V(z_2)}=\de^8(z_1-z_2)\,,
\quad
\dfud{\phi(z_1)}{\phi(z_2)}=\bD^2\de^8(z_1-z_2)\,,\quad
\dfud{\bphi(z_1)}{\bphi(z_2)}=D^2\de^8(z_1-z_2)\,.
\eeq

Finally, in order to separate the trivial cocycles from the anomaly in
\re{sprimod}, it can be useful to switch to components. Then for
the non-supersymmetric YM sector the anomaly is proportional to
$\eps_{\m\n\r\s}$, which is generated by the following trace
$$
\tr\lq
\s^\m\bs^\n\s^\r\bs^\ta\rq = 2\lp g^{\m\n}g^{\r\ta} + g^{\n\r}g^{\m\ta} -
 g^{\m\r}g^{\n\ta}
  - i\eps^{\m\n\r\ta}\rp \,.
$$  
\end{appendix}

\end{document}